\documentclass[useAMS,usenatbib]{mn2e}

\usepackage{verbatim} 
\usepackage{natbib} 
\usepackage{amsmath} 
\usepackage{amsbsy}
\usepackage{amssymb}
\usepackage{mathrsfs} 
\usepackage{lscape} 
\usepackage{graphicx}
\usepackage{epstopdf}
\usepackage{deluxetable} 
\usepackage{fixltx2e} 

\title[Feedback-regulated star formation]{Feedback-regulated star formation in molecular clouds and galactic discs}

\author[Faucher-Gigu\`ere, Quataert \& Hopkins]{Claude-Andr\'e Faucher-Gigu\`ere\thanks{Miller Fellow;
cgiguere@berkeley.edu}, Eliot Quataert, Philip Hopkins \vspace*{6pt}  \\ 
Department of Astronomy and Theoretical Astrophysics Center, University of California, Berkeley, CA 94720-3411, USA}

\begin{document}
\maketitle


\begin{abstract}
We present a two-zone theory for feedback-regulated star formation in galactic discs, consistently connecting the galaxy-averaged star formation law with star formation proceeding in giant molecular clouds (GMCs). 
Our focus is on galaxies with gas surface density $\Sigma_{\rm g} \gtrsim 100$ M$_{\odot}$ pc$^{-2}$, where the interstellar medium (ISM) can be assumed to be fully molecular. 
This regime includes most star formation in the Universe and our basic framework can be extended to other galaxies. 
In our theory, the galactic disc consists of Toomre-mass GMCs embedded in a volume-filling ISM. 
Radiation pressure on dust disperses GMCs and most supernovae explode in the volume-filling medium. 
A galaxy-averaged star formation law is derived by balancing the momentum input from supernova feedback with the vertical gravitational weight of the disc gas. 
This star formation law is in good agreement with observations for a CO conversion factor depending continuously on $\Sigma_{\rm g}$. 
We argue that the galaxy-averaged star formation efficiency per free fall time, $\epsilon_{\rm ff}^{\rm gal}$, is only a weak function of the efficiency with which GMCs convert their gas into stars, $\epsilon_{\rm int}^{\rm GMC}$. 
This is possible because the rate limiting step for star formation is the rate at which GMCs form: for large 
efficiency of star formation in GMCs, the Toomre $Q$ parameter obtains a value slightly above unity so that the GMC formation rate is consistent with the galaxy-averaged star formation law.  
We contrast our results with other theories of turbulence-regulated star formation and discuss predictions of our model. 
Using a compilation of data from the literature, we show that the galaxy-averaged star formation efficiency per free fall time is non-universal and increases with increasing gas fraction, as predicted by our model. 
We also predict that the fraction of the disc gas mass in bound GMCs decreases for increasing values of the GMC star formation efficiency. This is qualitatively consistent with the smooth molecular gas distribution inferred in local ultra-luminous infrared galaxies and the small mass fraction in giant clumps in high-redshift galaxies. 
\end{abstract}

\begin{keywords} 
Galaxies: formation, evolution, ISM, starburst, high-redshift -- stars: formation
\end{keywords}

\section{Introduction}

\subsection{Observations of star formation in galactic discs and giant molecular clouds}
Galaxies turn their gas into stars remarkably slowly.  
Averaged over galactic discs, only $\epsilon_{\rm ff}^{\rm gal} \sim 0.01$ of the gas mass is converted into stars per free fall time in ordinary galaxies (Kennicutt 1998; Genzel et al. 2010)\nocite{1998ApJ...498..541K, 2010MNRAS.407.2091G}. 
A similar result is found when the average is limited to spatial scales $\gtrsim 200$ pc within galaxies \citep[e.g.,][]{2007ApJ...671..333K, 2008AJ....136.2846B, 2009ApJ...704..842B}. 
However, star formation is not this slow down to arbitrary scales. 

In their study of local molecular clouds, \cite{2009ApJS..181..321E} found that the clouds lay a factor of $\sim20$ above the relationship between star formation rate surface density ($\dot{\Sigma}_{\star}$) and gas surface density ($\Sigma_{\rm g}$) measured by \cite{1998ApJ...498..541K} on galactic scales (hereafter, the Kennicutt-Schmidt [KS] law).\footnote{We note that different slopes have been inferred observationally for the star formation law \citep[e.g.,][]{2008AJ....136.2846B}. Unless otherwise noted, in this paper we use the term KS law to refer generically to the star formation law averaged over large portions of galactic discs, rather than a specific measurement.} 
More generally, observations of nearby Galactic clouds indicate that such an elevated $\dot{\Sigma}_{\star} - \Sigma_{\rm g}$ relation is commonly found above a gas surface density threshold $\Sigma_{\rm g} \approx 125$ M$_{\odot}$ pc$^{-2}$ \citep[e.g.,][]{2010ApJ...724..687L, 2010ApJ...723.1019H}. 
Interestingly, this gas surface density threshold is comparable to the characteristic surface density of molecular clouds in the Milky Way and other nearby dwarf and spiral galaxies \citep[e.g.,][]{1981MNRAS.194..809L, 1987ApJ...319..730S, 2008ApJ...686..948B}. 
It is thus likely that these departures from the standard KS relation on small scales 
correspond to the transition to gravitationally-bound objects. 
This interpretation, rather than one directly tied to $\Sigma_{\rm g}$, is appealing because the disc-averaged $\dot{\Sigma}_{\star} - \Sigma_{\rm g}$ relation for external galaxies with $\Sigma_{\rm g} \gg 125$ M$_{\odot}$ pc$^{-2}$ (including local and high-redshift starburst galaxies) does not show the same elevation.\footnote{In \S \ref{comparison observations}, we compile observations of $\epsilon_{\rm ff}^{\rm gal}$ versus $\Sigma_{\rm g}$, showing that there is no significant trend.} 

Focusing on the most luminous Galactic free-free sources, \cite{2011ApJ...729..133M} inferred the star formation rate per free fall time 
\begin{align}
\label{eps ff GMC def}
\epsilon_{\rm ff}^{\rm GMC} \equiv \frac{t_{\rm ff} \dot{M}_{\star}}{M_{\rm GMC}}
\end{align}
in giant molecular clouds (GMCs), where $t_{\rm ff}$ is the free fall time of the GMC, $\dot{M}_{\star}$ is the star formation rate, and $M_{\rm GMC}$ is the total GMC mass (including stars). 
Weighted by ionizing luminosity, \cite{2011ApJ...729..133M} finds $\langle \epsilon_{\rm ff}^{\rm GMC} \rangle_{\rm ion} = 0.14-0.24$.\footnote{This average does not include a large number of more quiescent GMCs, which can have a significantly lower instantaneous  $\epsilon_{\rm ff}^{\rm GMC}$. One interpretation for the wide range of $\epsilon_{\rm ff}^{\rm GMC}$ values is that it is time-dependent, with $\epsilon_{\rm ff}^{\rm GMC}$ peaking near the end of the life of the GMC as it is being disrupted by stellar feedback \citep[][]{2012ApJ...746...75M}. Regardless of the precise interpretation, the observations described by \cite{2011ApJ...729..133M} demonstrate that GMCs hosting $\sim 1/3$ of the current star formation rate in the Milky Way have $\epsilon_{\rm ff}^{\rm GMC}$ well in excess of the galaxy-averaged value. 
}  
This is much larger than the Galactic average, $\langle \epsilon_{\rm ff} \rangle_{\rm MW} = 0.006$. 
The difference is significant because most star formation in the Milky Way is observed to occur in a relatively small number of massive GMCs \citep[][]{1997ApJ...476..166W, 2010ApJ...709..424M}. 
In general, star formation correlates strongly with molecular gas but not atomic gas \citep[e.g.,][]{2002ApJ...569..157W, 2011AJ....142...37S} and the mass spectrum of molecular clouds in nearby galaxies is sufficiently shallow that the mass is concentrated in the most massive objects \citep[e.g.,][]{1997ApJ...476..166W, 2005PASP..117.1403R}. 

Thus, a central problem in understanding what regulates star formation in galaxies is to reconcile the slowness of star formation as observed on galactic scales with the higher rate at which (at least some) massive GMCs are inferred to form stars. 
Theoretical studies of star formation in galaxies have generally not addressed this because they have focused either on averages over entire or substantial portions of galactic discs (as in the KS law) or on the physics of individual GMCs. 

\subsection{Theories of galactic star formation}
We review in this section some of the previous results most directly relevant to our work. 

\cite{2005ApJ...630..167T} developed a one-zone model of starburst discs supported by radiation pressure on dust. 
\cite{2010ApJ...709..191M} presented a complementary study of the disruption of GMCs by stellar feedback, showing that radiation pressure is likely the main process by which GMCs are disrupted, at least in high-density galaxies where competing mechanisms like photoionization and supernovae (SNe) become inefficient. 
In particular, \cite{2010ApJ...709..191M} showed that in the limit in which GMCs are optically thick to reprocessed far infrared (FIR) radiation, a fraction 
\begin{align}
\label{eps int GMC}
\epsilon^{\rm GMC}_{\rm int} \equiv \frac{M_{\star}}{M_{\rm GMC}}
\end{align}
as high as $\approx 0.35$ of the initial GMC gas mass is converted into stars in its lifetime. 
\cite{2010ApJ...709..191M} noted that this large integrated GMC efficiency needed to be reconciled with galaxy-averaged constraints from the observed KS law.

\cite{2011ApJ...731...41O} also focused on the starburst case. These authors argued that while radiation pressure may in fact disrupt GMCs, SNe provide the dominant vertical support in galactic discs, except in the very innermost regions.
By balancing the gravitational weight of the disc gas with the momentum flux from SN feedback, they derived a simple expression for the KS law in good agreement with measurements, provided that the CO intensity ($I_{\rm CO}$) to molecular gas surface density ($\Sigma_{\rm H_{2}}$) conversion factor ($\alpha_{\rm CO} \equiv \Sigma_{\rm H_{2}} / I_{\rm CO}$) varies continuously with $I_{\rm CO}$. 
Earlier work by \cite{2010ApJ...721..975O} considered the complementary regime where $\Sigma_{\rm g} \lesssim 100$ M$_{\odot}$ pc$^{-2}$, which we do not consider in this work. 

\cite{2009ApJ...699..850K} presented a theory of the KS law, covering both $\Sigma_{\rm g} \lesssim 100$ M$_{\odot}$ pc$^{-2}$ and $\Sigma_{\rm g} \gtrsim 100$ M$_{\odot}$ pc$^{-2}$, predicting the star formation rate as a function of observed galaxy properties. 
This theory is based on the fact that star formation proceeds in molecular gas and builds on the theory of \cite{2005ApJ...630..250K} in which the slowness of star formation in molecular clouds follows from the properties of supersonic turbulence. 
However, the model of \cite{2009ApJ...699..850K} leaves some important questions open. 
In particular, it assumes but does not explain how the gas is assembled into GMCs with virial parameter $\alpha_{\rm vir} \sim 1$. 
Furthermore, the theory of \cite{2005ApJ...630..250K} does not specify how the turbulence is driven and maintained. 
According to the theory of \cite{2009ApJ...699..850K}, the slowness of star formation averaged over galaxies derives from its slowness within GMCs. 
The predictions of that theory also depend on the assumed fraction of the galaxy mass bound into GMCs. 
In this paper, we derive the GMC mass fraction and argue that the star formation rate of galaxies does not depend sensitively on how rapidly or efficiently gas turns into stars once it is assembled in GMCs. 
Rather, the critical factor is the strength of stellar feedback relative to the gravity of the galactic disc; the formation rate of GMCs from the disc adjusts itself so that the disc-averaged star formation law is realized largely independent of sub-GMC physics. 

\subsection{This work}
A key element of many star formation theories is the gravitational instability of galactic discs. 
Gravitational instability, however, can only be a part of a successful theory of star formation in galaxies. Left unimpeded, gravitational collapse would convert all of the gas in a galaxy into stars in about one dynamical time, $\epsilon_{\rm ff}^{\rm gal} \sim 1$, in stark contrast with observations. 
In dense galaxies that are primarily molecular, we believe that stellar feedback is the main factor regulating star formation. 
Our goal here is thus to develop a model for star formation in GMCs and the resulting galaxy-averaged star formation law based on self-regulation by stellar feedback. 

Our work builds on previous models of galactic discs supported by feedback-driven turbulence \citep[e.g.,][]{1997ApJ...481..703S, 2005ApJ...630..167T, 2011ApJ...731...41O} and of GMC evolution \citep[e.g.,][]{2002ApJ...566..302M, 2009ApJ...703.1352K, 2010ApJ...709..191M}. 
In particular, our principal contribution is to show explicitly how the disc-averaged and GMC feedback regulations, which appear to imply different star formation efficiencies on galactic and GMC scales, can be consistently understood in a unified framework. 
This allows us to generalize the disc-averaged star formation law predicted by feedback regulation to the case in which the disc has a Toomre $Q$ parameter deviating from unity (\S \ref{disk averaged SF law}) and to predict for the mass fraction of gas in gravitationally-bound clouds as a function of galaxy surface density (\S \ref{GMC consistency}). 
We also compile observational evidence that the disc-averaged star formation efficiency scales with gas mass fraction of the galaxy (\S \ref{star formation efficiency observations}), supporting the predictions of our feedback-regulated theory, but in tension with models in which the star formation efficiency is a nearly universal constant \citep[e.g.,][]{2005ApJ...630..250K, 2009ApJ...699..850K}.

We concentrate on the high gas surface density case ($\Sigma_{\rm g} \gtrsim 100$ M$_{\odot}$ pc$^{-2}$) to avoid effects related to the conversion of atomic to molecular gas.  
Since gas is expected to be essentially purely molecular in such discs, we treat the interstellar medium (ISM) as a single-phase turbulent medium. 
This regime includes local merging galaxies \citep[e.g.,][]{1998ApJ...507..615D}, ordinary star-forming galaxies at redshift $z\gtrsim1$ \citep[where the star formation rate is elevated at a fixed stellar mass; e.g.,][]{2006ApJ...647..128E, 2007ApJ...670..156D, 2010MNRAS.407.2091G}, and high-redshift sub-millimeter galaxies \citep[e.g.,][]{2008ApJ...680..246T}. 
These systems are among the primary targets for new and upcoming observatories sensitive to the gas content of galaxies, including the Herschel Space Observatory,\footnote{http://herschel.esac.esa.int} the Jansky Very Large Array (JVLA),\footnote{http://www.vla.nrao.edu} the Atacama Large Millimeter Array (ALMA),\footnote{http://www.almaobservatory.org} and the Cerro Chajnantor Atacama Telescope (CCAT).\footnote{http://www.ccatobservatory.org} 
Since approximately half of all stars formed before $z\sim2$ and in halos of mass comparable to those currently probed by observation \citep[e.g.,][]{2012arXiv1207.6105B}, our model in fact applies to most star formation in the Universe. 
The high gas surface density regime is also relevant for the inner regions of ordinary galaxies like the Milky Way and to the fueling of central massive black holes. 

In \S \ref{two zone disk}, we define a two-zone galactic disc model, consisting of GMCs and a volume-filling inter-cloud medium. 
In \S \ref{connecting disk and GMCs}, we balance the momentum returned by stellar feedback and the weight of the disc to derive an average $\dot{\Sigma}_{\star} - \Sigma_{\rm g}$ relation as a function of the momentum per stellar mass formed ($P_{\star}/m_{\star}$) and the $Q$ parameter of the disc. 
In this section, we also derive consistency requirements between the efficiencies of star formation in GMCs and the overall galactic disc, $\epsilon_{\rm int}^{\rm GMC}$ and $\epsilon_{\rm ff}^{\rm gal}$. 
We show that $\epsilon_{\rm ff}^{\rm gal}$ is primarily a function of the gas fraction, the circular velocity of the galaxy, and the effective $P_{\star}/m_{\star}$ in the volume-filing medium. 
There is also a dependence on $\epsilon_{\rm int}^{\rm GMC}$, but it is generally surprisingly weak. 
This is because the disc-averaged $Q$ regulates itself to a value that can significantly exceed the threshold for marginal stability ($Q=1$). This enables the GMC formation rate to adjust itself to yield a consistent relationship between $\epsilon_{\rm ff}^{\rm gal}$ and $\epsilon_{\rm int}^{\rm GMC}$. 
A comparison with observations in \S \ref{comparison observations} shows that our model agrees well with measurements of the galaxy-averaged KS law, the turbulent gas velocity dispersions in galaxies, and the fraction of the disc gas mass collapsed in GMCs. 
We conclude in \S \ref{conclusions}. Table \ref{symbol summary} summarize symbols used in this paper. 

While the high surface density limit simplifies the theoretical treatment by avoiding explicit consideration of a multiphase atomic and molecular ISM, the main points of this paper regarding the interplay between GMCs and the rest of the galactic disc likely apply to galaxies with lower surface density as well. 
In particular, while lower gas surface density galaxies can maintain important multiphase structure in their ISM and are subject to a different set of feedback processes, we expect that the formation rate of GMCs is also the rate limiting step in other galaxies that are supported stellar feedback. 
This is supported by numerical simulations of galaxies including Milky Way and Small Magellanic Cloud analogs in which the disc-averaged star formation law is insensitive to the small-scale star formation prescription \citep[][]{2011MNRAS.417..950H}. 
Similarly, we carry out our numerical calculations assuming that supernovae dominate feedback in the volume-filling medium and that radiation pressure on dust disrupts GMCs, our current best model for galaxies with gas surface density $100 < \Sigma_{\rm g} < 10^{4}$ M$_{\odot}$ pc$^{-2}$ (Appendix \ref{model closure appendix}). 
However, many of our results can be extended to other feedback mechanisms (such as photoionization or cosmic rays; e.g. McKee 1989, Socrates et al. 2008\nocite{1989ApJ...345..782M, 2008ApJ...687..202S}) by appropriate choices for $P_{\star} / m_{\star}$ and $\epsilon_{\rm int}^{\rm GMC}$. 

\begin{figure}
\begin{center}
\includegraphics[width=0.9\columnwidth]{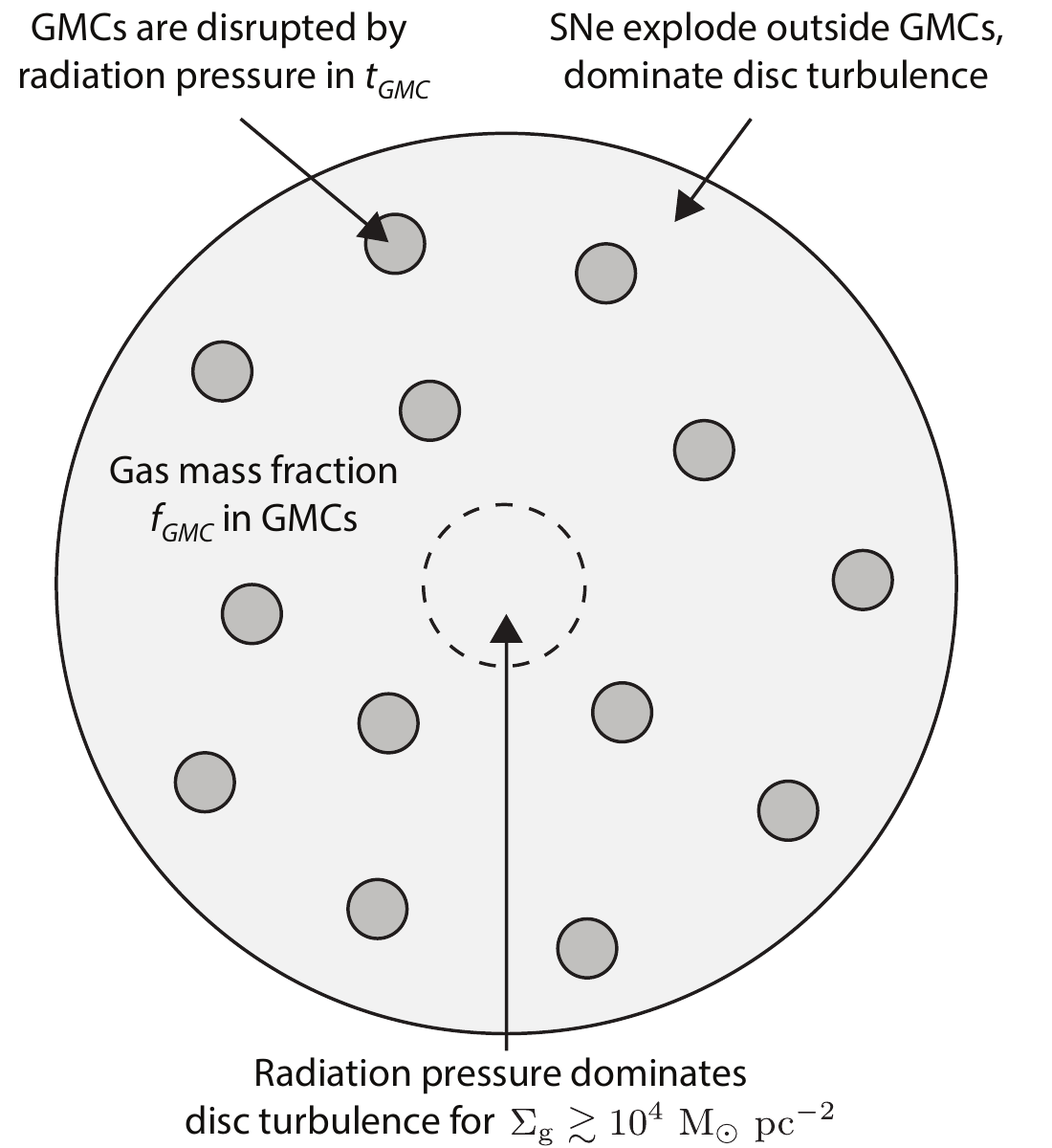}
\end{center}
\caption{Summary of the two-zone disc models used in this work and described in \S \ref{two zone disk}. Star formation proceeds in GMCs, which are dispersed by radiation pressure on dust. Most supernovae then explode in the volume-filling medium and drive turbulence in it, except at $\Sigma_{\rm g} \gtrsim 10^{4}$ M$_{\odot}$ pc$^{-2}$, where radiation pressure dominates the turbulence in the volume-filling ISM (Appendix \ref{model closure appendix}).   
}
\label{fig:2zone_disk}
\vspace*{0.1in}
\end{figure}

\section{Two-zone disc model}
\label{two zone disk}
We consider a two-zone model, consisting of a volume-filling background disc in which gravitationally-bound GMCs are embedded.\footnote{In the Milky Way, gravitationally-bound clouds roughly coincide with molecular clouds. In denser galaxies which are the focus of this paper, the gas is assumed to be completely molecular. We use the term GMC to refer to gravitationally-bound clouds even though their chemical composition does not necessarily differ from the volume-filling medium.}  
According to our definition, the self-gravity of GMCs exceeds their internal pressure. 
For any effective equation of state in which the pressure increases with average density, this implies that the gravity of GMCs also exceeds the external pressure acting on them. 
The GMCs are thus treated as collapsed entities that are hydrodynamically decoupled from the background disc before they are dispersed by stellar feedback. 
Figure \ref{fig:2zone_disk} shows a schematic of our two-zone disc model. 

When comparing our results to observations, it is important to distinguish between GMCs as gravitationally-bound clouds and the stage in their evolution during which they may exist as more compact quasi-virialized objects. The mean density of the quasi-virialized state, which follows an initial period of gravitational collapse, exceeds the mean density of the cloud when it first becomes gravitationally bound. 
Similarly, the lifetime as a gravitationally-bound cloud is longer than that in the quasi-virialized state. In this work, we generally define GMC properties in terms of their longer existence as gravitationally-bound clouds. 
This frees the model of assumptions regarding how the properties of the quasi-virial equilibrium are related to galaxy properties. 
For purely molecular galaxies, gravitational boundedness is also a well-defined criterion for GMCs. 
However, GMCs defined in this way do not in general correspond exactly to the density peaks traced by molecular gas in the Milky Way \citep[e.g.,][]{2001ApJ...547..792D}, which can have an envelope of atomic gas that is gravitationally-bound but not fully virialized \citep[e.g.,][]{2010ApJ...715.1302V}. 

\begin{table*}
\centering
\caption{Summary of symbols used in this work\label{symbol summary}}
\begin{tabular}{|lll|}
\hline\hline
Symbol                              &  Definition  & Eq.  \\ 
\hline
$\epsilon_{\rm ff}^{\rm GMC}$ & Star formation efficiency per free fall time of GMCs & \ref{eps ff GMC def} \\ 
$\epsilon_{\rm int}^{\rm GMC}$ & Integrated star formation efficiency of GMCs & \ref{eps int GMC} \\ 
$\epsilon_{\rm ff}^{\rm gal}$ & Star formation efficiency per free fall time in the galactic disc & \ref{eps ff gal def} \\ 
$\sigma$ & Velocity dispersion of isothermal galactic potential, corresponding to circular velocity $v_{\rm c} = \sqrt{2} \sigma$ & \ref{Sigma tot} \\
$\kappa$ & Epicyclic frequency of the disc & \ref{Q def} \\
$h$ & Gas disc scale height & \ref{h vs cT Omega} \\
$M_{\rm g}$ & Gas mass in the disc & \ref{Sigma tot} \\
$\Sigma_{\rm tot}$, $\Sigma_{\rm g}$ & Total and gas mass surface densities & \ref{Sigma tot} \\
$\dot{\Sigma}_{\star}$ & Star formation rate surface density & \ref{pT star}, \ref{star formation law momentum balance} \\
$f_{\rm g}$ & Gas mass fraction $M_{\rm g}/M_{\rm tot}$ & \ref{h over r fg}, \ref{cT sigma} \\
$\bar{\rho}$, $\bar{n}_{\rm H}$ & Mean gas density & \ref{vertical eq}, \ref{gas density} \\
$c_{\rm T}$ & Turbulent gas velocity dispersion & \ref{cT sigma} \\
$p_{\rm T}$ & Turbulent gas pressure & \ref{vertical eq} \\
$\mathcal{M}$, $\mathcal{M}(k)$ & Mach number of the turbulence on the outer scale $\sim h$ and for mode $k$ & \ref{sigma k} \\
$Q$, $\bar{Q}(h)$ & Toomre stability parameter of the disc and smoothed on scale $h$ & \ref{Q def} \\
$\phi$ & Factor $\sim1$ if potential is dominated by stars+dark matter and $\sim 1/Q$ if the gas disc dominates & \ref{vertical eq} \\
$t_{\rm ff}$, $t_{\rm ff}^{\rm disc}$ & Free fall time and free fall time evaluated using mean gas density in the disc & \ref{free fall time def}, \ref{disk free fall time} \\
$P_{\star}/m_{\star}$ & Momentum returned by stellar feedback in ISM per stellar mass formed & \ref{pT star}, \ref{star formation law momentum balance} \\
$\mathcal{F}$ & Dimensionless parameters encapsulating $\sim1$ uncertainties in $\dot{\Sigma}_{\star}-\Sigma_{\rm g}$ relation & \ref{F def} \\ 
$t_{\rm GMC}$, $\tilde{t}_{\rm GMC}$ & GMC lifetime and dimensionless GMC lifetime in units of $t_{\rm ff}^{\rm disc}$ & \ref{Mstar GMC view}, \ref{tilde t GMC} \\
$f_{\rm GMC}$ & Fraction of disc gas mass in GMCs & \ref{Mstar GMC view} \\
$f_{\rm coll}$ & Instantaneous fraction of disc gas mass in gravitationally-unstable density fluctuations & \ref{g def} \\
$\alpha$, $\beta$ & Parameters of power-law approximation for $\dot{m}_{\rm GMC}^{\rm tot}$ & \ref{g def} \\ 
$\alpha_{\rm CO}$, $X_{\rm CO}$ & Conversion from CO intensity to gas surface density and column density & \ref{OS alpha CO} \\
\hline
\end{tabular}
\end{table*}

\subsection{Background disc}
\label{background disk}
Our background disc model is similar to the one introduced by \cite{2005ApJ...630..167T}. The disc is modeled using a radius-dependent mean gas density $\bar{\rho}$ (including the GMC contribution). 
As a simple model for galaxies with flat rotation curve, we assume that the disc is in radial centrifugal balance in an isothermal potential with velocity dispersion $\sigma$ and angular frequency $\Omega = \sqrt{2} \sigma / r$. 
The circular velocity is then $v_{\rm c} = \sqrt{2} \sigma$. 
The total mass enclosed within radius $r$ is $M_{\rm tot}(r) = 2 \sigma^{2} r /G$ and the corresponding surface density is
\begin{equation}
\label{Sigma tot}
\Sigma_{\rm tot} = \frac{\sigma^{2}}{\pi G r}. 
\end{equation}
The gas mass fraction is $f_{\rm g} \equiv M_{\rm g} / M_{\rm tot} = \Sigma_{\rm g} / \Sigma_{\rm tot}$. 

In general, the vertical component of the hydrostatic equilibrium equation in cylindrical coordinates is $(1 / \rho) \partial p / \partial z = \partial \Phi_{\rm g} / \partial z$, where $\Phi_{\rm g}$ is the gravitational potential. 
For a thin disc in a spherical gravitational potential, this is $\partial p / \partial z = - \rho \Omega^{2} z$, where $p$ is the effective gas pressure. This can be approximated as
\begin{equation}
\label{vertical eq}
p \approx \phi \bar{\rho} h^{2} \Omega^{2},
\end{equation}
where $h$ is the disc scale height and $\phi \approx 1$ is a constant.
In the limit in which the self-gravity of the thin disc is dominant and $f_{\rm g}=1$, the solution is similar but with $\phi \sim 1 / Q$. 
Assuming that turbulence dominates the effective gas pressure,\footnote{For a molecular medium at temperature $T \sim 100$ K, the sound speed is only $\sim 1$ km s$^{-1}$, much smaller than observed turbulent velocity dispersions $c_{\rm T} \sim 10-100$ km s$^{-1}$ \citep[e.g.,][]{1998ApJ...507..615D, 2011ApJ...733..101G}. In the Milky Way, magnetic fields and cosmic rays each contribute comparably to turbulence to the total pressure of the ISM \citep[e.g.,][]{1990ApJ...365..544B}. 
Using observations of radio supernova remnants (SNR), \cite{2009MNRAS.397.1410T} show that the magnetic pressure is likely a small fraction of the total ISM pressure for starbursts with $\Sigma_{\rm g} \gtrsim 100$ M$_{\odot}$ pc$^{-2}$. Cosmic ray models predict and $\gamma-$ray observations indicate that cosmic rays are also dynamically unimportant for the gas surface densities considered here \citep[][]{2010ApJ...717....1L, 2011ApJ...734..107L}.} $p \approx p_{\rm T} \approx \bar{\rho} c_{\rm T}^{2}$, where $c_{\rm T}$ is the turbulent velocity on the scale $h$. 
The corresponding Mach number is $\mathcal{M}\equiv c_{\rm T}/c_{\rm s}$, where $c_{\rm s}$ is the sound speed of the gas. 
Equation (\ref{vertical eq}) implies that 
\begin{align}
\label{h vs cT Omega}
h \approx \frac{c_{\rm T}}{\phi^{1/2} \Omega}.
\end{align}

There is strong observational evidence that star-forming discs self-regulate to sustain a \cite{1964ApJ...139.1217T} parameter 
\begin{equation}
\label{Q def}
Q = \frac{\kappa c_{\rm T}}{\pi G \Sigma_{\rm g}}  = \frac{2 \sigma c_{\rm T}}{\pi G \Sigma_{\rm g} r} \sim 1,
\end{equation}
where $\kappa\equiv \sqrt{4 \Omega^{2} + d\Omega^{2}/d\ln{r}} = 2 \sigma / r$ is the epicyclic frequency
\citep[e.g.,][]{1972ApJ...176L...9Q, 1989ApJ...344..685K, 2001ApJ...555..301M}. 
This can be understood intuitively from the fact that galactic discs tend to cool until $Q \approx 1$, at which point they become gravitationally unstable and form stars. 
Stellar feedback then drives turbulence in the disc, ensuring that $Q$ does not drop significantly below unity. 
We will show in this paper that even when stellar feedback can regulate the disc to $Q \sim 1$, significant deviations such that $Q>1$ are expected and are important for regulating the galactic star formation rate. 
Note that in our model, $Q$ is evaluated using quantities averaged over the entire disc, including the gas in GMCs. 

Real galactic discs consist not only of gas, but also of stars and dark matter, each of which has a different surface density and velocity dispersion. 
When the gas turbulent velocity dispersion is much smaller than the stellar velocity dispersion, the results of \cite{2001MNRAS.323..445R} imply that the effective $Q$ criterion for the gas only in equation (\ref{Q def}) is a good approximation to the stability of the disc. 
In the limit in which gas and stars have comparable velocity dispersions and surface densities, gas and stars contribute similarly to the disc instability. 
This limit may be realized in some cases of interest to us, for example in the central regions of galaxy mergers \citep[][]{1998ApJ...507..615D}. 
The stability criterion would then be modified by a factor of $\sim2$. 
A more accurate theory would take this into account, but at the level of our analysis we prefer to use the simpler gas criterion in equation (\ref{Q def}). 

Using $\Sigma_{\rm g} \equiv 2 h \bar{\rho}$ and eliminating $h$ using equation (\ref{h vs cT Omega}), it follows that
\begin{align}
\label{gas density}
\bar{n}_{\rm H}
& = \frac{\sqrt{2} \phi^{1/2} \sigma^{2}}{\pi G Q r^{2} m_{\rm p}} \\ \notag
& \approx 1.7 \times 10^{4}~{\rm cm^{-3}}~Q^{-1} \phi^{1/2} \left( \frac{\sigma}{\rm 200~km~s^{-1}} \right)^{2} \\ \notag 
&~~~~~~~~~~~~~~~~~~~~~~~~~~~~~~~~~~~~~~~~~~~~~\times \left( \frac{r}{\rm 100~pc} \right)^{-2},
\end{align}
where $\bar{n}_{\rm H} \equiv \bar{\rho}/m_{\rm p}$. 
From the definitions of $Q$ and $f_{\rm g}$, using equation (\ref{h vs cT Omega}), we also have the simple relations
\begin{align}
\label{h over r fg}
\frac{h}{r} = \frac{Q}{2^{3/2} \phi^{1/2}} f_{\rm g}
\end{align}
and
\begin{align}
\label{cT sigma}
\frac{c_{\rm T}}{\sigma} = \frac{Q}{2} f_{\rm g} .
\end{align}
Therefore, 
\begin{align}
\label{h over r}
\frac{h}{r} = \frac{1}{\sqrt{2 \phi}} \frac{c_{\rm T}}{\sigma}.
\end{align}

\subsection{Giant molecular clouds}
\label{GMC section}
As mentioned in the introduction, most star formation in galaxies occurs in a relatively small number of massive GMCs. 
We identify the mass of these most massive GMCs with the Toomre mass of the disc \citep[e.g.,][]{1965MNRAS.130...97G},
\begin{align}
M_{\rm GMC} & \approx \pi h^{2} \Sigma_{\rm g} \\ \notag
& \approx 1.3\times10^{6}~{\rm M_{\odot}} \left( \frac{h}{\rm 20~pc} \right)^{2} \left( \frac{\Sigma_{\rm g}}{\rm 10^{3}~M_{\odot}~pc^{-2}} \right),
\end{align}
where we have scaled the parameters to values characteristic of local ULIRGs. 
The Toomre-mass GMCs are initially self-gravitating but are eventually dispersed by feedback from massive stars forming in them \citep[e.g.,][]{2002ApJ...566..302M, 2010ApJ...709..191M}. 
We assume that all stars form in GMCs of Toomre mass and denote the fraction of the disc gas mass stored in gravitationally-bound GMCs at any given time by $f_{\rm GMC}$: $M_{\rm GMC}^{\rm tot} \equiv f_{\rm GMC} M_{\rm g}$. 

In general, the free fall time in a region of mean density $\bar{\rho}$ is
\begin{align}
\label{free fall time def}
t_{\rm ff} & \equiv \sqrt{ \frac{3 \pi}{32 G \bar{\rho}} } \\ \notag
& \approx 5\times10^{7}~{\rm yr} \left( \frac{\bar{n}_{\rm H} }{\rm 1~cm^{-3}} \right)^{-1/2}. 
\end{align}
Since the mean density of a GMC in quasi-virial equilibrium exceeds the mean density in the disc, the free fall time of the GMC in that state is shorter than the free fall time in the disc. In the Milky Way, quasi-virialized GMCs are over-dense with respect to the disc by a factor $\sim30$, so that their internal free fall time is shorter than the free fall time in the disc by a factor $\sim5$. 
The lifetime of quasi-virialized GMCs before they are dispersed by feedback is likely a few internal free fall times \citep[e.g.,][]{2012ApJ...746...75M}, comparable to or less than the free fall time at mean disc density, $t_{\rm ff}^{\rm disc}$. 
Since $t_{\rm ff}^{\rm disc}$ is also the time scale for formation of the GMC from the disc (for $Q\sim1$), the GMC lifetime as a gravitationally-bound entity $t_{\rm GMC} \sim t_{\rm ff}^{\rm disc}$.

Using equation (\ref{gas density}) for the mean gas density in the disc, 
\begin{align}
\label{disk free fall time}
t_{\rm ff}^{\rm disc} & = \sqrt{ \frac{3 \pi^{2} Q}{32\times 2^{1/2} \phi^{1/2}}} \frac{r}{\sigma} \\ \notag
& \approx 4 \times 10^{5}{\rm~yr}~\frac{Q^{1/2}}{\phi^{1/4}} \left( \frac{r}{\rm 100~pc} \right) \left( \frac{\sigma}{\rm 200~km~s^{-1}} \right)^{-1}.
\end{align}
The short dynamical times in dense starburst discs imply that massive stars can outlive their parent GMC. 
We expand on this point in Appendix \ref{model closure appendix} in the context of which feedback processes disrupt GMCs and which processes drive turbulence in the volume-filling ISM.

\section{Connecting disc and GMC star formation}
\label{connecting disk and GMCs}
We now connect star formation in GMCs and the disc-averaged star formation law. 
We first derive a general expression for the disc-averaged KS law based on vertical hydrostatic equilibrium in the disc (\S \ref{disk averaged SF law}), discuss the relationship to the turbulent gas velocity dispersion (\S \ref{turbulent velocity theory}), and show how the implied disc-averaged star formation efficiency is related to the integrated star formation efficiency and lifetime of GMCs (\S \ref{GMC consistency}). 

We express our results here in terms of a fiducial value for $P_{\star}/m_{\star}=3,000$ km s$^{-1}$ (the effective momentum injected by stellar feedback per stellar mass formed) appropriate for SN feedback under typical conditions. 
We motivate this choice and discuss how $P_{\star}/m_{\star}$ depends on the ambient conditions (including the importance of radiation pressure) further in Appendix \ref{model closure appendix}. 

\subsection{The disc-averaged star formation law}
\label{disk averaged SF law}
The general requirement that the turbulent pressure balances the gravitational weight of the overlying gas in the disc is
\begin{align}
\label{pressure balances gravity}
p_{\rm T} = \frac{\pi G \Sigma_{\rm g}^{2}}{2^{3/2}} Q \phi.
\end{align}
This expression follows from hydrostatic equilibrium and the definition of $Q$, and the pre-factor depends on the assumption of an isothermal potential. Equation (\ref{pressure balances gravity}) holds for any value of $Q$ or $f_{\rm g}$. A similar equilibrium relation was derived by \cite{2005ApJ...630..167T} and \cite{2011ApJ...731...41O}. 

In general, the total turbulent pressure results from a combination of stellar feedback and all other processes that drive turbulence:
\begin{align}
\label{pT tot}
p_{\rm T} = p_{\rm T,\star} + p_{\rm T,other}.
\end{align}
The second term, $p_{\rm T,other}$, may include contributions from the magneto-rotational instability \citep[MRI; e.g.,][]{2005ApJ...629..849P},  the thermal instability \citep[e.g.,][]{2002ApJ...569L.127K}, the release of gravitational potential energy as gas accretes onto the galaxy from the intergalactic medium \citep[e.g.,][]{2012MNRAS.425..788G} or as gas is transported inward by internal galactic torques \citep[e.g.,][]{2010ApJ...724..895K}. 

We focus on the case in which stellar feedback dominates turbulence, $p_{\rm T} \approx p_{\rm T,\star}$. 
Then $p_{\rm T,\star}$ can be derived by equating the rate of energy injection per unit volume, $\dot{e}_{\rm in,\star}$, with the total turbulent energy dissipation rate, $\dot{e}_{\rm diss}$. 
The energy injection rate $\dot{e}_{\rm in,\star} \approx \dot{\Sigma}_{\star} (P_{\star}/m_{\star}) v_{\star}/2$, where $v_{\star}$ is a velocity term used to convert momentum to kinetic energy injection. 
If isolated SNe dominate the feedback, $v_{\star}$ is the supernova remnant (SNR) velocity \emph{at the stage used to evaluate $P_{\star}/m_{\star}$}. 
We define this stage such that $v_{\star} = c_{\rm T}$, which for SNRs corresponds to when the remnants effectively merge with the ISM. 
In general, $P_{\star}/m_{\star}$ can also be determined by other processes, such as photoionization in lower-$\Sigma_{\rm g}$ galaxies \citep[e.g.,][]{1989ApJ...345..782M, 2010ApJ...721..975O} or radiation pressure on dust in higher-$\Sigma_{\rm g}$ galaxies (Thompson et al. 2005\nocite{2005ApJ...630..167T} and Appendix \ref{model closure appendix}). 

Numerical simulations show that turbulence dissipates in approximately one flow crossing time,
\begin{align}
\label{t diss}
t_{\rm diss} \approx \gamma t_{\rm flow};~~~~~t_{\rm flow} \equiv \frac{L}{c_{\rm T}},
\end{align}
where $\gamma \approx 1$ and $L$ is the size of the largest eddies \citep[e.g.,][]{1998ApJ...508L..99S, 1999ApJ...524..169M}. 
This expression is valid for hydrodynamic and magnetohydrodynamic turbulence, both subsonic and supersonic. 
The turbulent energy dissipation rate is thus $\dot{e}_{\rm diss} \approx \bar{\rho} c_{\rm T}^{2}/t_{\rm diss} = \bar{\rho} c_{\rm T}^{3}/(\gamma L)$ and we can solve for $p_{\rm T,\star} \approx \bar{\rho} c_{\rm T}^{2}$:
\begin{align}
\label{pT star}
p_{\rm T,\star} \approx \frac{f_{\rm P}(1-f_{\rm w}) f_{h} \gamma}{8} \dot{\Sigma}_{\star} \left( \frac{P_{\star}}{m_{\star}} \right).
\end{align}
In the above equation, we defined $f_{h} \equiv L/h$ and introduced a term $f_{\rm P} (1 - f_{\rm w})/4$, where the factor of $1/4$ accounts for cancelation of momentum in the disc plane and $f_{\rm P}$ parameterizes the uncertainty in it \citep[][]{2011ApJ...731...41O}. We also defined $f_{\rm w}$ as the fraction of the input momentum that is lost to a galactic wind, rather than contributing directly to the vertical pressure support in the disc. 

The dependence on $f_{h} = L/h$ in equation (\ref{pT star}) illustrates the dependence on the details of how turbulence is driven. 
For example, if turbulence is driven by individual SNe and the ambient density is very large, then it is possible that $L \ll h$. 
In this limit the turbulent energy is efficiently radiated away and SNe are relatively inefficient. 
On the other hand, SNe could cluster and merge in super-bubbles of scale $\sim h$ before achieving pressure equilibrium with the ambient ISM. 

Equating equations (\ref{pressure balances gravity}) and (\ref{pT star}), we obtain the star formation law
\begin{align}
\label{star formation law momentum balance}
\dot{\Sigma}_{\star} & = \frac{2 \sqrt{2} \pi G Q \phi}{\mathcal{F}} \left( \frac{P_{\star}}{m_{\star}} \right)^{-1} \Sigma_{\rm g}^{2} \\ \notag
& \approx 13{\rm~M_{\odot}~yr^{-1}~kpc^{-2}}~\frac{Q \phi}{\mathcal{F}} \\ \notag 
& ~~~~~~~~~~\times \left( \frac{P_{\star}/m_{\star}}{\rm 3,000~km~s^{-1}} \right)^{-1} \left( \frac{\Sigma_{\rm g}}{\rm 10^{3}~M_{\odot}~pc^{-2}} \right)^{2},
\end{align}
where
\begin{align}
\label{F def}
\mathcal{F} \equiv f_{\rm P}(1-f_{\rm w}) f_{h} \gamma 
\end{align}
encapsulates uncertain factors of order unity. 
This star formation law derives from feedback-driven turbulence support, as in \cite{2005ApJ...630..167T}, but we find a different normalization because of how we treat supernova feedback (see Appendix \ref{model closure appendix}). 
Equation (\ref{star formation law momentum balance}) is consistent with the derivation of \cite{2011ApJ...731...41O}.

\subsection{Relation between turbulent gas velocity dispersion and star formation efficiency}
\label{turbulent velocity theory}
We can also write the star formation law by defining a disc-averaged star formation efficiency per free fall time, $\epsilon_{\rm ff}^{\rm gal}$, as in the usual KS law:
\begin{align}
\label{eps ff gal def}
\dot{\Sigma}_{\star} \equiv \epsilon_{\rm ff}^{\rm gal} \frac{\Sigma_{\rm g}}{t_{\rm ff}^{\rm disc}}.
\end{align}
Using the definition of $t_{\rm ff}^{\rm disc}$ (eq. \ref{free fall time def}) and expressing the mean density $\bar{\rho}$ using the vertical balance equation (\ref{pressure balances gravity}), the star formation law becomes
\begin{align}
\dot{\Sigma}_{\star} = \frac{2^{7/4} G Q^{1/2} \phi^{1/2} \epsilon_{\rm ff}^{\rm gal}}{\sqrt{3}} \frac{\Sigma_{\rm g}^{2}}{c_{\rm T}}. 
\end{align}
We can further use equation (\ref{star formation law momentum balance}) to eliminate $\Sigma_{\rm g}$ and solve for the turbulent velocity:
\begin{align}
\label{cT vs eta}
c_{\rm T} & = \frac{2^{1/4}}{\sqrt{3} \pi} \frac{\mathcal{F} \epsilon_{\rm ff}^{\rm gal}}{Q^{1/2} \phi^{1/2}} \left( \frac{P_{\star}}{m_{\star}} \right) 
\\ \notag
& \approx {\rm 6.6~km~s^{-1}}~\frac{\mathcal{F}}{Q^{1/2} \phi^{1/2}} \left( \frac{\epsilon_{\rm ff}^{\rm gal}}{0.01} \right) \left( \frac{P_{\star}/m_{\star}}{\rm 3,000~km~s^{-1}} \right)
\end{align}
\citep[see also][]{2005ApJ...630..167T, 2011ApJ...731...41O}. 
This result implies that the turbulent velocity dispersion is constant at fixed $\epsilon_{\rm ff}^{\rm gal}$, even if the supernova rate varies by orders of magnitude with the star formation rate, in agreement with the numerical simulations of \cite{2009ApJ...704..137J}. 
In \S \ref{velocity dispersion observations}, we show that the elevated turbulent velocity dispersions observed in local ULIRGs and in high-redshift star-forming galaxies can be explained by the dependence of $\epsilon_{\rm ff}^{\rm gal}$ on the product $f_{\rm g} \sigma$, which we discuss next. 

\begin{figure*}
\mbox{
\includegraphics[width=0.5\textwidth]{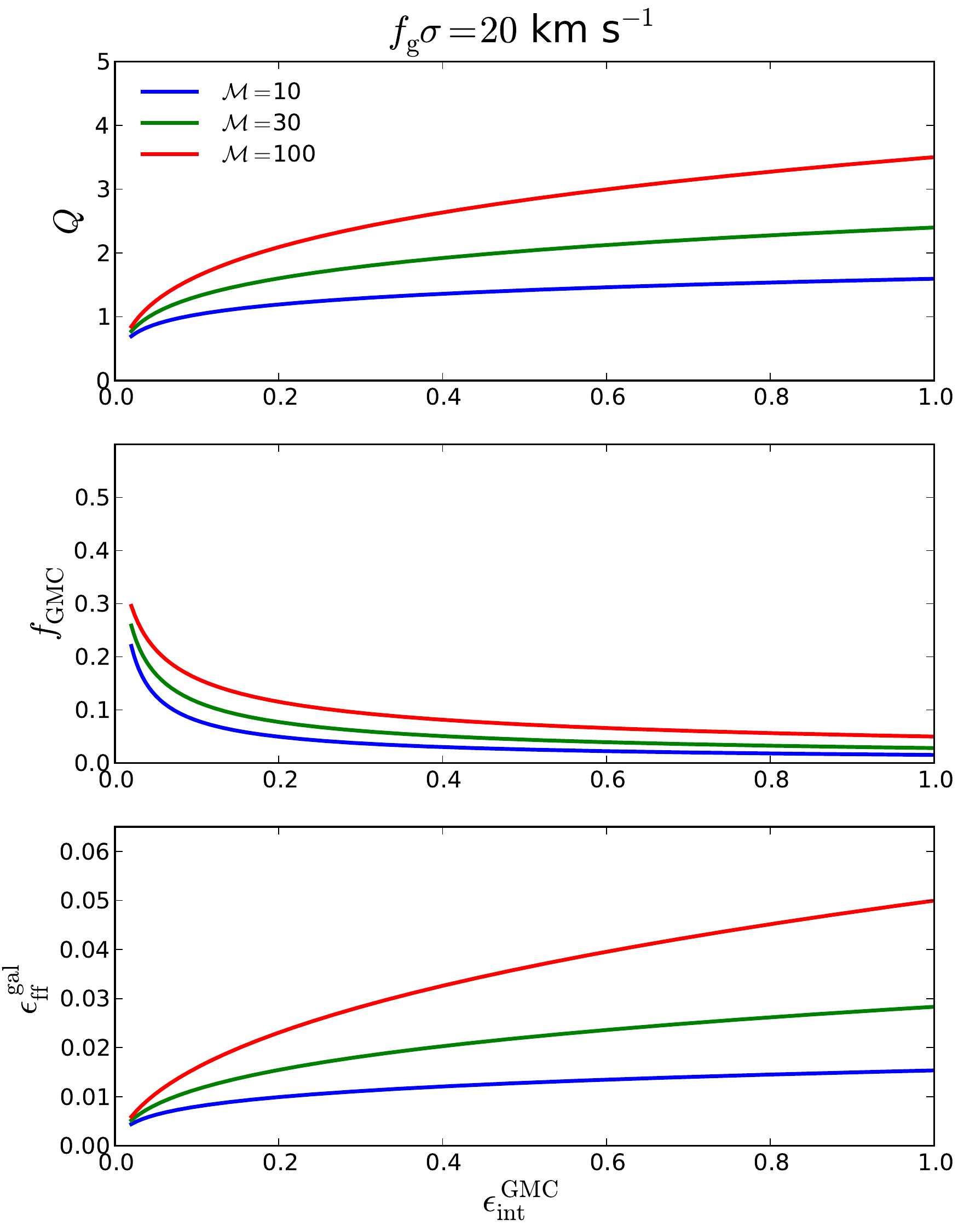}
\includegraphics[width=0.5\textwidth]{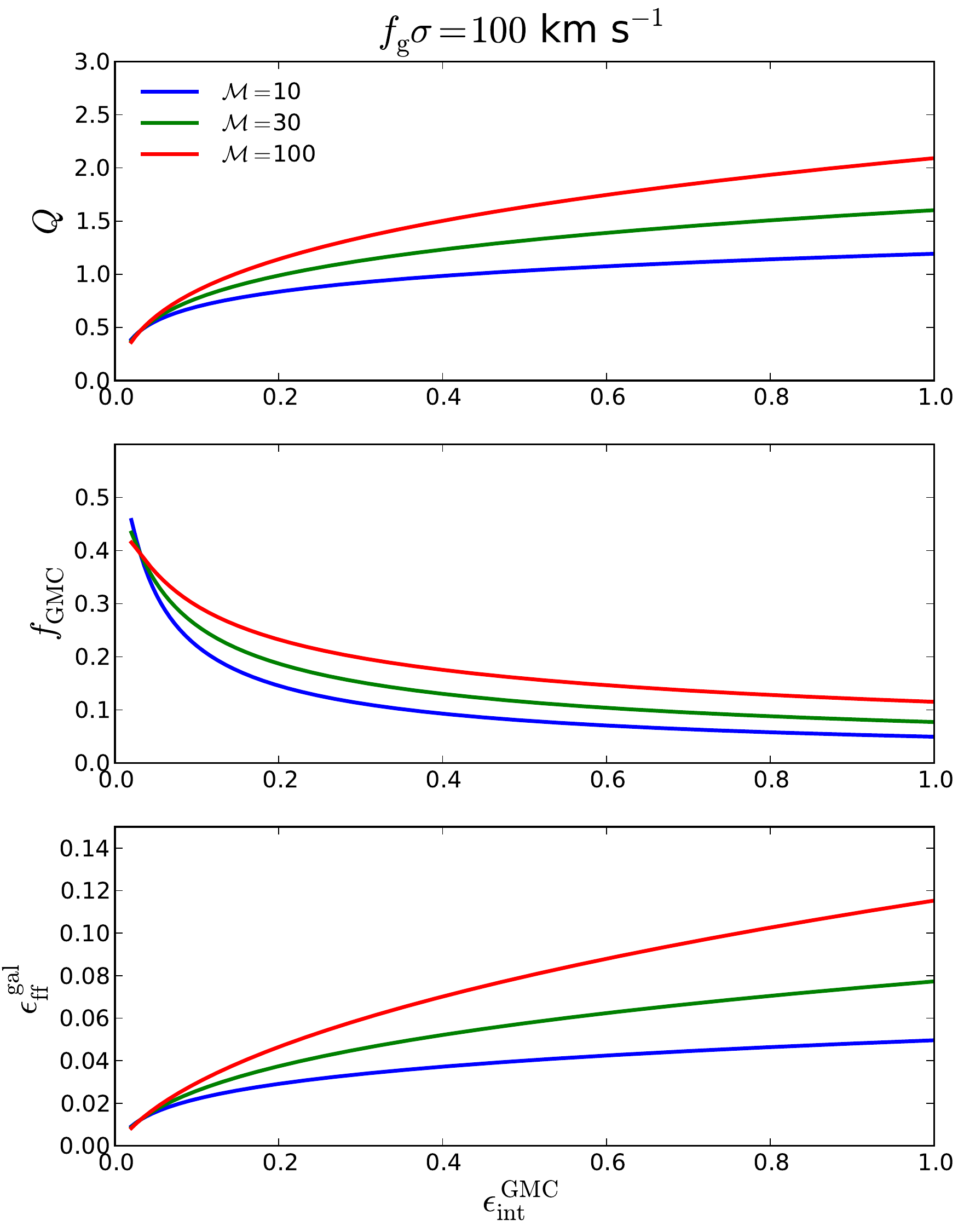}
}
\caption{
Equilibrium solutions for the disc-averaged Toomre $Q$ parameter, fraction of the gas mass in gravitationally-bound GMCs ($f_{\rm GMC}$), and the disc-averaged star formation efficiency per free fall time ($\epsilon_{\rm ff}^{\rm gal}$) as a function of the integrated efficiency with which GMCs convert their gas into stars ($\epsilon_{\rm int}^{\rm GMC}$). Each panel shows the curves for different Mach numbers $\mathcal{M}$ of the molecular ISM. The left panel is representative of a low-mass or gas-poor galaxy with $f_{\rm g} \sigma=20$ km s$^{-1}$ and the panel on the right is illustrative of a massive gas-rich system (such as an ULIRG or a high-redshift star-forming galaxy) with $f_{\rm g} \sigma=100$ km s$^{-1}$. 
The disc-averaged star formation efficiency per free fall time increases slowly with the integrated GMC star formation efficiency, while $f_{\rm GMC}$ decreases. 
This is realized by the global $Q$ parameter increasing above the classical stability threshold of unity, so that GMCs form only where turbulent fluctuations cause the self-gravity of the gas to exceed its turbulent and rotational support. 
These examples assume $P_{\star} / m_{\star} = 3,000$ km s$^{-1}$, $\mathcal{F}=2$ (shown to yield a good fit to the observed disc-averaged $\dot{\Sigma}_{\star}-\Sigma_{\rm g}$ relation in \S \ref{star formation law observations}), $\phi=1$, and $\tilde{t}_{\rm GMC}=1$. 
The equilibrium solutions were obtained using equation (\ref{fcoll EPS}) for the collapsed fraction as a function of $Q$, with values $(a,~b,~\chi)=(1.5,~0.5,~1)$ for the parameters defined in Appendix \ref{collapsed fraction appendix}. 
The curves start at $\epsilon_{\rm int}^{\rm GMC}=0.02$ in each panel.}
\label{fig:equilibrium_quantities}
\vspace*{0.1in}
\end{figure*}

\begin{figure*}
\mbox{
\includegraphics[width=0.5\textwidth]{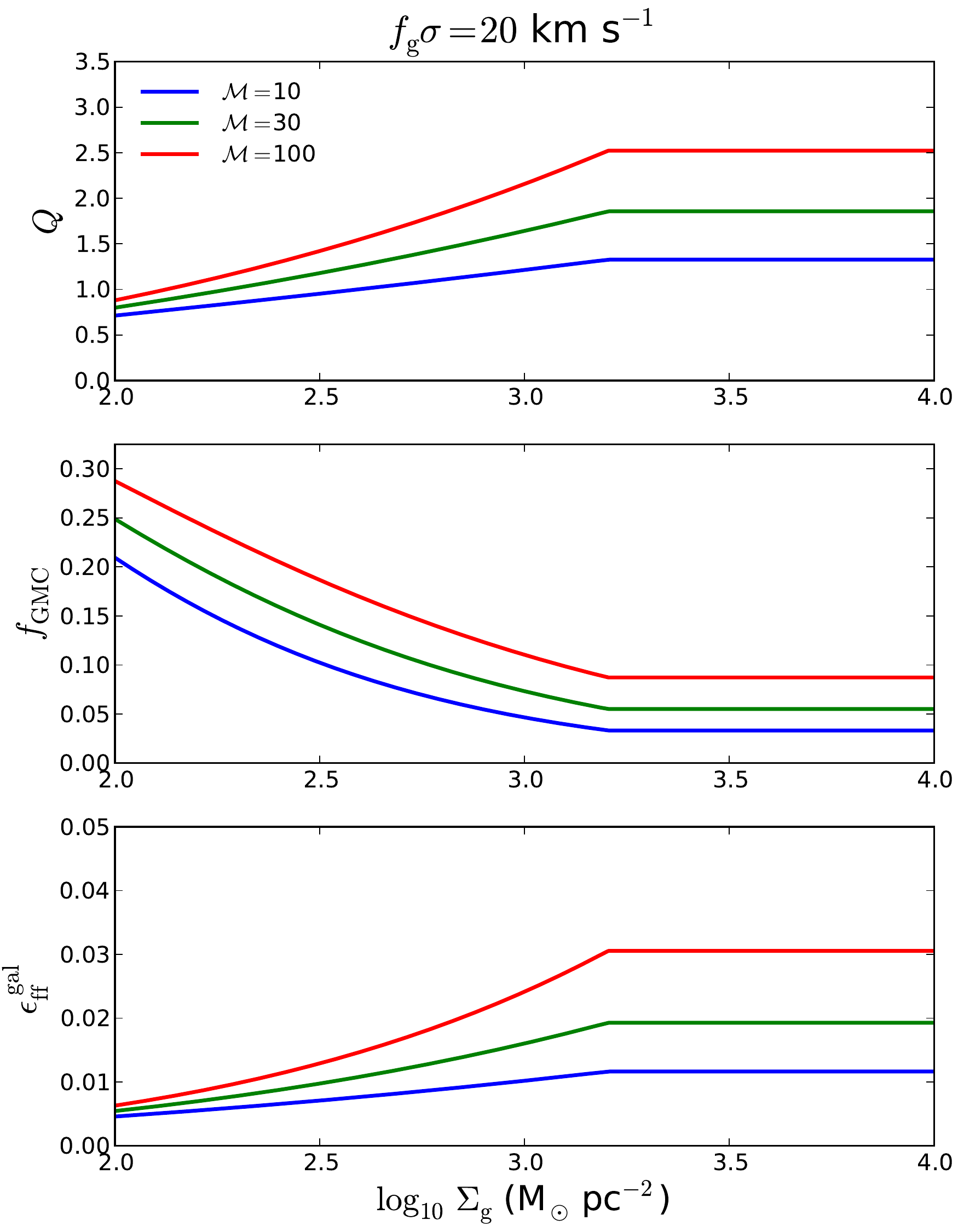}
\includegraphics[width=0.5\textwidth]{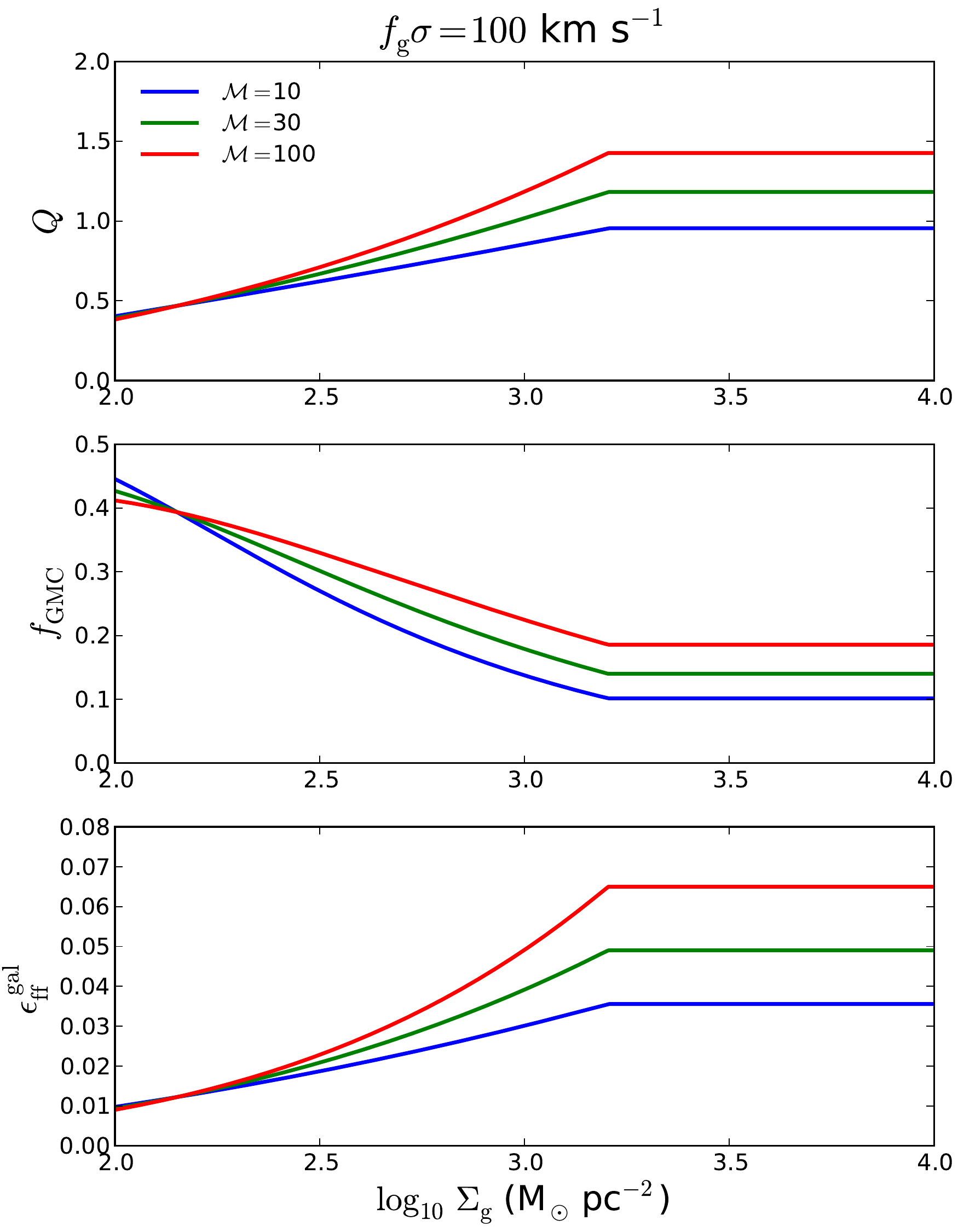}
}
\caption{Equilibrium solutions for the disc-averaged Toomre $Q$ parameter, fraction of the gas mass in gravitationally-bound GMCs ($f_{\rm GMC}$), and the disc-averaged star formation efficiency per free fall time ($\epsilon_{\rm ff}^{\rm gal}$) as a function of the galaxy gas surface density ($\Sigma_{\rm g}$). Each panel shows the curves for different Mach numbers $\mathcal{M}$ of the molecular ISM. The left panel is representative of a low-mass or gas-poor galaxy with $f_{\rm g} \sigma=20$ km s$^{-1}$ and the panel on the right is illustrative of a massive gas-rich system (such as an ULIRG or a high-redshift star-forming galaxy) with $f_{\rm g} \sigma=100$ km s$^{-1}$. 
We assume a simple model of GMC disruption by radiation pressure on dust  to evaluate $\epsilon_{\rm int}^{\rm GMC}$ as a function of $\Sigma_{\rm g}$ (eq. ~\ref{epsGMC Murray}).
The model predicts that $Q$ and $\epsilon_{\rm ff}^{\rm gal}$ increase modestly with $\Sigma_{\rm g}$, while $f_{\rm GMC}$ decreases. 
The flattening at $\Sigma_{\rm g}=1,600$ M$_{\odot}$ pc$^{-2}$ is caused by the saturation of $\epsilon_{\rm int}^{\rm GMC}$ in the optically thick limit. 
These examples assume $P_{\star} / m_{\star} = 3,000$ km s$^{-1}$, $\mathcal{F}=2$ (shown to yield a good fit to the observed disc-averaged $\dot{\Sigma}_{\star}-\Sigma_{\rm g}$ relation in \S \ref{star formation law observations}), $\phi=1$, and $\tilde{t}_{\rm GMC}=1$. 
The equilibrium solutions were obtained using equation (\ref{fcoll EPS}) for the collapsed fraction as a function of $Q$, with values $(a,~b,~\chi)=(1.5,~0.5,~1)$ for the parameters defined in Appendix \ref{collapsed fraction appendix}. 
}
\label{fig:equilibrium_quantities_vs_Sigma_g}
\vspace*{0.1in}
\end{figure*}

\subsection{Consistency with star formation in GMCs}
\label{GMC consistency}
Consider the total mass of stars formed in a time interval of duration $t_{\rm ff}^{\rm disc}$, $M_{\star}(t_{\rm ff}^{\rm disc})$. 
This can be written in two ways:
\begin{align}
\label{Mstar GMC view}
M_{\star}(t_{\rm ff}^{\rm disc}) = \left( \frac{t_{\rm ff}^{\rm disc}}{t_{\rm GMC}} \right) M_{\rm g} f_{\rm GMC} \epsilon_{\rm int}^{\rm GMC}~~~~{\rm (GMC~view)}
\end{align}
and
\begin{align}
\label{Mstar disk view}
M_{\star}(t_{\rm ff}^{\rm disc}) = \epsilon_{\rm ff}^{\rm gal} M_{\rm g}~~~~~~~~~~~~~~~~~{\rm (disc~average~view)}.
\end{align}
The right-hand side of equation (\ref{Mstar GMC view}) is simply the number of GMC generations in one $t_{\rm ff}^{\rm disc}$ times the stellar mass formed in GMCs in one generation, while equation (\ref{Mstar disk view}) follows directly from the definition of $\epsilon_{\rm ff}^{\rm gal}$.\footnote{In equations (\ref{Mstar GMC view}) and (\ref{Mstar disk view}), the time interval is arbitrary and $t_{\rm ff}^{\rm disc}$ was chosen to simplify the algebra that follows. It is assumed that the GMC lifetime is at least one free fall time at the density threshold for gravitational instability, which ensures that the turbulence re-arranges itself rapidly enough that the number of GMC generations in a given time interval is determined by the GMC lifetime, rather than the time scale for new density fluctuations to begin gravitational collapse to a GMC.} 
As before, $f_{\rm GMC}$ and $t_{\rm GMC}$ are the disc gas mass fraction in GMCs and the total GMC lifetime, where GMCs are defined as gravitationally-bound clouds (see \S \ref{two zone disk}). 
Since we assume all star formation occurs in GMCs, we can equate these two expressions to find
\begin{align}
\label{eta consistency}
\epsilon_{\rm ff}^{\rm gal} = \frac{f_{\rm GMC}}{\tilde{t}_{\rm GMC}} \epsilon_{\rm int}^{\rm GMC},
\end{align}
where
\begin{align}
\label{tilde t GMC}
\tilde{t}_{\rm GMC} \equiv \left( \frac{t_{\rm GMC}}{t_{\rm ff}^{\rm disc}} \right)
\end{align}
is the GMC lifetime as a fraction of the disc free fall time. We argued in \S \ref{GMC section} that $\tilde{t}_{\rm GMC} \sim 1$.  
By definition,
\begin{align}
\label{fGMC over tilde t GMC}
\frac{f_{\rm GMC}}{\tilde{t}_{\rm GMC}} = 
\frac{M_{\rm GMC}^{\rm tot}/t_{\rm GMC}}{M_{\rm g}/t_{\rm ff}^{\rm disc}}.
\end{align}

The numerator is the rate at which gas is processed by GMCs, i.e. either turned into stars or returned to the ISM. 
The denominator is proportional to the rate at which gas from the disc is incorporated into bound GMCs, $\dot{M}_{\rm GMC}^{\rm tot}$. 
More precisely, the GMC formation rate is determined by the rate at which turbulent fluctuations become gravitationally unstable,
\begin{align}
\label{g def}
\dot{M}^{\rm tot}_{\rm GMC} \equiv \frac{M_{\rm g} f_{\rm coll}}{\chi t_{\rm flow}},
\end{align}
where $f_{\rm coll}$ is the fraction of the gas mass in the disc that is unstable to gravitational collapse at any time.
We have identified the time scale for the turbulence to rearrange itself and for new GMCs to collapse with the flow crossing time and introduced the dimensionless parameter $\chi \sim 1$ to parameterize the uncertainty in the scaling.  
Using equations (\ref{h over r}) and (\ref{disk free fall time}), we find that 
$t_{\rm flow} =  s(Q) t_{\rm ff}^{\rm disc}$, where
\begin{align}
s(Q) \equiv \frac{4 \times 2^{1/4}}{\pi} \frac{1}{\phi^{1/4} \sqrt{3 Q}},
\end{align}
and thus
\begin{align}
\dot{M}^{\rm tot}_{\rm GMC} = \frac{M_{\rm g} f_{\rm coll}}{\chi s(Q) t_{\rm ff}^{\rm disc}}.
\end{align}

The collapse fraction $f_{\rm coll}$ depends on the Mach number of the turbulence (through the PDF of gas density fluctuations), the epicyclic frequency of the disc (which determines the rotational support on scales $\sim h$), and the disc $Q$ parameter (which quantifies the stability to gravitational collapse). 
It also in general depends on the excitation of spiral structure and other large-scale disc disturbances by internal dynamics or interactions with external perturbers. 

If $Q < 1$, the disc is Toomre-unstable and $f_{\rm coll} \sim 1$. 
For $Q>1$, a quiescent disc supported by thermal pressure would be stable and $f_{\rm coll} \to 0$. 
However, a realistic galactic disc is subject to turbulent fluctuations, so that even if $Q>1$ when evaluated using averaged properties (e.g., over a ring of finite width in the galaxy), random fluctuations imply a certain probability that $Q<1$ at some locations. 
Spatially-resolved spectroscopy of $z\sim2$ star-forming galaxies in fact indicate that the locations of star-forming clumps correspond to local minima in $Q$, where $Q<1$, while $Q$ often exceeds unity outside the clumps \citep[][]{2011ApJ...733..101G}.\footnote{Since the H$\alpha$ observations trace star formation, these maps likely miss a substantial area where $Q>1$.} 

In Appendix \ref{collapsed fraction appendix}, we provide a simple heuristic derivation for $f_{\rm coll}(Q)$ and summarize the more detailed calculation from \cite{2012arXiv1210.0903H}. 
However, these results are sensitive to the amplitude of fluctuations on a smoothing scale $\sim h$ (identified with GMCs) and are subject to significant uncertainties because the properties of the large-scale turbulence depend on the mechanism driving the turbulence and the effects of disc rotation. 
Here, we carry out analytic estimates using a parameterization of $f_{\rm coll}(Q)$ that captures its essential behavior:
\begin{align}
\label{g Q}
f_{\rm coll}(Q) \equiv \beta Q^{-\alpha}~~~~~(Q>1).
\end{align}
The results of Appendix \ref{collapsed fraction appendix} suggest that reasonable parameters are $\beta \approx 0.5$ and $\alpha \approx 5-3$, for Mach numbers $\mathcal{M}=10-100$, respectively. 
 
If GMCs are in steady state in the sense that they process gas at the same rate as they are supplied with gas, $\dot{M}_{\rm GMC}^{\rm tot} = M_{\rm GMC}^{\rm tot} / t_{\rm GMC}$. 
This is likely usually a reasonable assumption in practice, because we expect the lifetimes of GMCs to be comparable to the free fall time of their host disc ($\tilde{t}_{\rm GMC} \sim 1$), and the latter is a lower bound for the time scale over which the host disc changes. 
Then 
\begin{align}
\label{mdot GMC}
f_{\rm coll}(Q) = \chi s(Q) \frac{f_{\rm GMC}}{\tilde{t}_{\rm GMC}} 
\end{align}
and thus equation (\ref{eta consistency}) implies
\begin{align}
\label{eta with g}
\epsilon_{\rm ff}^{\rm gal} = \frac{f_{\rm coll}(Q)}{\chi s(Q)} \epsilon_{\rm int}^{\rm GMC}.
\end{align}
Using $c_{\rm T} = Q f_{\rm g} \sigma / 2$ (eq. \ref{cT sigma}), equations (\ref{cT vs eta}), (\ref{g Q}) and (\ref{eta with g}) can be combined to eliminate $\epsilon_{\rm ff}^{\rm gal}$ and solve for $Q$:
\begin{align}
Q = \left[ \frac{\beta \mathcal{F}}{2 \chi \phi^{1/4}} \frac{(P_{\star} / m_{\star})}{f_{\rm g} \sigma} \epsilon_{\rm int}^{\rm GMC} \right]^{1/(\alpha+1)}
\end{align}
(This solution assumes that $\phi$ is a constant; for the case of a pure gas disc with no external potential, $\phi \sim 1/Q$ (\S \ref{two zone disk}) and the solution involves an additional power of $Q$.) 
We can use this solution to directly relate the galaxy-averaged and GMC star formation efficiencies:
\begin{align}
\label{eta vs eps}
\epsilon_{\rm ff}^{\rm gal} & = \frac{\sqrt{3} \pi \phi^{1/4}}{4\times2^{1/4} \chi} 
\left[ 
\frac{2 \chi \phi^{1/4}}{\mathcal{F} \beta} \frac{f_{\rm g} \sigma}{(P_{\star}/m_{\star})}
\right]^{(\alpha-1/2)/(\alpha+1)} \\ \notag
& ~~~~~~~~~~~~~~~~~~~~~~~~~~~~~~~~~~~~~~~~~~~\times  (\beta \epsilon_{\rm int}^{\rm GMC})^{3/2(\alpha+1)}
\end{align}

The limit $\alpha \to \infty$, in which the probability of GMC formation is suppressed very rapidly as $Q$ exceeds unity, is particularly enlightening. 
Then,
\begin{align}
Q \stackrel{\alpha \to \infty}{\to} 1
\end{align}
and
\begin{align}
\label{eps ff gal alpha infty}
\epsilon_{\rm ff}^{\rm gal} \stackrel{\alpha \to \infty}{\to} \frac{\sqrt{3} \pi \phi^{1/2}}{2^{5/4} \mathcal{F}} \frac{f_{\rm g} \sigma}{(P_{\star} / m_{\star})} \sim \frac{f_{\rm g} \sigma}{(P_{\star} / m_{\star})} \approx 2 \frac{c_{\rm T}}{(P_{\star} / m_{\star})}.
\end{align}
In this limit, in which $Q$ approaches exactly unity, the galaxy-averaged star formation efficiency per free fall time is set by the dimensionless ratio $f_{\rm g} \sigma / (P_{\star}/m_{\star})$, a proxy for the ratio of the strength of gravity to the strength of feedback. 
Perhaps most importantly, in the $\alpha \to \infty$ limit, $\epsilon_{\rm ff}^{\rm gal}$ is completely independent of how efficiently gas is converted into stars once collapsed in GMCs. 
The rate limiting step for star formation in this case is the rate at which GMCs form in the galactic disc. 
For $Q=1$, $c_{\rm T} = f_{\rm g} \sigma / 2$ (eq. \ref{cT sigma}) and thus $\epsilon_{\rm ff}^{\rm gal}$ is also proportional to the ratio $c_{\rm T} / (P_{\star} / m_{\star})$. 

For finite $\alpha \gtrsim 1$, $Q$ is regulated to a value that can exceed unity and there is a weak dependence of $\epsilon_{\rm ff}^{\rm gal}$ on $\epsilon_{\rm int}^{\rm GMC}$. 
This is illustrated in Figure \ref{fig:equilibrium_quantities}, in which we show equilibrium solutions for $Q$, $f_{\rm GMC}$, and $\epsilon_{\rm ff}^{\rm gal}$ (simultaneously satisfying eq. \ref{cT sigma}, \ref{cT vs eta}, and \ref{eta with g}) as a function of $\epsilon_{\rm int}^{\rm GMC}$ obtained using equation (\ref{fcoll EPS}) for the collapsed fraction as a function of $Q$ instead of the power-law approximation in eq. (\ref{g Q}). 
The equilibrium solutions assume a constant $t_{\rm GMC} = t_{\rm ff}^{\rm disc}$ ($\tilde{t}_{\rm GMC}=1$; see \S \ref{GMC section}). 
The left panel in Figure \ref{fig:equilibrium_quantities} is representative of a low-mass or gas-poor galaxy with $f_{\rm g} \sigma=20$ km s$^{-1}$ and the panel on the right is illustrative of a massive gas-rich system (such as an ULIRG or a high-redshift star-forming galaxy) with $f_{\rm g} \sigma=100$ km s$^{-1}$. 
In the limit $\epsilon_{\rm int}^{\rm GMC} \to 0$, GMCs do not form any stars and the assumptions of our model break down. 
We thus show the equilibrium solutions only for $\epsilon_{\rm int}^{\rm GMC} \geq 0.02$, the minimum efficiency implied by equation (\ref{epsGMC Murray}) below for $\Sigma_{\rm g} \geq 100$ M$_{\odot}$ pc$^{-2}$.   
The disc-averaged star formation efficiency per free fall time increases slowly with the integrated GMC star formation efficiency, while $f_{\rm GMC}$ decreases. 
For example, for a Mach number $\mathcal{M}=30$, $\epsilon_{\rm ff}^{\rm gal}$ increases by a factor $\sim3$ when $\epsilon_{\rm int}^{\rm GMC}$ increases by a factor of $20$ from $0.05$ to $1$. 
This is realized by the global $Q$ parameter increasing above the classical stability threshold of unity, so that GMCs form only where turbulent fluctuations cause the self-gravity of the gas to exceed its turbulent and rotational support. 
Figure \ref{fig:equilibrium_quantities} also shows that for small $\epsilon_{\rm int}^{\rm GMC}$, the star formation rate is too low to support the disc to $Q\gtrsim1$ and the disc settles to a $Q<1$.

In Figure \ref{fig:equilibrium_quantities_vs_Sigma_g}, we show the same equilibrium solutions but as a function of $\Sigma_{\rm g}$ for a simple model of how $\epsilon_{\rm int}^{\rm GMC}$ depends on $\Sigma_{\rm g}$ for GMC disruption by radiation pressure on dust.  
Specifically, we assume
\begin{align}
\label{epsGMC Murray}
\epsilon_{\rm int}^{\rm GMC}  = \min \left\{ \frac{\pi G \Sigma_{\rm g} c}{2 (L/M_{\star})},~0.35 \right\}
\end{align}
based on the scaling arguments and 1-D numerical models of \cite{2010ApJ...709..191M}. 
These authors showed that $\epsilon_{\rm int}^{\rm GMC} \sim (\pi G \Sigma_{\rm GMC} c )/ [2 (L/M_{\star})]$ for GMCs that are optically thin to far-infrared radiation (where $\Sigma_{\rm GMC}$ is the GMC surface density), reaching a constant $\sim 0.35$ as GMCs become optically thick to far-infrared photons. 
Here, we identify $\Sigma_{\rm GMC}$ with $\Sigma_{\rm g}$, the mean gas surface density in the galaxy, and adopt a light-to-mass ratio $L/M_{\star} = 3,000$ cm$^{2}$ s$^{-3}$. 
In this case, Figure \ref{fig:equilibrium_quantities_vs_Sigma_g} shows that 
$\epsilon_{\rm ff}^{\rm gal}$ increases by a factor of a few with $\Sigma_{\rm g}$ ranging from $100$ to $1,600$ M$_{\odot}$ pc$^{-2}$, but that $f_{\rm GMC}$ decreases with $\Sigma_{\rm g}$ over this range. 
The flattening at $\Sigma_{\rm g}=1,600$ M$_{\odot}$ pc$^{-2}$ is caused by the saturation of $\epsilon_{\rm int}^{\rm GMC}$ in the optically thick limit. 
We return to the prediction of decreasing $f_{\rm GMC}$ with increasing $\Sigma_{\rm g}$ in \S \ref{fGMC observations}. 

\section{COMPARISON WITH OBSERVATIONS}
\label{comparison observations}

\subsection{CO conversion factor}
\label{alpha CO section}
One complication in comparing star formation models to observations arises because the molecular gas mass is usually estimated using CO emission lines that are optically thick. 
Furthermore, the CO mass is a small fraction of the total molecular mass and is therefore only an indirect tracer. 
The $X_{\rm CO} \equiv N_{\rm H_{2}} / I_{\rm CO}$ conversion factor between CO intensity and molecular gas column density depends on several factors, including the gas density, gas metallicity, gas temperature, gas turbulent velocity dispersion, and the ambient radiation field \citep[e.g.,][]{1988ApJ...325..389M, 2011MNRAS.412..337G, 2011MNRAS.415.3253S,  2012ApJ...746...69G, 2012MNRAS.421.3127N, 2012ApJ...747..124F}.  
An equivalent factor $\alpha_{\rm CO} \equiv \Sigma_{\rm H_{2}} / I_{\rm CO}$ is related to $X_{\rm CO}$ by $X_{\rm CO} = 6.3 \times 10^{19}~\alpha_{\rm CO}$, where $X_{\rm CO}$ has units of cm$^{-2}$/(K km s$^{-1}$) and $\alpha_{\rm CO}$ has units of M$_{\odot}$ pc$^{-2}$/(K km s$^{-1}$). 

In the local Universe, observations find a nearly constant $\alpha_{\rm CO}=3.2$ for ordinary galaxies, including the Milky Way \citep[e.g.,][]{1996A&A...308L..21S, 2001ApJ...547..792D, 2007prpl.conf...81B}. 
For local ULIRGs associated with galaxy mergers, a significantly lower conversion factor $\alpha_{\rm CO} \lesssim 1$ is inferred on the basis that a Milky Way-like factor would imply a gas mass exceeding the dynamical mass of the galaxy \citep[][]{1997ApJ...478..144S}. 
On the other hand, $\alpha_{\rm CO}$ rises at low metallicities such as in the Small Magellanic Cloud \citep[][]{2007prpl.conf...81B} and for high-redshift galaxies below a critical stellar mass \citep[][]{2012ApJ...746...69G}, probably because CO molecules are photo-dissociated by the UV radiation field as the attenuation by dust decreases \citep[e.g.,][]{1986ApJS...62..109V, 2010ApJ...716.1191W}. 

The CO conversion factor is important for our comparison to observations because different assumptions lead to different $\dot{\Sigma}_{\star} - \Sigma_{\rm g}$ relations and galaxy-averaged star formation efficiencies $\epsilon_{\rm ff}^{\rm gal}$. 
One common assumption, adopted for example in the observational study of \cite{2010MNRAS.407.2091G}, is that of a bimodal conversion factor $\alpha_{\rm CO}=3.2$ for non-mergers and $\alpha_{\rm CO}=1$ for mergers. 
This assumption leads to a $\dot{\Sigma}_{\star} - \Sigma_{\rm g}$ relation that is also bimodal, with mergers occupying an elevated locus. 
A similar result was found by \cite{2010ApJ...714L.118D}.  
Theoretical considerations have led other authors to propose an $\alpha_{\rm CO}$ factor that depends smoothly on galaxy properties. 
\cite{2011ApJ...731...41O} suggested that $\alpha_{\rm CO}$ should depend primarily on $\Sigma_{\rm g}$. 
Interpolating between $\alpha_{\rm CO}=3.2$ (non-merger value) at $\Sigma_{\rm g}=100$ M$_{\odot}$ pc$^{-2}$ and $\alpha_{\rm CO}=1$ (merger value) at $\Sigma_{\rm g}=1,000$ M$_{\odot}$ pc$^{-2}$ with a power-law, they propose that
\begin{align}
\label{OS alpha CO}
\alpha_{\rm CO}^{\rm OS} = 1~\left( \frac{\Sigma_{\rm g}}{\rm 1,000~M_{\odot}~pc^{-2}} \right)^{-0.52}.
\end{align}
Using this expression, they showed that the merger and non-merger galaxies in the \cite{2010MNRAS.407.2091G} sample align on an uni-modal $\dot{\Sigma}_{\star} - \Sigma_{\rm g}$ relation. 
In this picture, mergers have lower $\alpha_{\rm CO}$ because they have higher characteristic $\Sigma_{\rm g}$. 

\cite{2012MNRAS.421.3127N} performed radiative transfer calculations on simulations of merging and non-merging galaxies and predicted the dependence of $\alpha_{\rm CO}$ on galaxy properties. 
Based on these calculations, these authors also advocate an $\alpha_{\rm CO}$ that varies smoothly with galaxy properties, rather than a bimodal distribution between mergers and non-mergers. 
For solar metallicity, the calculations of \cite{2012MNRAS.421.3127N} yield a dependence on $\Sigma_{\rm g}$ close to the empirical fit of \cite{2011ApJ...731...41O} in equation (\ref{OS alpha CO}). 
\cite{2012MNRAS.421.3127N} also show, in agreement with \cite{2011ApJ...731...41O}, that assuming a smoothly-varying $\alpha_{\rm CO}$ as predicted by their analysis leads to a $\dot{\Sigma}_{\star} - \Sigma_{\rm g}$ relation that has a smaller scatter than when assuming a bimodal model. 
They further argue that the scatter is reduced even for a range of galaxies excluding mergers, so that the smaller scatter is not simply due to using an unimodal $\alpha_{\rm CO}$, indicating that the dependence on $\Sigma_{\rm g}$ is important. 

A recent observational study using dust masses to estimate the gas masses of $z\sim2$ galaxies \citep{2012arXiv1210.1035M} infers CO conversion factors consistent with the metallicity-dependent theoretical prediction of \cite{2012MNRAS.421.3127N} within uncertainties. 
This is the case even though the $\alpha_{\rm CO}$ factor of ordinary high-redshift star forming galaxies is also formally consistent with a standard Milky Way value, owing to two factors. First, while ordinary high-redshift star forming galaxies have elevated star formation rates relative to local galaxies, their star formation is typically more spatially-extended than that in local mergers, so that the gas surface densities are often not as high as those of local mergers. 
Second, high-redshift galaxies generally have lower metallicity $Z$, which \cite{2012MNRAS.421.3127N} predict implies a larger $\alpha_{\rm CO} \propto 1/Z$ at fixed $\Sigma_{\rm g}$. 

In light of these considerations, we use an $\alpha_{\rm CO}$ depending smoothly on $\Sigma_{\rm g}$ as in equation (\ref{OS alpha CO}) in our analysis. 

\begin{figure}
\includegraphics[width=0.98\columnwidth]{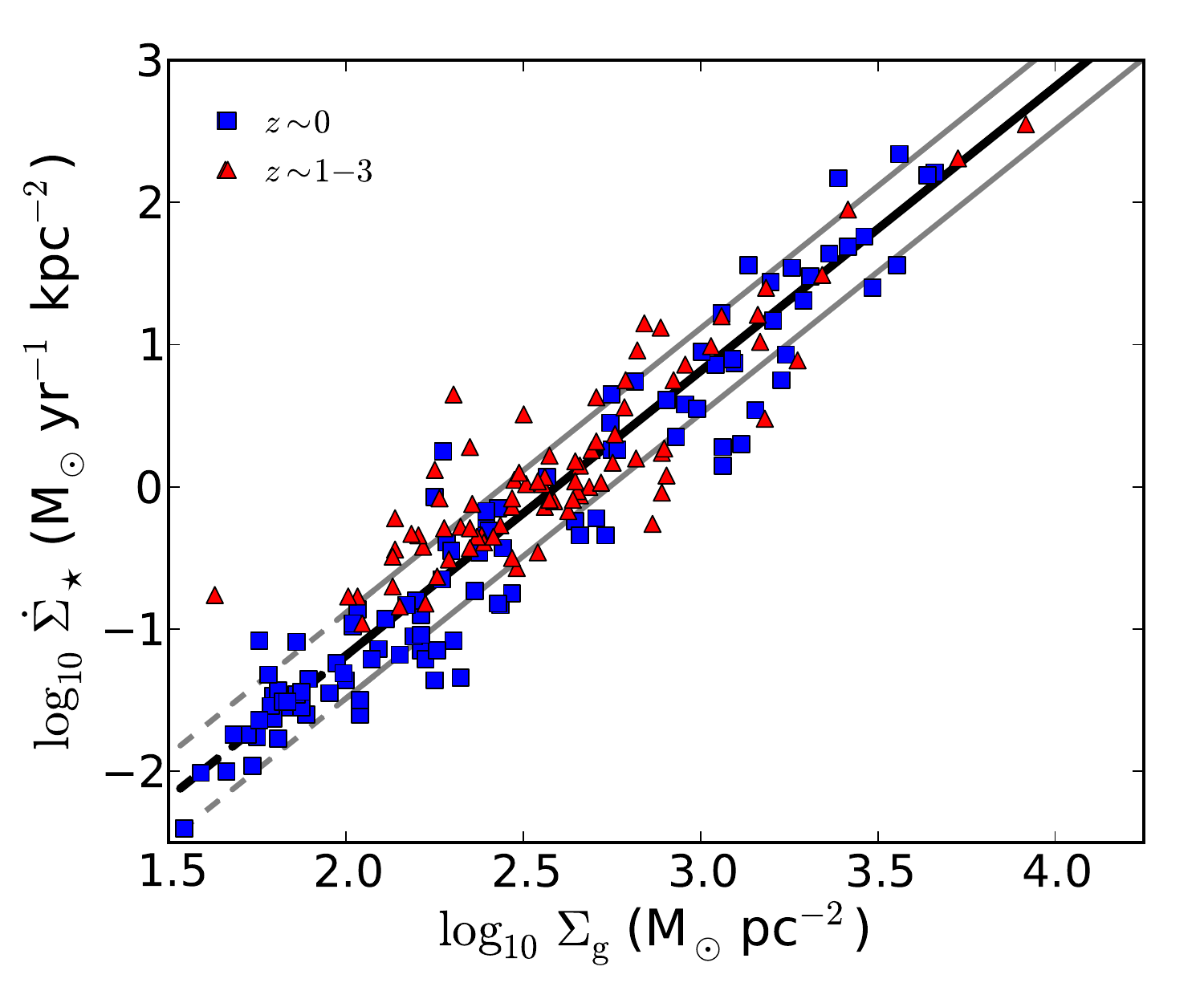}
\caption{
Disc-averaged star formation law. 
All data are from the compilations of star-forming galaxies of Genzel et al. (2010) and Tacconi et al. (2012). 
We distinguish between local $z\sim0$ galaxies (blue squares) and high-redshift $z\sim1-3$ galaxies (red triangles). 
For each redshift interval, the data include both merging and non-merging galaxies. 
The gas surface densities were converted from the values reported by Genzel et al. (2010) and Tacconi et al. (2012) using an $\alpha_{\rm CO}$ factor interpolating smoothly between standard merger and non-merger values using equation (\ref{OS alpha CO}). 
We assume that all gas is molecular. 
The solid black line shows the theoretical prediction $\dot{\Sigma}_{\star} \propto \Sigma_{\rm g}^{2}$ in equation (\ref{star formation law momentum balance}) obtained by balancing the momentum input from stellar feedback with the vertical weight of the disc gas. 
The line assumes $Q=\phi=1$, $P_{\star}/m_{\star} = 3,000$ km s$^{-1}$, and the dimensionless normalization $\mathcal{F}=2$. 
The parallel grey lines indicate the range $Q\approx0.5-1.5$ expected for gas-rich galaxies (see Fig. \ref{fig:equilibrium_quantities}) and the dashed segments show extrapolations to $\Sigma_{\rm g}<100$ M$_{\odot}$ pc$^{-2}$.  
}
\label{fig:sf_law}
\vspace*{0.1in}
\end{figure}

\begin{figure*}
\mbox{\includegraphics[width=0.98\columnwidth]{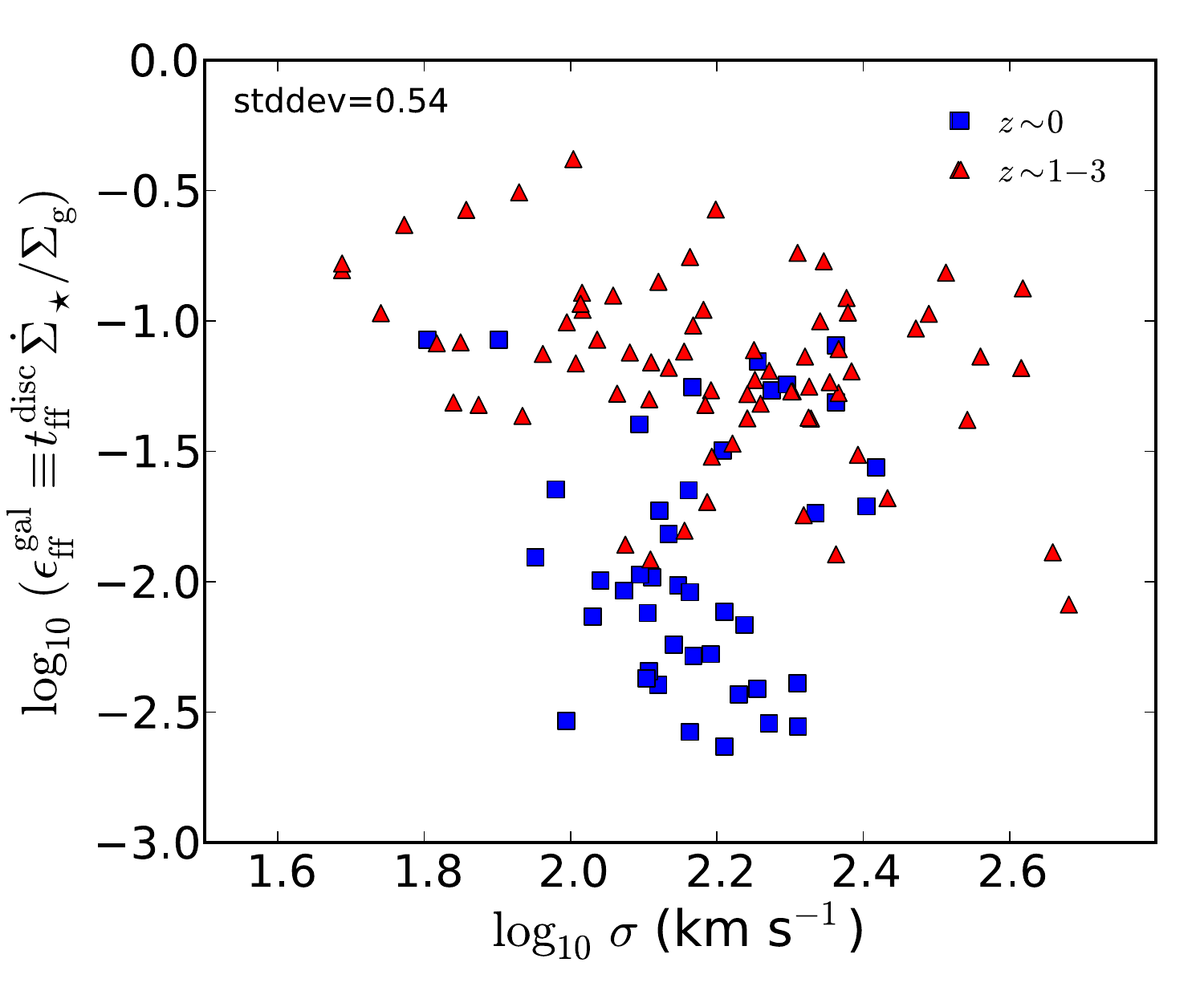}
\includegraphics[width=0.98\columnwidth]{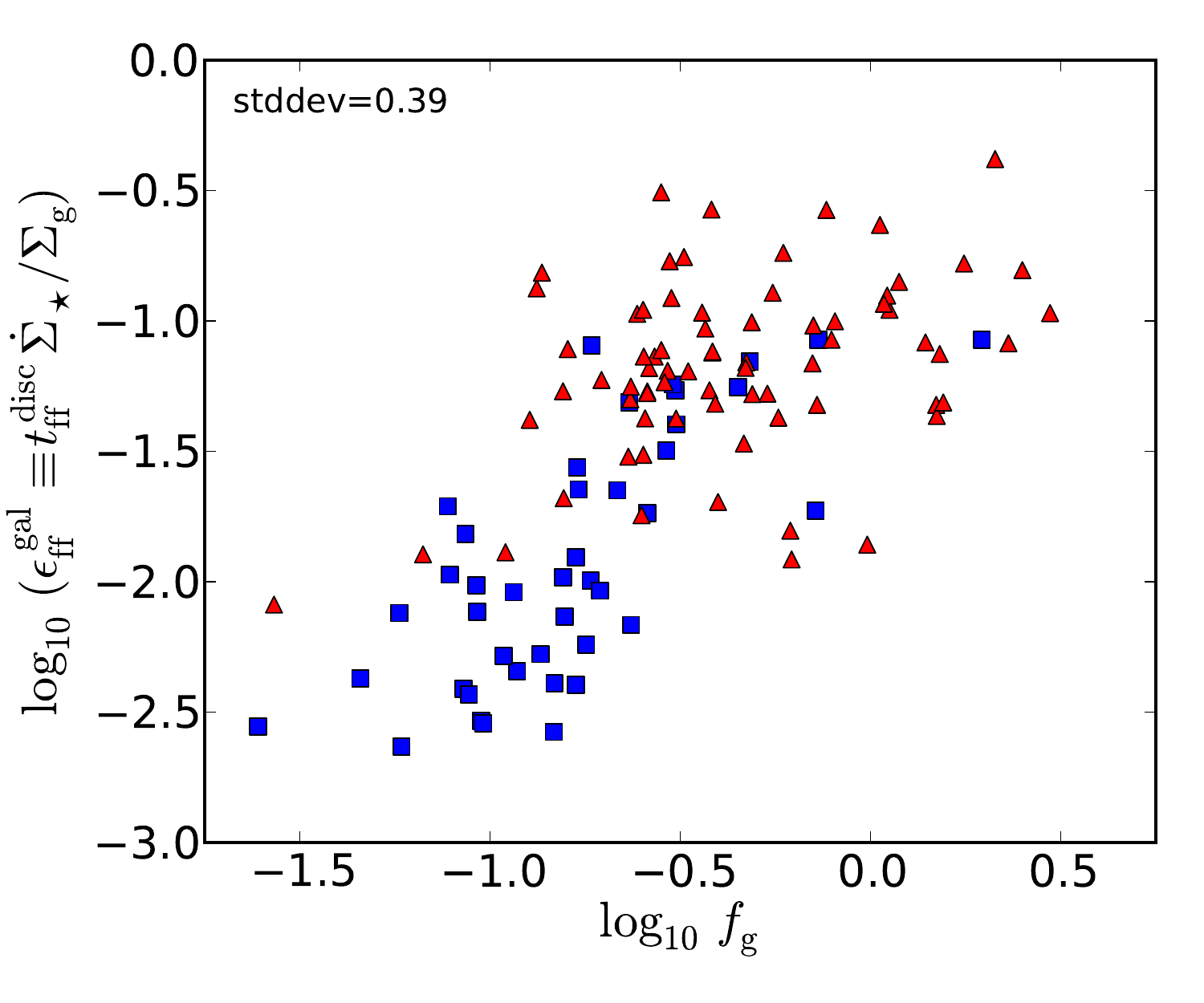}}
\mbox{
\includegraphics[width=0.98\columnwidth]{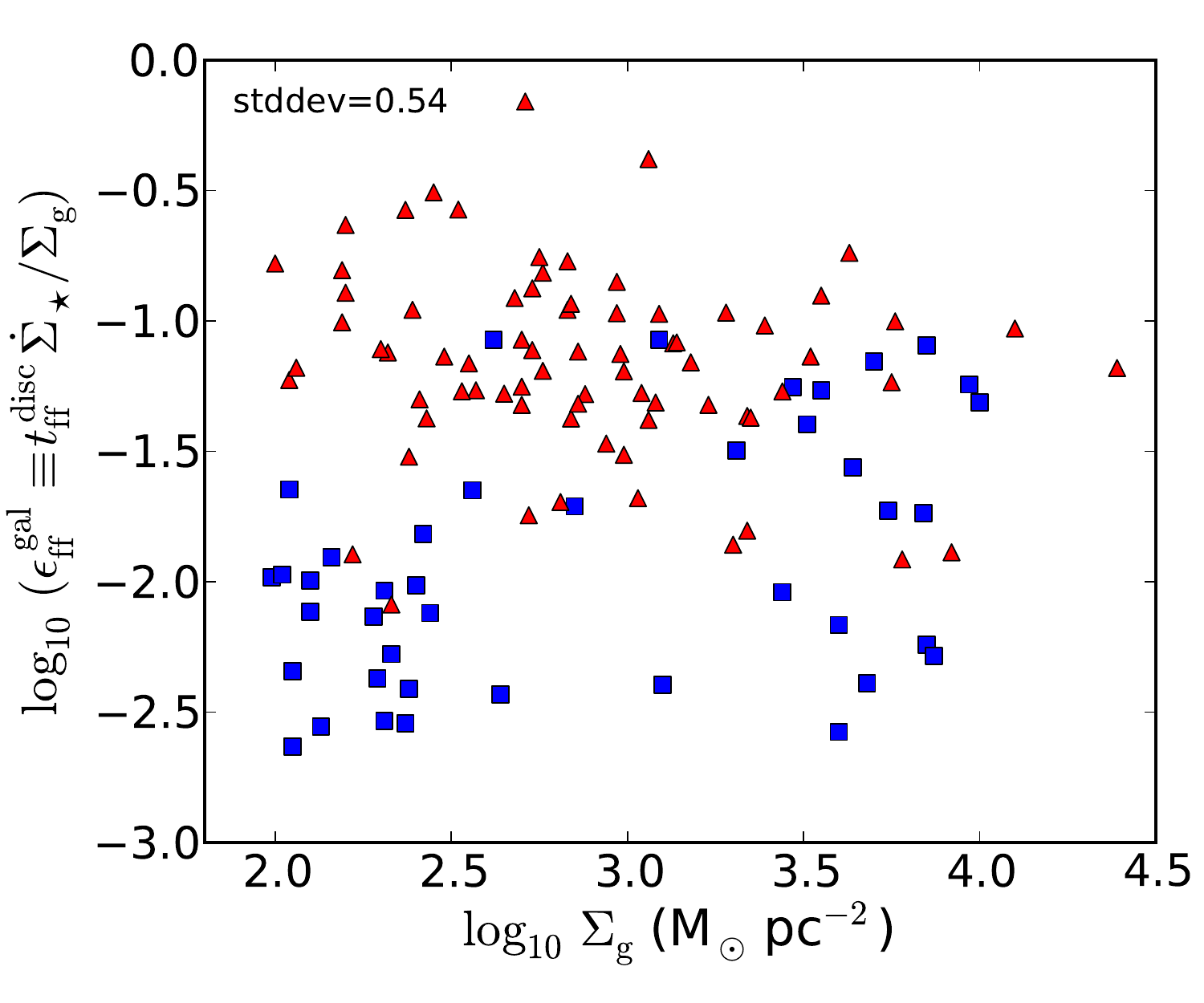}
\includegraphics[width=0.98\columnwidth]{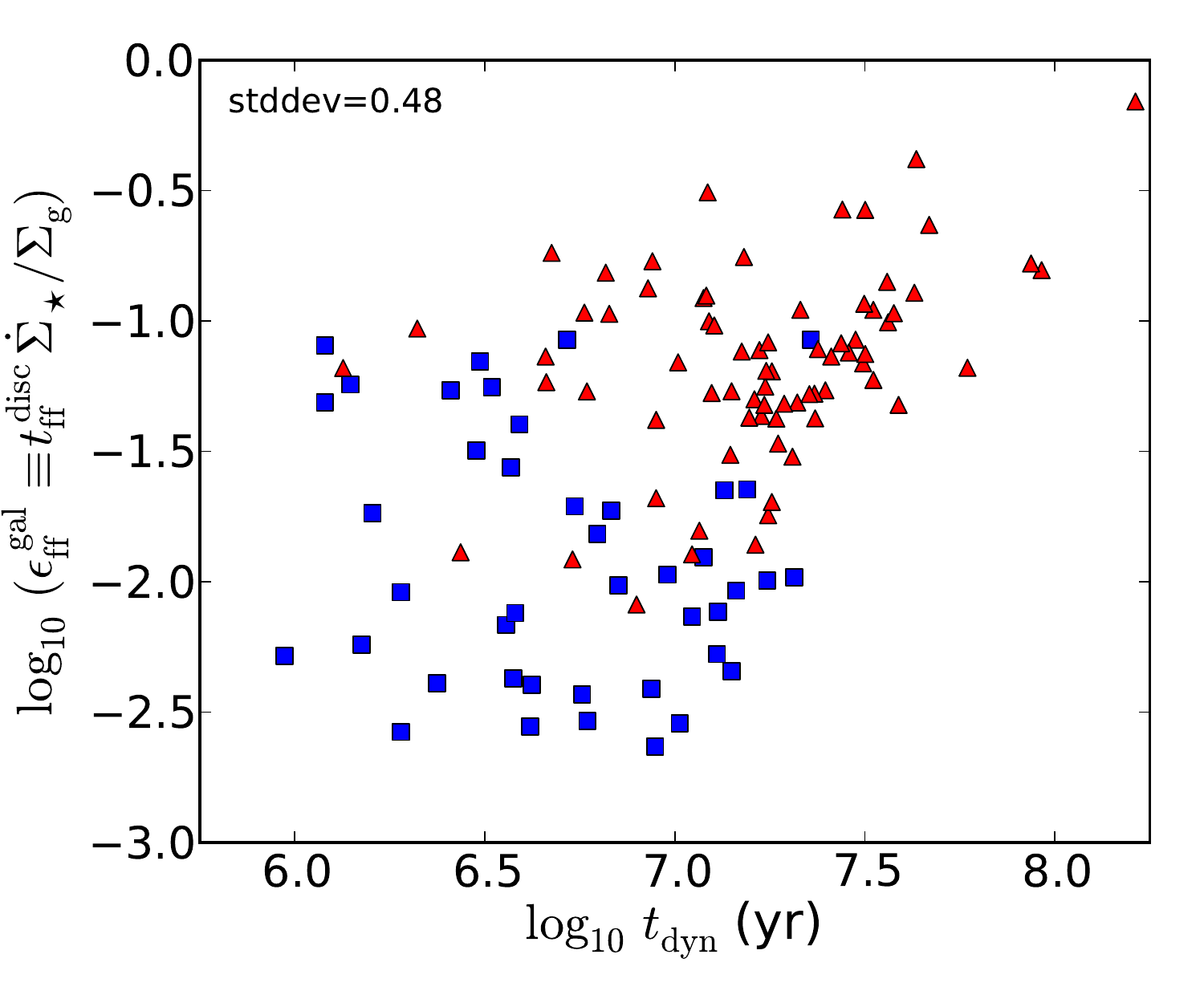}}
\caption{Disc-averaged star formation efficiency per free fall time as a function of isothermal potential velocity dispersion ($\sigma=v_{\rm c}/\sqrt{2}$, where $v_{\rm c}$ is the maximum circular velocity), gas mass fraction $f_{\rm g}$, gas surface density $\Sigma_{\rm g}$, and dynamical time in the disc ($t_{\rm dyn} \equiv R_{\rm 1/2} / v_{\rm c}$, where $R_{\rm 1/2}$ is the half-light radius). 
The observed disc-averaged star formation efficiency per free fall time is non-universal and increases with increasing gas fraction (upper right panel), as predicted by our feedback model (eq. \ref{eps ff gal alpha infty}). 
The overall scatter in the star formation efficiency from galaxy to galaxy is a factor $\sim100$. 
In each panel, the scatter relative to the best-fit linear relation in log-log space is quantified by the {\tt stddev} statistic described in \S \ref{star formation efficiency observations}. 
The data are taken from the observational compilations of Genzel et al. (2010) and Tacconi et al. (2012) but converted to a smoothly-varying CO conversion as described in \S \ref{comparison observations}. 
We distinguish between local $z\sim0$ galaxies (blue squares) and high-redshift $z\sim1-3$ galaxies (red triangles) as in Figure \ref{fig:sf_law}. 
In accordance with the assumptions of our modeling, we only show data for galaxies with $\Sigma_{\rm g} > 100$ M$_{\odot}$ pc$^{-2}$. 
Some estimated gas masses exceed the galaxy dynamical mass ($f_{\rm g}>1$) by a factor up to $\sim3$, but this does not significantly affect our conclusions (see \S \ref{star formation efficiency observations}). 
}
\label{fig:eta_comparison}
\vspace*{0.1in}
\end{figure*}

\begin{figure}
\includegraphics[width=0.98\columnwidth]{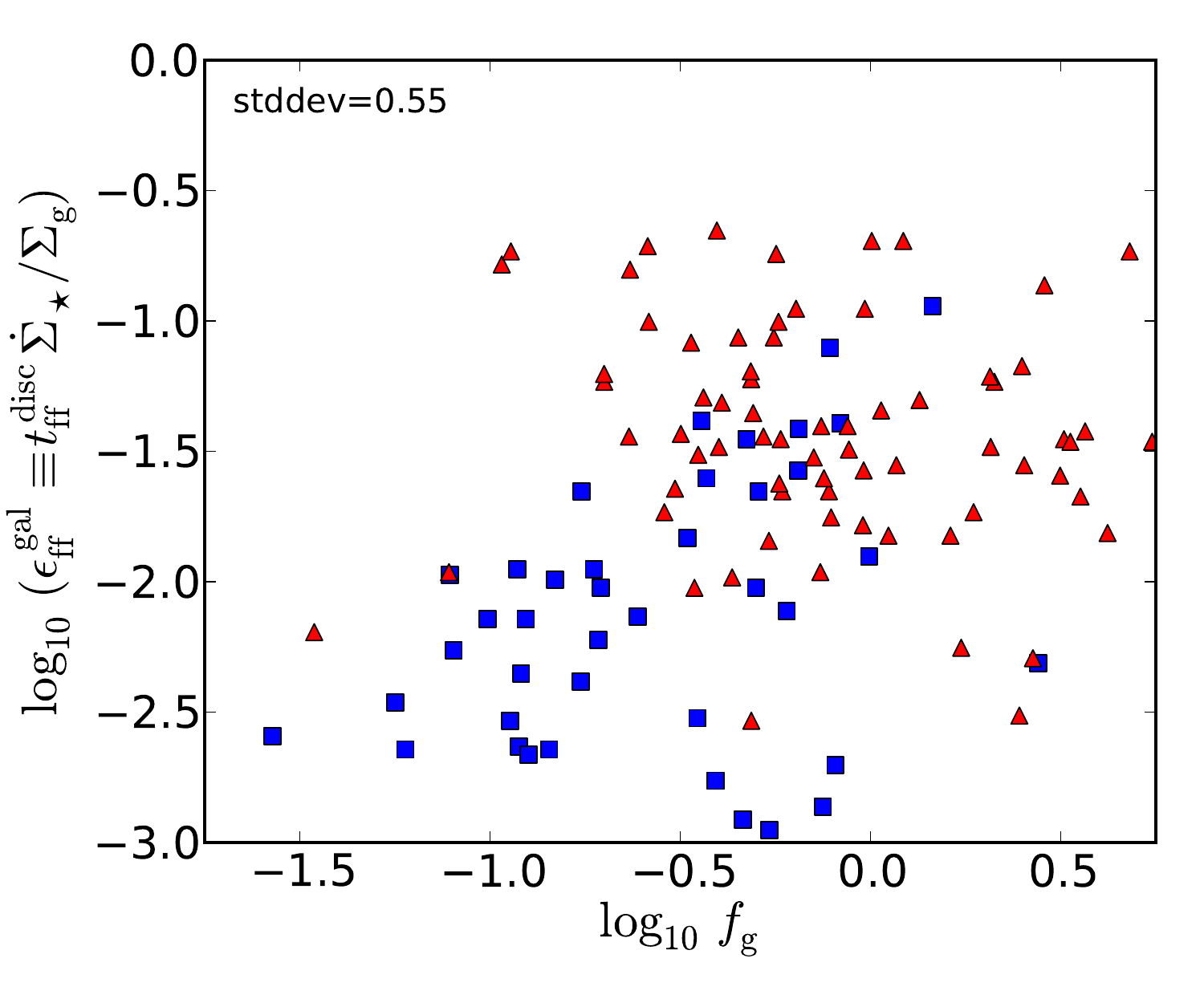}
\caption{
Disc-averaged star formation efficiency per free fall time as a function of gas mass fraction, assuming a bimodal $\alpha_{\rm CO}$ conversion factor. 
The data are the same as in the upper right panel of Figure \ref{fig:eta_comparison} but assume a bi-modal $\alpha_{\rm CO}$ for mergers and non-mergers, as in Genzel et al. (2010) and Tacconi et al. (2012), rather than the continuous model in equation (\ref{OS alpha CO}).  
This results in a larger scatter in the data and a larger number of inferred gas masses exceeding the dynamical mass of the galaxy. 
As in the case of the continuously-varying $\alpha_{\rm CO}$ assumption, the data do not strongly support the existence of a universal value of $\epsilon_{\rm ff}^{\rm gal}$.
}
\label{fig:eta_vs_fg_bimodal}
\vspace*{0.1in}
\end{figure}

\begin{figure*}
\mbox{
\includegraphics[width=0.98\columnwidth]{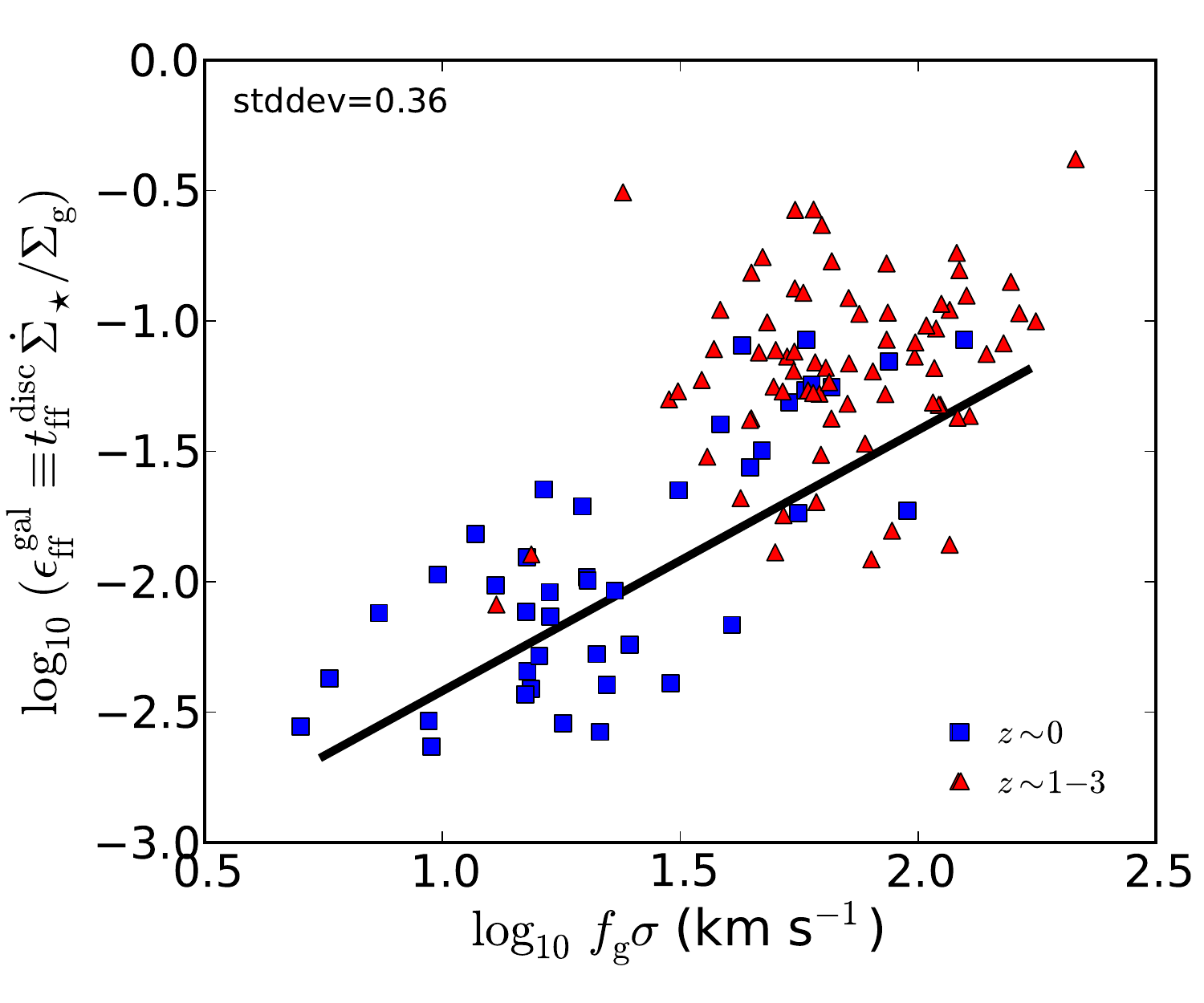}
\includegraphics[width=0.98\columnwidth]{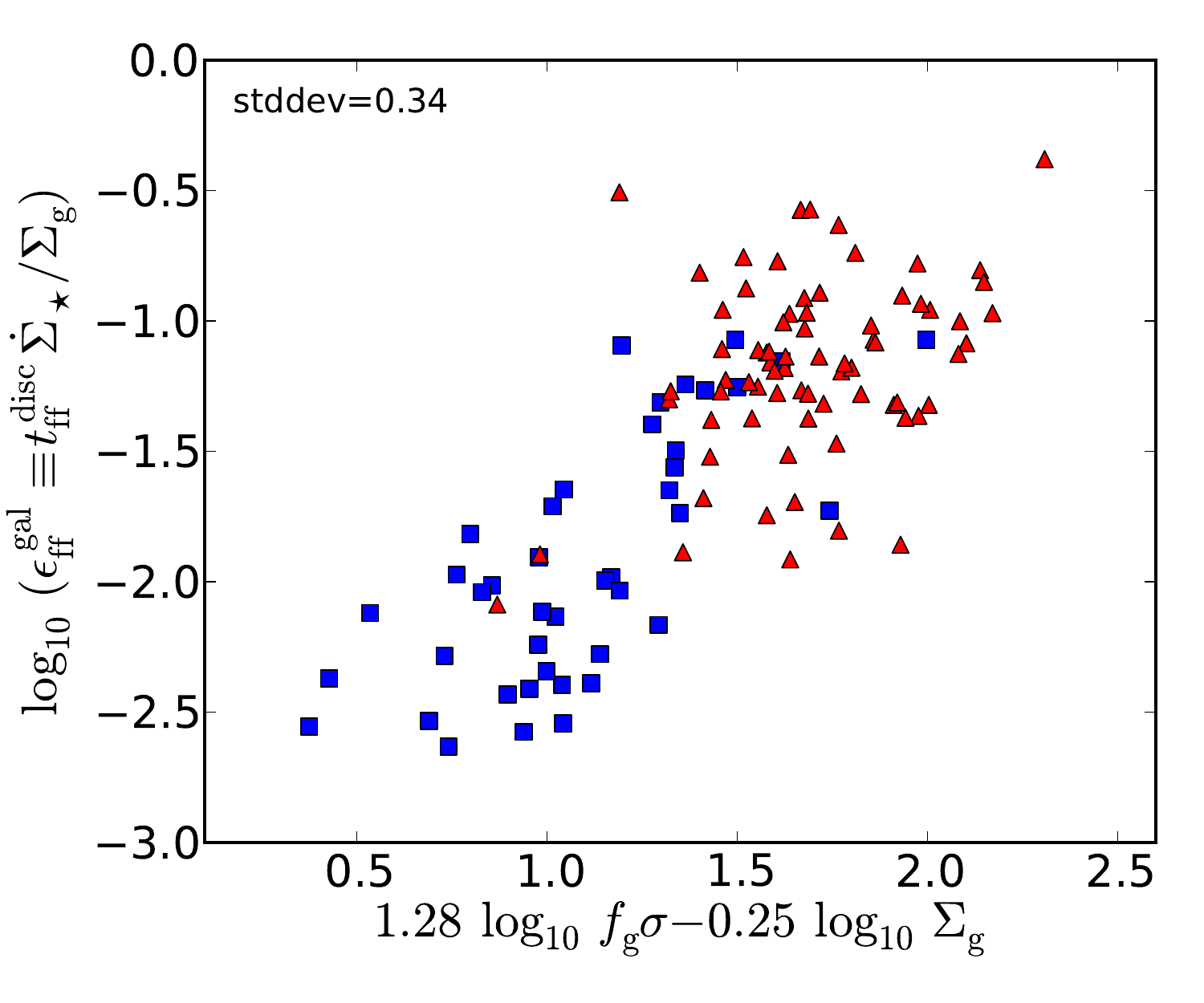}
}
\caption{Same as in Figure \ref{fig:eta_comparison}, but for the disc-averaged star formation efficiency per free fall time as a function of combinations of $f_{\rm g}$, $\sigma$, and $\Sigma_{\rm g}$. 
The solid black line in the left panel shows the theoretical prediction $\epsilon_{\rm ff}^{\rm gal} \propto f_{\rm g} \sigma$ of our feedback model in equation (\ref{eps ff gal alpha infty}), valid in the limit $\alpha \to \infty$, for $\phi=1$, $P_{\star}/m_{\star} = 3,000$ km s$^{-1}$, and $\mathcal{F}=2$ (as in Figure \ref{fig:sf_law}). 
The observed disc-averaged star formation efficiency per free fall time is non-universal and increases with $f_{\rm g} \sigma$, as predicted by our feedback model (eq. \ref{eps ff gal alpha infty}). 
This agreement emphasizes that the large scatter in $\epsilon_{\rm ff}^{\rm gal}$ versus $\Sigma_{\rm g}$ in Figure \ref{fig:eta_comparison} is not simply scatter about a universal value of $\epsilon_{\rm ff}^{\rm gal}$, since the scatter is significantly reduced when the dependence on $f_{\rm g} \sigma$ is accounted for. 
The horizontal axis in the right hand panel is the best-fit linear combination of $\log_{10}{f_{\rm g} \sigma}$ and $\log_{10}{\Sigma_{\rm g}}$. 
The scatter is not significantly reduced by allowing a dependence of the disc-averaged star formation efficiency on $\Sigma_{\rm g}$, indicating that $f_{\rm g} \sigma$ is the most important parameter.}
\label{fig:eta_comparison2}
\vspace*{0.1in}
\end{figure*}

\subsection{Disc-averaged star formation law}
\label{star formation law observations}
In Figure \ref{fig:sf_law}, we plot $\dot{\Sigma}_{\star}$ versus $\Sigma_{\rm g}$ for the data compiled by \cite{2010MNRAS.407.2091G}, supplemented by a larger sample of $z\sim1-3$ star-forming galaxies from \cite{2012arXiv1211.5743T}. 
This compilation includes both ordinary and merging galaxies. 
In accordance with the range of applicability of our model, we focus on data for galaxies with $\Sigma_{\rm g} \gtrsim 100$ M$_{\odot}$ pc$^{-2}$ and assume that all gas is molecular. 
We convert the gas surface densities from those reported by \cite{2010MNRAS.407.2091G} and \cite{2012arXiv1211.5743T} (who assume a bimodal $\alpha_{\rm CO}$) to values obtained using equation (\ref{OS alpha CO}). 
Figure \ref{fig:sf_law} shows the data points converted in this way lie on a well-defined unimodal $\dot{\Sigma}_{\star} - \Sigma_{\rm g}$ relation. 

On this plot, we also show the model prediction $\dot{\Sigma}_{\star} \propto \Sigma_{\rm g}^{2}$ in equation (\ref{star formation law momentum balance}). 
We set $Q=\phi=1$ in this comparison and also hold $P_{\star}/m_{\star}$ fixed to the fiducial value of $3,000$ km s$^{-1}$. Departures from these assumptions are encapsulated in the dimensionless factor $\mathcal{F}$; a good match to the data is obtained for $\mathcal{F} = 2$. 
The fact that a good fit is found for a constant $P_{\star}/m_{\star}$ is consistent with the weak dependence of the momentum input by SNe on ambient density and turbulent gas velocity dispersion discussed in Appendix \ref{model closure appendix}. 
Interestingly, the star formation law that fits the $\Sigma_{\rm g}\geq100$ M$_{\odot}$ pc$^{-2}$ data also fits the data well for $\Sigma_{\rm g}$ as low as $30$ M$_{\odot}$ pc$^{-2}$, suggesting that our model may apply somewhat more broadly. 
Parallel grey lines indicate the range $Q\approx0.5-1.5$ expected for gas-rich galaxies (see Fig. \ref{fig:equilibrium_quantities}).

At very high gas surface densities, supernovae become increasingly ineffective due to radiative losses while radiation pressure on dust becomes more important in the volume-filling ISM (Appendix \ref{model closure appendix}). 
In the limit of the disc being optically thick to reprocessed infrared photons, $P_{\star}/m_{\star} \propto \Sigma_{\rm g}$, so we expect a flattening to $\dot{\Sigma}_{\star} \propto \Sigma_{\rm g}$. 
This is not visible in Figure \ref{fig:sf_law} because of the paucity of the data points for $\Sigma_{\rm g} \gtrsim 10^{4}$ M$_{\odot}$ pc$^{-2}$ (eq. \ref{Sigma g rad}) but is relevant for galactic nuclei \citep[][]{2005ApJ...630..167T}. 
Observations of a sample of local active galactic nuclei by \cite{2009ApJ...696..448H} in fact support such a flattening in the $\dot{\Sigma}_{\star}-\Sigma_{\rm g}$ relation at $\Sigma_{\rm g} \sim 10^{4}$ M$_{\odot}$ pc$^{-2}$. 

\subsection{Disc-averaged star formation efficiency}
\label{star formation efficiency observations}
It is also useful to compare the star formation efficiency per free fall time in the disc predicted by equation (\ref{eta vs eps}) to the observations. 
We again compare with the galaxies compiled by \cite{2010MNRAS.407.2091G} and \cite{2012arXiv1211.5743T}, but restrict ourselves to systems with $\Sigma_{\rm g} > 100$ M$_{\odot}$ pc$^{-2}$ in the rest of this section. 
\cite{2010MNRAS.407.2091G} and \cite{2012arXiv1211.5743T} calculate the star formation efficiency relative to the dynamical time of the disc, defined as $t_{\rm dyn} \equiv R_{\rm 1/2}/v_{\rm c}$, where $R_{\rm 1/2}$ is the half-light radius and $v_{\rm c}$ is the maximum circular velocity. 
Since our star formation efficiency is defined relative to the free fall time in the disc (\ref{eps ff gal def}), we multiply the values reported by these authors by a factor of 1.14 to use a consistent time scale (this is the ratio $t_{\rm ff}^{\rm disc} / t_{\rm dyn}$ evaluated using eq. (\ref{disk free fall time}), for $Q=\phi=1$). 

We compare the observationally inferred $\epsilon_{\rm ff}^{\rm gal}$ against several independent variables in order to investigate the relative scatter of different relations. 
In Figure \ref{fig:eta_comparison}, we show $\epsilon_{\rm ff}^{\rm gal}$ as a function of the observationally-inferred $\sigma$, $f_{\rm g}$, $\Sigma_{\rm g}$, and $t_{\rm dyn}$. 
The isothermal potential velocity dispersion is estimated as $\sigma = v_{\rm c} / \sqrt{2}$, where $v_{\rm c}$ is the maximum observed circular velocity. 
The gas mass fraction is estimated from the observations as
\begin{align}
\label{fg dynamical}
f_{\rm g} = \frac{\pi G \Sigma_{\rm g} R_{\rm 1/2}}{2 \sigma^{2}},
\end{align} 
where $R_{\rm 1/2}$ is the half-light radius. 
This follows from the definition $f_{\rm g} \equiv M_{\rm g} (<R_{\rm 1/2}) / M_{\rm tot}(<R_{\rm 1/2})$ and the simple dynamical model in \S \ref{background disk}. 
This approach is not guaranteed to yield an observationally-estimated $f_{\rm g}<1$, as must be the case physically. 
Figure \ref{fig:eta_comparison} shows that it results in a subset of galaxies with inferred $f_{\rm g} \approx 1-3$. 
We attribute this primarily to scatter due to observational uncertainties in the parameters entering in equation (\ref{fg dynamical}) and to the simplified dynamical model used. 
This small subset of galaxies with $f_{\rm g}>1$ does not significantly affect our discussion. 
We emphasize that it is \emph{not} due to our assumption of an $\alpha_{\rm CO}$ factor varying continuously with $\Sigma_{\rm g}$. 
Adopting a bimodal $\alpha_{\rm CO}$ with ordinary star-forming galaxies at high redshift having a Milky Way-like value as in some other studies \citep[e.g.,][]{2010ApJ...713..686D, 2010MNRAS.407.2091G, 2012arXiv1211.5743T} increases the inferred gas mass for such galaxies and thus \emph{increases} the number of systems with $f_{\rm g}>1$. 
This is shown in Figure \ref{fig:eta_vs_fg_bimodal}, which shows the same data as in the upper right panel of Figure \ref{fig:eta_comparison} but assuming a bi-modal $\alpha_{\rm CO}$ for mergers and non-mergers, as in Genzel et al. (2010) and Tacconi et al. (2012), rather than the continuous $\alpha_{\rm CO}$ from equation (\ref{OS alpha CO}).  

For each relation in Figure \ref{fig:eta_comparison}, we show in the top left corner of the panel the statistic {\tt stddev} calculated by first fitting a linear relation in log-log space, then evaluating the standard deviation of the data points from the best fit. 
The uncertainty on the data points (not shown) is dominated by systematics; since it is difficult to estimate accurately, we assign equal weight to each data point. 

The smallest scatter for the comparison of $\epsilon_{\rm ff}^{\rm gal}$ as a function of a single independent variable in Figure \ref{fig:eta_comparison} is relative to $f_{\rm g}$, for which there is a monotonically increasing relation with {\tt stddev}=0.39.  
Since $\epsilon_{\rm ff}^{\rm gal}$ depends on $\Sigma_{\rm g}$, one may worry that a spurious trend between $\epsilon_{\rm ff}^{\rm gal}$ and $f_{\rm g}$ could result from scatter in the observational estimates. 
In Appendix \ref{dependence of eps ff on fg}, we consider a model prediction equivalent to equation (\ref{eps ff gal alpha infty}) (showing that $\epsilon_{\rm ff}^{\rm gal} \propto f_{\rm g} \sigma$ for $Q\to1$), but expressed in terms of quantities that are measured independently of each other. 
The agreement between the data and the model prediction confirms that the observed trend of increasing $\epsilon_{\rm ff}^{\rm gal}$ with increasing $f_{\rm g}$ is physical. 
For the other variables ($\sigma$, $\Sigma_{\rm g}$, and $t_{\rm dyn}$), the scatter is larger {\tt stddev}=0.48-0.54. 
Figure \ref{fig:eta_vs_fg_bimodal} also shows that the scatter in $\epsilon_{\rm ff}^{\rm gal}$ versus $f_{\rm g}$ is substantially increased when a bimodal $\alpha_{\rm CO}$ is assumed ({\tt stddev}=0.55).  This is, we believe, independent evidence that a continuous $\alpha_{\rm CO}$ (as in eq. \ref{OS alpha CO}) is a better approximation.   
Regardless of whether a continuous or bimodal $\alpha_{\rm CO}$ is adopted, the inferred values of $\epsilon_{\rm ff}^{\rm gal}$ vary by more than a factor of 100 for the molecule-rich galaxies in the sample considered. 
We return to this point in \S \ref{role of turbulence}.

In Figure \ref{fig:eta_comparison2}, we show similar comparisons but as a function of combinations of $f_{\rm g}$, $\sigma$, and $\Sigma_{\rm g}$. 
The left panel shows $\epsilon_{\rm ff}^{\rm gal}$ as a function of $f_{\rm g} \sigma$ ({\tt stddev}=0.36). 
The solid line in this panel is the theoretical prediction $\epsilon_{\rm ff}^{\rm gal} \propto f_{\rm g} \sigma$ in equation (\ref{eps ff gal alpha infty}), valid in the limit $\alpha \to \infty$, for $\phi=1$, $P_{\star}/m_{\star} = 3,000$ km s$^{-1}$, and $\mathcal{F}=2$ (as in Figure \ref{fig:sf_law}). 
The agreement between the prediction and the data (both in normalization and in slope) is reasonable given the uncertainties in the data and the simplifications made in deriving this theoretical prediction. 
Moreover, this agreement emphasizes that the large scatter in $\epsilon_{\rm ff}^{\rm gal}$ versus $\Sigma_{\rm g}$ in Figure \ref{fig:eta_comparison} is not simply scatter about a universal value of $\epsilon_{\rm ff}^{\rm gal}$. Rather, the scatter is significantly reduced when the dependence of $\epsilon_{\rm ff}^{\rm gal}$ on $f_{\rm g} \sigma$ is accounted for. 
For finite $\alpha$, as must be the case in reality, the upturn from this simple prediction at large $f_{\rm g} \sigma$ suggested by the data might be explained by an increase of $\epsilon_{\rm int}^{\rm GMC}$ with $\Sigma_{\rm g}$ (see Fig. \ref{fig:equilibrium_quantities_vs_Sigma_g}) or a decrease of $P_{\star}/m_{\star}$ with gas density, both of which are expected theoretically. 
The right hand panel of Figure \ref{fig:eta_comparison2} shows $\epsilon_{\rm ff}^{\rm gal}$ as a function of the best-fit linear combination of $\log_{\rm 10}{f_{\rm g} \sigma}$ and $\log_{\rm 10}{\Sigma_{\rm g}}$. 
The best fit depends weakly on $\log_{\rm 10}{\Sigma_{\rm g}}$ and the {\tt stddev}=0.34 statistic is marginally improved relative to the fit with respect to $f_{\rm g} \sigma$ alone, indicating that $f_{\rm g} \sigma$ is the most important parameter.

While the simple GMC disruption model in equation (\ref{epsGMC Murray}) predicts a modest increase of $\epsilon_{\rm ff}^{\rm gal}$ with $\Sigma_{\rm g}$ at fixed $f_{\rm g} \sigma$ (see Fig.~\ref{fig:equilibrium_quantities_vs_Sigma_g}), a trend of $\epsilon_{\rm ff}^{\rm gal}$ with $\Sigma_{\rm g}$ is not apparent in Figure \ref{fig:eta_comparison}. 
However, the predicted trend as a function of $\Sigma_{\rm g}$ is  weak relative to the scatter in the data points in Figure \ref{fig:eta_comparison}. 
More importantly, the trend is only predicted at fixed $f_{\rm g} \sigma$, while the data points cover $\approx1.5$ dex in $f_{\rm g} \sigma$. 
The prediction of increasing $\epsilon_{\rm ff}^{\rm gal}$ with $\Sigma_{\rm g}$ also relies on the accuracy of the GMC disruption model, which equation (\ref{epsGMC Murray}) undoubtedly oversimplifies. 
The more robust and general prediction of our feedback-regulated model of star formation is that $\epsilon_{\rm ff}^{\rm gal}$ scales with $f_{\rm g} \sigma$ (eq. \ref{eps ff gal alpha infty}), which the data support (Fig. \ref{fig:eta_comparison2}).

\subsection{Turbulent velocity dispersion}
\label{velocity dispersion observations}
Our theory also predicts the turbulent velocity dispersion $c_{\rm T}$ as a function of galaxy properties. 
In the limit $\alpha \to \infty$, $Q \to 1$ and equation (\ref{cT sigma}) implies that
\begin{align}
\label{cT Q 1}
c_{\rm T} = \frac{f_{\rm g} \sigma}{2}~~~~~~~~~~(Q=1).
\end{align}
Since different galaxies have different $f_{\rm g} \sigma$, we expect $c_{\rm T}$ to vary from galaxy to galaxy. 
This is supported by observations. 
In particular, local ULIRGs have $\sigma \approx 200$ km s$^{-1}$ and $f_{\rm g} \approx 0.5$, so that equation (\ref{cT Q 1}) implies that $c_{\rm T} \approx 50$ km s$^{-1}$ in such systems, comparable to what is inferred observationally \citep[][]{1998ApJ...507..615D}.  High-redshift ($z\sim2$) star-forming galaxies have comparably large gas fractions \citep[e.g.,][]{2010ApJ...713..686D, 2010Natur.463..781T} and are also inferred to have velocity dispersions that are elevated \citep[][]{2009ApJ...697..115C, 2011ApJ...733..101G} relative to local spirals, which have $c_{\rm T} \approx 10$ km s$^{-1}$ \citep[][]{2006ApJ...638..797D}.

In the limit $Q \to 1$, our model also predicts $c_{\rm T} \propto \epsilon_{\rm ff}^{\rm gal}$ (eq. \ref{cT vs eta}), a result previously noted by \cite{2011ApJ...731...41O}. 
However, \cite{2011ApJ...731...41O} assumed a constant value $\epsilon_{\rm ff}^{\rm gal} \sim 0.01$ motivated by a combination of observations \citep[][]{2007ApJ...654..304K, 2008AJ....136.2846B, 2009ApJS..181..321E, 2009ApJ...704..842B} and numerical simulations of turbulent gas \citep[][]{2005ApJ...630..250K, 2012ApJ...754....2S}. 
In contrast, we predict that both $c_{\rm T}$ and $\epsilon_{\rm ff}^{\rm gal}$ scale with $f_{\rm g} \sigma$ and that neither quantity is universal, in agreement with \cite{2005ApJ...630..167T}.

\subsection{Fraction of disc gas mass in GMCs}
\label{fGMC observations}
A key aspect of the physical picture presented here is that, viewed from the GMC perspective, the star formation rate is determined primarily by the rate at which GMCs form in galactic discs and is nearly independent of sub-GMC scale physics. In particular,  provided that $\alpha \gg 1$ (see eq. \ref{g Q}), the galactic star formation efficiency per free fall time $\epsilon_{\rm ff}^{\rm gal}$ depends only weakly on the efficiency with which GMCs turn their gas into stars ($\epsilon_{\rm int}^{\rm GMC}$). 
This is possible because for any $\epsilon_{\rm int}^{\rm GMC}$ the GMC formation rate adjusts so as to maintain the value of $\epsilon_{\rm ff}^{\rm gal}$  
set by the balance between momentum injection from stellar feedback and the gravitational weight of the disc gas at the mid-plane (eq. \ref{star formation law momentum balance}). 
For any model of the dependence of the GMC lifetime $t_{\rm GMC} = \tilde{t}_{\rm GMC} t_{\rm ff}^{\rm disc}$ and $\epsilon_{\rm int}^{\rm GMC}$ on galaxy properties, this translates into a prediction for $f_{\rm GMC}$, the mass fraction of the disc gas collapsed into gravitationally-bound GMCs at any given time (see Fig. \ref{fig:equilibrium_quantities} and \ref{fig:equilibrium_quantities_vs_Sigma_g} for specific examples). 

As discussed in \S \ref{two zone disk}, GMCs are defined in our model as gravitationally-bound clouds. 
This definition does not necessarily coincide with clouds identified based on their molecular gas content in the Milky Way or other Local Group galaxies. 
The properties of GMCs defined in this way also differ from clouds identified in quasi-virial equilibrium. 
In general, the mass fraction of gas in gravitationally-bound GMCs is greater than the mass fraction in quasi-virialized clouds, and their lifetime as gravitationally-bound clouds is also longer than the lifetime as a quasi-virialized clouds. 
In a purely molecular ISM, gravitational boundedness provides a well-defined criterion for identifying GMCs. 
When observations only permit measurement of the mass fraction in a more advanced state of gravitational collapse, our predictions for $f_{\rm GMC}$ should however be interpreted as upper limits to the mass fraction in such dense clouds. 
It is worth noting that $f_{\rm GMC}$ (or a variant) is often an input into analytic models, but in this work we derive $f_{\rm GMC}$ and show how it depends on galaxy properties.\footnote{Other models make predictions for the partition of the gas mass in atomic and molecular phases \citep[e.g.,][]{2009ApJ...693..216K}, but do not distinguish between gravitationally bound and unbound gas in purely molecular media as our model does \citep[see also][]{2010ApJ...721..975O}.}

We consider now the simple scenario described more quantitatively in \S \ref{GMC consistency} and Figure \ref{fig:equilibrium_quantities_vs_Sigma_g}, in which the GMC lifetime scales with the disc free fall time and GMCs are dispersed by radiation pressure on dust.  
In this scenario, $\epsilon_{\rm int}^{\rm GMC}$ saturates at a value $\sim 0.35$ as GMCs become optically thick to far-infrared photons. 
Quantitatively, this argument depends on the detailed properties of GMCs and how effectively re-processed far-infrared radiation is trapped by scattering on dust grains. 
Thanks to its scaling with $\Sigma_{\rm GMC}$, radiation pressure on dust is the only known stellar feedback process potentially capable of disrupting GMCs in the densest starbursts such as Arp 220. 
Thus, if radiation pressure on dust were less effective than estimated by \cite{2010ApJ...709..191M} \citep[e.g.,][]{2012arXiv1203.2926K}, then GMCs in luminous starbursts would likely have even larger $\epsilon_{\rm int}^{\rm GMC} \sim 1$.\footnote{Powerful outflows from star-forming clumps have been detected in high-redshift star-forming galaxies \citep[][]{2011ApJ...733..101G, 2012ApJ...761...43N}, suggesting that stellar feedback is effective at dispersing them. Winds driven by young super star clusters are also observed to unbind cluster gas in the prototypical  merging galaxies NGC 4038/4039 \citep[the Antennae;][]{2007ApJ...668..168G}.} 
A robust prediction is therefore that in luminous starbursts $\epsilon_{\rm int}^{\rm GMC}$ should exceed the Milky Way value $\epsilon_{\rm int}^{\rm GMC} \sim 0.1$ \citep[e.g.,][]{1997ApJ...476..166W, 2011ApJ...729..133M} by a factor $\sim 3-10$.

Consider the example of the local ULIRG Arp 220 ($\Sigma_{\rm g} \sim 10^{4}$ M$_{\odot}$; Scoville et al. 1997\nocite{1997ApJ...484..702S}). 
We fiducially assume $\tilde{t}_{\rm GMC} \approx 1$
and estimate $\epsilon_{\rm ff}^{\rm gal} \approx 0.05$ using the data compiled in \cite{2010MNRAS.407.2091G}. For this galaxy, the GMC disruption theory of \cite{2010ApJ...709..191M} predicts $\epsilon_{\rm int}^{\rm GMC} \approx 0.35$. 
Therefore, equation (\ref{eta consistency}) implies $f_{\rm GMC} \approx 0.14$, in agreement with the more accurate numerical solutions in Figure \ref{fig:equilibrium_quantities_vs_Sigma_g}. 

Interestingly, high-resolution aperture synthesis CO observations of the nuclear gas disc in Arp 220 indicate a high area filling factor $\sim0.25$, suggesting that the molecular gas is more uniformly distributed than in less extreme systems \citep[][]{1997ApJ...484..702S}. 
In contrast, the volume-filling factor of molecular clouds in the inner Galaxy is $\sim0.005$ \citep[][]{2010ApJ...723..492R, 2012ARA&A..50..531K} and is much lower in the rest of the Milky Way. 
The apparently-smooth CO-emitting gas distribution in ULIRGs in general has been interpreted to be connected to the smaller $\alpha_{\rm CO}$ conversion factor in those systems \citep[e.g.,][]{1997ApJ...478..144S, 1998ApJ...507..615D}, in qualitative agreement with the prediction that $f_{\rm GMC}$ is relatively small. 
It must be noted, though, that there is a possible alternative interpretation. 
In the central $\sim 100$ pc of ULIRGs, the mean gas density is typically $n_{\rm H} \sim 10^{4}$ cm$^{-3}$ (see eq. \ref{gas density}) and nearly all the gas is in molecular form. 
Thus, even if a dominant mass fraction of the gas is collapsed in gravitationally-bound GMCs ($f_{\rm GMC} \sim 1$), it is conceivable that the CO-emitting area filling factor appears unusually large simply because of the inter-cloud medium is sufficiently dense to be optically thick. 
More detailed modeling 
is needed to distinguish these possibilities observationally. 

High-redshift ($z\sim2$) star-forming galaxies, with circular velocities and gas fractions comparable to local ULIRGs, also appear to have relatively low $f_{\rm GMC}$. 
Generally, $f_{\rm GMC}$ is not directly measured since most existing observations instead trace the star formation rate or stellar mass of high-redshift galaxies. 
Estimates based on cosmological simulations suggest $f_{\rm GMC} \approx 0.1-0.2$ \citep[][]{2010MNRAS.404.2151C}, although in our picture this quantity is sensitive to feedback parameters, which are highly uncertain in existing simulations. 
Recent observations with the Hubble Space Telescope however indicate that at most a fraction $\approx 0.02-0.07$ of the total stellar mass of such galaxies is contained in clumps. 
Integral field measurements by \cite{2011ApJ...733..101G} furthermore find no prominent kinematical imprint at the locations of star-forming clumps, indicating that they cannot be dominant by mass. 

The nuclei of local gas-rich spirals offer another promising opportunity to test our $f_{\rm GMC}$ predictions. 
Galactic nuclei can have gas surface densities $\Sigma_{\rm g}>100$ M$_{\odot}$ pc$^{-2}$ and have the advantage of being numerous in the nearby Universe. 
They can thus be observed at spatial resolution sufficient to resolve their molecular clouds. 
\cite{2005ApJ...623..826R} observed the nucleus of M64, where $\Sigma_{\rm g} \approx 160$ M$_{\odot}$ pc$^{-2}$ over a $\sim 2$ kpc diameter region. 
In spatially-resolved studies like this one, optically thick emission from $^{12}$CO can be used to infer to total gas mass, while transitions of molecules such as $^{13}$CO (only marginally optically thick) and HCN (with a critical density for the $J=1\to0$ transition of $n_{\rm H}\sim10^{5}$ cm$^{-3}$) can be used to identify the locations of GMCs. 
The masses of GMCs can then be derived from measurements of their velocity dispersions and radii. 
In the future, such studies could be extended to statistical samples using existing interferometers such as the Combined Array for Research in Millimeter-wave Astronomy (CARMA)\footnote{http://www.mmarray.org/} and ALMA. 

Independent support for our physical picture is provided by numerical simulations. 
\cite{2011MNRAS.417.1318D} performed simulations of galactic discs with supernova feedback and varied the feedback energy per molecular gas mass formed via their $\epsilon$ parameter. 
In an experiment in which $\epsilon$ was increased from 0.05 to 0.2, the fraction of disc gas mass contained in GMCs is reduced by a factor of approximately two. 
For a fixed gas mass in which the feedback kinetic energy is injected, the momentum returned per GMC mass formed scales as $\propto \sqrt{\epsilon}$ in the simulations of \cite{2011MNRAS.417.1318D}. Thus, the experiment of \cite{2011MNRAS.417.1318D} is consistent with the prediction in equation (\ref{eta consistency}) that $f_{\rm GMC}$ should be reduced $\propto (\epsilon_{\rm int}^{\rm GMC})^{-1}$.  
\section{DISCUSSION}
\label{conclusions}

\subsection{Summary of the main results}
We  presented a physical picture connecting  star formation in giant molecular clouds with the global star formation rate in galaxies. 
We focused on galaxies with gas surface density $\Sigma_{\rm g} \gtrsim 100$ M$_{\odot}$ pc$^{-2}$ to simplify the theoretical treatment by avoiding explicit consideration of a multiphase (atomic and molecular) ISM; we expect, however, that our qualitative points apply to galaxies with lower surface density as well. 
The high gas surface density regime includes local starbursts, most star-forming galaxies at redshift $z \gtrsim 1$, and the nuclei of nearby ordinary galaxies. 
Our theory builds on previous models of star formation which focused on feedback on scales of galactic discs \citep[e.g.,][]{1997ApJ...481..703S, 2005ApJ...630..167T, 2011ApJ...731...41O} or GMCs \citep[e.g.,][]{2002ApJ...566..302M, 2009ApJ...703.1352K,2010ApJ...709..191M}. 
In particular, our model explicitly determines the relationship between the formation and destruction of GMCs and the galaxy-averaged star formation rate. 

In our theory, the star formation rate of a galaxy and thus the star formation efficiency per free fall time $\epsilon_{\rm ff}^{\rm gal}$ is set largely by the balance between the gravity acting on the disc gas and the strength of stellar feedback in the bulk of the ISM (see also Thompson et al. 2005\nocite{2005ApJ...630..167T}; Ostriker \& Shetty 2011\nocite{2011ApJ...731...41O}). 
If supernovae dominate the turbulence in the volume-filling ISM, this yields a star formation law $\dot{\Sigma}_{\star} \propto \Sigma_{\rm g}^{2}$ (eq. \ref{star formation law momentum balance}). 
The integrated efficiency with which GMCs convert their gas into stars, $\epsilon_{\rm int}^{\rm GMC}$, also enters into the galaxy-averaged star formation rate but only weakly (in spite of the fact that we assume that all star formation occurs in GMCs). 
This is possible because the \emph{formation rate of GMCs} adjusts itself to ensure that a $\dot{\Sigma}_{\star}-\Sigma_{\rm g}$ relation set by vertical hydrostatic equilibrium in the disc is satisfied, for any reasonable scenario of sub-GMC physics. 
This is realized via the Toomre $Q$ parameter, which can obtain values above the classical stability threshold of unity so that the GMC formation rate is regulated. 
Our physical model is consistent with numerical simulations that find that the  star formation rate of galaxies is not sensitive to the small-scale star formation law \citep[e.g.,][]{2011MNRAS.417..950H, 2012ApJ...754....2S}. 

A key implication of our results is that in very dense galaxies where stellar feedback in GMCs is relatively ineffective and high efficiencies of star formation in GMCs, $\epsilon_{\rm int}^{\rm GMC} \sim 0.35-1$, are expected \citep[e.g.,][]{2010ApJ...709..191M}, the star formation efficiency of GMCs can exceed the galaxy-averaged star formation efficiency $\epsilon_{\rm ff}^{\rm gal}$ by a substantial factor. 
\emph{As a result, the low observed star formation efficiency of galaxies does not depend fundamentally on star formation being slow within GMCs} (although this can be the case in some galaxies). 
While this fact is implicit in feedback-regulated models on galactic scales \citep[e.g.,][]{2005ApJ...630..167T, 2011ApJ...731...41O}, in this work we explicitly derived the quantitative relationship between the disc-averaged and GMC star formation efficiencies.

Our theory makes several predictions. In the limit in which the GMC formation rate is rapidly suppressed as the Toomre $Q$ parameter of the disc exceeds unity ($\alpha \gg 1$; see eq. \ref{g Q}), they are: 
\begin{enumerate}
\item The galaxy-averaged star formation efficiency per free fall time $\epsilon_{\rm ff}^{\rm gal} \propto f_{\rm g} \sigma / (P_{\star} / m_{\star})$, as implied by hydrostatic equilibrium. 
\item The turbulent gas velocity dispersion $c_{\rm T} \approx f_{\rm g} \sigma / 2$ as a result of the regulation to $Q\approx1$. 
\item The fraction of the disc gas mass collapsed in gravitationally-bound GMCs $f_{\rm GMC} \propto (\epsilon_{\rm int}^{\rm GMC})^{-1}$ in order to satisfy the star formation law set by hydrostatic equilibrium in the disc. 
\end{enumerate}

In \S \ref{star formation efficiency observations}, we provide strong observational evidence that $\epsilon_{\rm ff}^{\rm gal} \propto f_{\rm g} \sigma$, in agreement with the first prediction (Fig. \ref{fig:eta_comparison2}). 
The second prediction regarding the velocity dispersion depends only on $Q \approx 1$ and is thus very robust. 
It explains why gas-rich systems like local ULIRGs and high-redshift star-forming galaxies have velocity dispersions $\sim 50-100$ km s$^{-1}$ \citep[e.g.,][]{1998ApJ...507..615D, 2011ApJ...733..101G}, significantly above more gas-poor local galaxies like the Milky Way (e.g., Dib et al. 2006\nocite{2006ApJ...638..797D}; \S \ref{velocity dispersion observations}). 
The third prediction of a relatively small $f_{\rm GMC}$ in dense galaxies where larger GMC star formation efficiencies
are expected theoretically tentatively explains the smooth gas distributions and low CO conversion factors inferred in local ULIRGs (e.g., Scoville et al. 1997, Downes \& Solomon 1998\nocite{1997ApJ...484..702S, 1998ApJ...507..615D}) and the small mass fraction in giant clumps in high-redshift galaxies (Genzel et al. 2011; Wuyts et al. 2012; \S \ref{fGMC observations}). 

We note that our main results can be generalized to feedback mechanisms other than supernovae and radiation pressure (such as photoionization or cosmic rays; e.g. McKee 1989, Socrates et al. 2008\nocite{1989ApJ...345..782M, 2008ApJ...687..202S}) by appropriate choices for $P_{\star} / m_{\star}$ and $\epsilon_{\rm int}^{\rm GMC}$. 

\subsection{Role of turbulence}
\label{role of turbulence}
Turbulence has been invoked to regulate star formation in many models. 
In our theory, galactic discs are supported vertically by supersonic turbulence and the formation of GMCs relies on turbulent density fluctuations. 
The role of turbulence in our model is, however, different than in other popular theories of star formation. 
In particular, our theory does not rely on supersonic turbulence being maintained within GMCs, an assumption that \cite{2005ApJ...630..250K} used to predict a nearly universal volumetric star formation law of $\sim 0.01$ of the gas mass turned into stars per free fall time in molecular gas \citep[see also][]{2011ApJ...730...40P, 2012arXiv1209.2856F}. 
As mentioned above, our theory allows for star formation efficiencies within GMCs as high as $\epsilon_{\rm int}^{\rm GMC} \sim \epsilon_{\rm ff}^{\rm GMC} \sim 1$. 
There is in fact some tentative evidence in the Milky Way that some GMCs have star formation efficiencies significantly larger than predicted by theories based on the small-scale properties of supersonic turbulence \citep[e.g.,][and references therein]{2011ApJ...729..133M}. 
We suspect that this is even more likely to be true in denser galaxies. 

Even if supersonic turbulence is maintained within GMCs, simulations show that in the absence of a mechanism to support the galactic disc on large scales (such as stellar feedback), self-gravitating regions collapse until their free fall time is very short. 
Therefore, even if the small-scale star formation efficiency is low, the star formation efficiency per galactic free fall time $\epsilon_{\rm ff}^{\rm gal}$ is $\sim1$ in the absence of global support \citep[e.g.,][]{2010MNRAS.409.1088B, 2012MNRAS.421.3488H}. 

Our theory predicts that the volumetric star formation law is not universal either on galaxy scales ($\epsilon_{\rm ff}^{\rm gal}$) or GMC scales ($\epsilon_{\rm ff}^{\rm GMC}$). 
This is seemingly in conflict with the observational evidence compiled by \cite{2012ApJ...745...69K} in favor of a volumetric star formation law with a universal efficiency of $\sim 0.01$ per free fall time. 
We note that the overall scatter in the $\epsilon_{\rm ff}^{\rm gal}$ values inferred by \cite{2012ApJ...745...69K} for the molecule-rich galaxies considered in this work is  comparable to the factor $\sim 100$ we find (Figs \ref{fig:eta_comparison}-\ref{fig:eta_comparison2}), regardless of whether we assume a continuously-varying or bimodal $\alpha_{\rm CO}$ conversion factor (Fig. \ref{fig:eta_vs_fg_bimodal}). 
The observational analysis of \cite{2012ApJ...745...69K} also includes ordinary local group galaxies and individual molecular clouds within the Milky Way, and thus covers a larger dynamic range than ours. The linear relationship between $\dot{\Sigma}_{\star}$ and $\Sigma_{\rm g} / t_{\rm ff}$ over this large dynamic range likely reflects the fact that the star formation rate on average does scale inversely  with the local free fall time, as advocated by those authors. 
Averaged over galaxies, however, Figure \ref{fig:eta_comparison2} shows that $\epsilon_{\rm ff}^{\rm gal}$ is \emph{not} well described by a constant value with large scatter. Instead, the star formation efficiency $\epsilon_{\rm ff}^{\rm gal}$ correlates well with $f_{\rm g} \sigma$ (particularly for our assumption of a continuously-varying $\alpha_{\rm CO}$), in agreement with our feedback-regulated theory (eq. \ref{eps ff gal alpha infty}) but inconsistent with a universal volumetric star formation efficiency. 

Finally, we note that our calculations suggest that stellar feedback can drive the turbulence required to explain the observed star formation law and gas velocity dispersions, without the need for additional sources such as cosmological accretion and radial inflow within galactic discs. 
Significant contributions from such sources cannot, however, be ruled out by our arguments.

\subsection{Opportunities for numerical modeling}
The analytic arguments presented here are subject to some uncertainties, which could be addressed by numerical modeling. 
One significant uncertainty is exactly how the GMC formation rate is regulated in a turbulent disc, especially when $Q>1$ in the disc-averaged sense. 
We parameterized this uncertainty by the dimensionless function $f_{\rm coll} \propto Q^{-\alpha}$ (eq. \ref{g Q}). 
For $\alpha \to \infty$, $Q \to 1$ and previous results based on the assumption that $Q=1$ are recovered (\S \ref{GMC consistency}), but this is not always accurate. 
Thus, it is important to better quantify the function $f_{\rm coll}$ in order to determine the accuracy of previous results. 
Analytic estimates suggest that $\alpha \approx 2-5$ (Appendix \ref{collapsed fraction appendix}; Hopkins 2012a\nocite{2012arXiv1210.0903H}). 
However, this is sensitive to density fluctuations on a scale comparable to the disc scale height. 
These fluctuations in turn depend on how turbulence is driven and the effects of disc rotation. 

Similarly, we parameterized stellar feedback processes by the effective momentum injected into the ISM per stellar mass formed, $P_{\star} / m_{\star}$, and carried out our numerical calculations using simple estimates for supernovae and radiation pressure on dust.   
These remain uncertain at the factor of few level. 
An important uncertainty for radiation pressure is the role radiation-hydrodynamic instabilities in facilitating the leakage of scattering photons. 
This problem has received some attention recently \citep[e.g.,][]{2012arXiv1203.2926K}, but direct simulations have only been performed in specific idealized settings, such as a laminar background disc. 

While the dynamics of individual supernova remnants have been the subject of many studies \citep[e.g.,][]{1988ApJ...334..252C, 1991ApJ...383..621D, 1998ApJ...500...95T} the collective impact of multiple SNRs in realistic galactic discs does not follow simply from these studies. 
The effective $P_{\star} / m_{\star}$ depends not only the ambient density, but also the stage of SNR evolution at which it is evaluated. 
It is not clear how to accurately calculate this quantity in a real galaxy with an inhomogeneous ISM. 
Fortunately, the exact value of $P_{\star}/m_{\star}$ appears to depend only weakly on these details (see Appendix \ref{model closure appendix}).  
On the other hand, we also showed in \S \ref{disk averaged SF law} that the turbulent pressure in the ISM depends on the spatial scale on which the turbulence is driven relative to the disc scale height, the dimensionless factor $f_{\rm h} = L / h$. 
Thus, the turbulent pressure depends on how SNRs merge with the ambient ISM and/or interact with one another, a problem which controlled numerical experiments could address. 

\section*{Acknowledgments}
We thank Reinhard Genzel and Linda Tacconi for providing their compilation of gas and star formation rate measurements in electronic form. 
Ken Shen performed simulations of supernova remnants that informed our discussion of the momentum input by supernovae. 
We are also grateful to Leo Blitz for a discussion on the possibility of measuring $f_{\rm GMC}$ in local galaxies and to Jacob Lynn for help with understanding large-scale turbulent fluctuations in supersonic turbulence.
CAFG was supported by a fellowship from the Miller Institute for Basic Research in Science and NASA grant 10-ATP10-0187. 
EQ was supported by a Simons Investigator award from the Simons Foundation, the David and Lucile Packard Foundation, and the Thomas Alison Schneider Chair in Physics at UC Berkeley. 
Support for PFH was provided by NASA through Einstein Postdoctoral Fellowship Award Number PF1-120083 issued by the Chandra X-ray Observatory Center, which is operated by the Smithsonian Astrophysical Observatory for and on behalf of the NASA under contract NAS8-03060.

\appendix

\section{Feedback parameters}
\label{model closure appendix}
An important parameter in the calculations of the main text is the effective momentum returned to the ISM per stellar mass formed, $P_{\star} / m_{\star}$. 
Here we summarize relevant results on the importance of supernovae and radiation pressure on dust. 
Other processes -- including HII regions, stellar winds, and proto-stellar winds -- can be important in dwarf or Milky Way-like galaxies but are not effective at the densities considered in this work \citep[][]{2002ApJ...566..302M, 2010ApJ...709..191M}. 

We distinguish between stellar feedback in the volume-filling ISM and in GMCs because the processes that dominate turbulence driving in the volume-filling ISM are generally different from those responsible for disrupting GMCs. 
In what follows, the superscript `disc' corresponds to a process operating directly in the volume-filling ISM, while the superscript `GMC' corresponds to a process operating within a GMC. 

\subsection{Supernovae}
\label{supernova feedback}
The first SNe explode a time $t_{\rm SN,1st}\approx 3.6$ Myr following a star formation event, which in dense starbursts is after the parent GMC has been dispersed ($t_{\rm SN,1st} > t_{\rm GMC}$; see eq. \ref{disk free fall time}). 
It follows that SNe typically explode in the volume-filling phase of the ISM. 
We are also interested in high-redshift, gas-rich star-forming galaxies, for which the lifetime of the massive star-forming clumps are inferred to be $\approx 100-200$ Myr \citep[][]{2012ApJ...753..114W}. 
In these systems, it is not obvious that most SNe explode outside their parent GMC. 
In the Milky Way, where GMC lifetimes also exceed $t_{\rm SN,1st}$, it is nonetheless observed that GMCs often start to be disrupted prior to the onset of SNe \citep[][]{2010ApJ...719.1104R}. 
By analogy to the Milky Way, we assume that this is also the case in high-redshift galaxies. 

In the Sedov and pressure-driven snow plow phases of supernova remnants (SNRs), the momentum of the swept up ambient medium can reach values exceeding that of the original supernova ejecta by a factor
\begin{align}
\label{SNe P boost}
\frac{P_{\rm SN}}{P_{\rm SN,0}} 
& \approx 
50~ \left( \frac{E_{\rm SN}}{\rm 10^{51}~erg~s^{-1}} \right)^{-1/14} \\ \notag
& ~~~~~~~~~~ \times \left( \frac{n_{\rm H}}{\rm 1~cm^{-3}} \right)^{-1/7} 
\left( \frac{v_{\rm SN}}{\rm 10,000~km~s^{-1}} \right)
\end{align}
\citep[][]{1988ApJ...334..252C}, where $E_{\rm SN}$ is the kinetic energy of the supernova ejecta, $v_{\rm SN}$ is their velocity, $n_{\rm H}$ is the density of the ambient medium, and the exact pre-factor depends weakly on metallicity. 
Noting that the total rate of kinetic energy injection by supernovae is $\dot{E}_{\rm SN} \approx 0.01 L_{\star}$ \citep[e.g.,][]{1999ApJS..123....3L}, where $L_{\star}$ is the stellar bolometric luminosity, 
\begin{align}
\label{PSNR over Lc}
\dot{P}_{\rm SN} \approx 30 \frac{L_{\star}}{c}~\left( \frac{E_{\rm SN}}{\rm 10^{51}~erg~s^{-1}} \right)^{-1/14} \left( \frac{n_{\rm H}}{\rm 1~cm^{-3}} \right)^{-1/7}.
\end{align}

We define $m_{\star}$ to be the total mass of stars formed per supernova event and fiducially adopt $m_{\star}=100$ M$_{\odot}$ \citep[e.g.,][]{2011ApJ...731...41O}. In this case, equation (\ref{PSNR over Lc}) corresponds to  
\begin{align}
\label{Pstar mstar SN disk}
\left( \frac{P_{\star}}{m_{\star}} \right)_{\rm SN}^{\rm disc} & = 4,800~{\rm km~s^{-1}}  \xi \left( \frac{E_{\rm SN}}{\rm 10^{51}~erg} \right)^{13/14} \\ \notag
& \times \left( \frac{n_{\rm H}}{\rm 1~cm^{-3}} \right)^{-1/7} 
\left( \frac{v_{\rm SN}}{\rm 10,000~km~s^{-1}} \right) f_{\rm SN}^{\rm disc},
\end{align}
where $f_{\rm SN}^{\rm disc}$
is the fraction of SNe exploding in the volume-filling ISM. 
We have introduced the dimensionless pre-factor $\xi \sim 1$ to parameterize the uncertainty in the normalization of equation (\ref{Pstar mstar SN disk}). 
In an inhomogeneous medium, the relevant density at which to evaluate equation (\ref{Pstar mstar SN disk}) is the typical density where SNe explode. 
When massive stars outlive their parent GMCs, we can assume that SNe explode in random locations in the ISM. 
For a supersonically turbulent medium with Mach number $\mathcal{M}$, we define the effective density $\rho_{\rm eff}$ such that 50\% of the volume has density $<\rho_{\rm eff}$. 
In Appendix \ref{supersonic turbulence}, we show that $\rho_{\rm eff} \approx 0.06 \bar{\rho} (\mathcal{M}/30)^{-1.2}$ for a lognormal density probability distribution function (PDF). 
Thus, the effectiveness of supernova feedback in a dense ISM is not as reduced as would be inferred by evaluating equation (\ref{Pstar mstar SN disk}) at the mean density. 

Equations (\ref{SNe P boost}-\ref{Pstar mstar SN disk}) assume a value for the momentum boost valid at infinity. 
\cite{1988ApJ...334..252C} show, however, that convergence to this asymptotic value is slow and that it is likely not fully realized in practice. 
In \S \ref{disk averaged SF law}, we argued that for our purposes $P_{\star} / m_{\star}$ should be evaluated when the SNR velocity equals the turbulent gas velocity in the ISM, $v_{\rm SNR} = c_{\rm T}$. 
We also evaluated $P_{\star} / m_{\star}$ using the semi-analytic approximations of \cite{1991ApJ...383..621D} and found that $P_{\star} / m_{\star}$ ranges from $1,500$ km s$^{-1}$ to $3,700$ km s$^{-1}$ for $n_{\rm H}=100-10^{4}$ cm$^{-3}$ and final $v_{\rm SNR}=10-50$ km s$^{-1}$. 
The density dependence of $(P_{\star} / m_{\star})_{\rm SN}^{\rm disc}$ is thus sufficiently weak that to a fair approximation it can be assumed constant for most observed galaxies, justifying our fiducial assumption of $(P_{\star} / m_{\star})_{\rm SN}^{\rm disc} \approx 3,000$ km s$^{-1}$. Uncertainty in the precise value is encapsulated in the dimensionless parameter $\mathcal{F}$ defined in equation (\ref{F def}). 

Using the swept up mass and outer shock velocity from the numerical simulation results of \cite{1998ApJ...500...95T}, \cite{2005ApJ...630..167T} estimated $P_{\star}/m_{\star} \approx 1,500~{\rm km~s^{-1}}~(n_{\rm H} / {\rm 1~cm^{-3}} )^{-0.25}$ for SNe (their eq. 11 and associated text, in our notation).  This scaling with density suggested that supernovae become subdominant relative to radiation pressure in the volume-filling dense ISM of starbursts for lower $\Sigma_{\rm g}$ than estimated in \S \ref{rad P volume filling}. 
The \cite{1998ApJ...500...95T} results for the late-time momentum of supernova remnants are not, however, consistent with our semi-analytic results based on \cite{1988ApJ...334..252C} and \cite{1991ApJ...383..621D}.  As described above, the latter give a somewhat larger late-time momentum and a weaker dependence on ambient density.   In particular, the \cite{1998ApJ...500...95T} results for the velocity of the supernova shock (their eq. 23, which is evaluated at a time of $13 t_0$, where $t_0$ is the time of peak remnant luminosity) do not appear consistent with those of \cite{1988ApJ...334..252C} and \cite{1991ApJ...383..621D} while other remnant quantities are (e.g., radius, the time $t_0$, etc.). 
New spherically-symmetric simulations confirm the results of \cite{1988ApJ...334..252C} and \cite{1991ApJ...383..621D}  (K. Shen 2013, private communication).  
These new results are also consistent with the SNR momentum inferred from the fits of \cite{1998ApJ...500...95T} using the swept up mass and kinetic energy to calculate the momentum, instead of using the reported shock velocity directly.

\subsection{Radiation pressure on dust}
\label{radiation pressure feedback}
Radiation pressure on dust is the only known mechanism capable of disrupting GMCs in the densest starbursts.  
This is because radiation pressure on dust scales with the surface density of GMCs, so that it remains important to arbitrarily high densities \citep[][]{2010ApJ...709..191M}. 
After the GMC is disrupted, radiation pressure from the surviving massive stars acts on the volume-filling ISM. 

\subsubsection{GMC disruption}
When evaluating the effective $P_{\star} / m_{\star}$ from GMC disruption, we must account for the fraction of the momentum input used to overcome the self-gravity of the GMC. 
As before, we define $\epsilon_{\rm int}^{\rm GMC}$ as the fraction of a GMC mass that is converted into stars and let $v_{\rm f}$ be the terminal velocity of the dispersed GMC gas (after being unbound but before being decelerated by mass loading of diffuse gas). 
Since the dispersed GMC has asymptotic momentum $M_{\rm GMC}(1-\epsilon_{\rm int}^{\rm GMC}) v_{\rm f}$, 
\begin{align}
\label{Pstar mstar rad GMC}
\left( \frac{P_{\star}}{m_{\star}} \right)_{\rm rad}^{\rm GMC} = \frac{(1-\epsilon_{\rm int}^{\rm GMC})}{\epsilon_{\rm int}^{\rm GMC}} v_{\rm f}.
\end{align}
The integrated GMC efficiency $\epsilon_{\rm int}^{\rm GMC}$ and the terminal velocity $v_{\rm f}$ depend on the properties of the GMC and of the embedded stellar clusters. 

\cite{2010ApJ...709..191M} report $\epsilon_{\rm int}^{\rm GMC}$ and $v_{\rm f}$ values for their 1-D models of GMC disruption by radiation pressure on dust. 
For the local ULIRG Arp 220 ($\Sigma_{\rm g} \sim 10^{4}$ M$_{\odot}$ pc$^{-2}$), the local starburst M82 ($\Sigma_{\rm g} \sim 500$ M$_{\odot}$ pc$^{-2}$), and the $z\sim2$ star-forming galaxy BX482 ($\Sigma_{\rm g} \sim 200$ M$_{\odot}$ pc$^{-2}$) equation (\ref{Pstar mstar rad GMC}) implies $(P_{\star}  / m_{\star})_{\rm rad}^{\rm GMC}=82,~32,~{\rm and}~135$ km s$^{-1}$, respectively. 
Assuming the scaling in equation (\ref{Pstar mstar SN disk}) for $(P_{\star} / m_{\star})_{\rm SN}^{\rm disc}$, $(P_{\star} / m_{\star})_{\rm SN}^{\rm disc} > (P_{\star}  / m_{\star})_{\rm rad}^{\rm GMC}$ unless $n_{\rm H} \gtrsim 10^{11}$ cm$^{-3}$, indicating that supernovae in the volume-filing medium dominate over GMC disruption in driving turbulence in the volume-filling ISM in essentially all realistic conditions, in agreement with the analytic argument given by \cite{2011ApJ...731...41O}. 

Some caveats are in order here, since the effective $P_{\star}/m_{\star}$ from GMC disruption depends somewhat on the internal structure of GMCs and their time-dependent evolution. We cannot rule out that GMC disruption may be more important for turbulence in the volume-filling medium than estimated above.    In addition, once the GMC starts to disrupt, the massive stars continue to input momentum into the ambient medium, as discussed in \S \ref{rad P volume filling}.   This may be more important than GMC disruption in driving turbulence in the ISM.\footnote{In their numerical simulations, \citet{2011MNRAS.417..950H} found that radiation pressure alone was capable of regulating the star formation law  to be comparable to that found in Kennicutt's (1989) observational sample.   We note, however, that the version of the KS law  \citet{2011MNRAS.417..950H} compared to -- from \citet{1989ApJ...344..685K} -- is higher in normalization at $\Sigma_{\rm g} = 100$ g cm$^{-2}$ by a factor of $\sim 3$ than the data utilized in this paper.   As a result, the  \citet{2011MNRAS.417..950H} radiation pressure feedback only simulations of high star formation rate galaxies (e.g., their Sbc and HiZ models) would have star formation rates larger than the data utilized in this paper by a factor of $\sim 3-5$.   However, the subsequent models by \citet{2012MNRAS.421.3488H} include additional feedback processes, including SNe, and are in much better agreement with the data for the KS law used here.}

Giant clumps in high-redshift star-forming galaxies occupy an interesting region of parameter space in which photoionization is ineffective \citep[][]{2002ApJ...566..302M, 2010ApJ...709..191M}, radiation pressure is only marginally capable of disrupting the clumps (with $\epsilon_{\rm int}^{\rm GMC} \sim 0.35$; \citealt{2010ApJ...709..191M}), but the free fall time in the clumps
\begin{align}
t_{\rm ff}^{\rm cl} \approx {\rm 16~Myr} \left( \frac{M_{\rm cl}}{\rm 10^{9}~M_{\odot}} \right)^{-1/2} \left( \frac{R_{\rm cl}}{\rm 1~kpc} \right)^{3/2}
\end{align}
is sufficiently long that supernovae may play an important role in disrupting them.     This should be studied in more detail in future work.

\subsubsection{Direct action on the volume-filling ISM}
\label{rad P volume filling}
The volume-filling medium can also be optically thick to reprocessed far infrared (FIR) radiation \citep[e.g.,][]{2003JKAS...36..167S, 2005ApJ...630..167T, 2011ApJ...727...97A}.\footnote{The ISM is always optically thick in the UV for the galaxies we consider.} The momentum flux provided by radiation from massive stars is then
\begin{align}
\dot{P}_{\rm rad} = (1+\tau_{\rm IR}^{\rm eff}) \frac{L_{\star}}{c},
\end{align}
where $\tau_{\rm IR}^{\rm eff}$ is the effective IR optical depth. 
The $\tau_{\rm IR}^{\rm eff}$ term enters because confinement of IR photons by multiple scatterings boosts the momentum flux, in a manner analogous to the hot gas confinement that gives rise to the momentum boost in SNRs. 

The IR opacity of dust is temperature-dependent, but peaks at $\kappa_{\rm IR} \sim 5$ cm$^{2}$ g$^{-1}$ in the model of \cite{2003A&A...410..611S} for $T\sim100-1,000$ K ($\kappa_{\rm IR}$ drops sharply for $T\gtrsim1,000$ K owing to dust sublimation). 
Defining the vertical optical depth of the disc as $\tau_{\rm IR} = \Sigma_{\rm g} \kappa_{\rm IR}/2$ for the case of a homogeneous gas distribution,
\begin{align}
\label{tau IR uni}
\tau_{\rm IR} & = \frac{\kappa_{\rm IR} \sigma^{2}}{2 \pi G r} \\ \notag 
& \approx 15.4 \left( \frac{\kappa_{\rm IR}}{{\rm 5~cm^{2}~g^{-1}}} \right) \left( \frac{Z}{Z_{\odot}} \right) \\ \notag 
& ~~~~~~~~~~~~~~~~~~~~ \times
\left( \frac{\sigma}{\rm 200~km~s^{-1}} \right)^{2} \left( \frac{r}{\rm 100~pc} \right)^{-1},
\end{align}
where we scale the result with gas-phase metallicity, $Z$, relative to solar. 
The effective $\tau_{\rm IR}^{\rm eff}$ differs from this estimate because anisotropies in the column density distribution can provide paths of least resistance through which the scattered photons can escape. 
The magnitude of this effect depends on the column density distribution, and also possibly on radiation-hydrodynamic instabilities \citep[e.g.,][]{2012arXiv1203.2926K, 2012arXiv1212.1742J}. 
We parameterize this effect using the dimensionless factor $f_{\rm eff}<1$, such that $\tau_{\rm IR}^{\rm eff} = f_{\rm eff} \tau_{\rm IR}$. 

Let $\epsilon_{\rm rad} \sim 5\times10^{-4}$ be the efficiency with which star formation converts rest mass into radiation. 
Neglecting confinement of the reprocessed radiation, the radiative momentum released is simply $(P_{\star} / m_{\star})_{\rm rad} = \epsilon_{\rm rad} c$. In general, we thus have
\begin{align}
\label{Pstar mstar rad disk}
\left( \frac{P_{\star}}{m_{\star}} \right)_{\rm rad}^{\rm disc} &= (1 + \tau_{\rm IR}^{\rm eff}) f_{\rm rad}^{\rm disc} \epsilon_{\rm rad} c \\ \notag
& = 150~{\rm km~s}^{-1}~(1 + \tau_{\rm IR}^{\rm eff}) f_{\rm rad}^{\rm disc} \left( \frac{\epsilon_{\rm rad}}{5\times10^{-4}}. \right)
\end{align}
where $f_{\rm rad}^{\rm disc}$ 
is the fraction of the radiation emitted in the disc. 
Equation (\ref{Pstar mstar rad disk}) equals $(P_{\star} / m_{\star})_{\rm SN}^{\rm disc}$ when
\begin{align}
\label{Sigma g rad}
\Sigma_{\rm g} & \sim 10^{4}~{\rm M_{\odot}~pc^{-2}} \left( \frac{f_{\rm eff}}{0.5} \right)^{-1} \left( \frac{\epsilon_{\rm rad}}{5\times10^{-4}} \right)^{-1} \left( \frac{Z}{Z_{\odot}} \right)^{-1}  \\ \notag
& ~~~~~~~~~~~~~~~~~~~~~~~~~~~~~~ \times \left( \frac{f_{\rm SN}^{\rm disc}}{f_{\rm rad}^{\rm disc}} \right) \left( \frac{(P_{\star} / m_{\star})_{\rm SN}^{\rm disc}}{\rm 3,000~km~s^{-1}} \right).
\end{align}
Above this gas surface density (corresponding to $\tau_{\rm IR} \sim 10$ for the fiducial parameters above), radiation pressure on dust dominates the vertical pressure support over supernova feedback. 
For this optical depth, the analytic estimates of \cite{2011MNRAS.417..950H} imply a modest inefficiency factor $f_{\rm eff} \sim 0.5$. 
In particular, equation (\ref{Sigma g rad}) implies that radiation pressure is particularly important in galactic nuclei for $r \lesssim 100$ pc (as in, e.g., the Thompson et al. 2005 model). 

\section{Density PDF for supersonic turbulence}
\label{supersonic turbulence}
We assume that the gas density distribution in the ISM is determined by the properties of isothermal supersonic turbulence. Simulations of driven turbulence then imply that the mass-weighted density PDF is well described by a lognormal distribution:
\begin{align}
\label{fM y}
f_{\rm M}(y) dy = \frac{1}{\sqrt{2\pi \sigma_{0}^{2}}} \exp{\left[ -\frac{(y - \mu)^{2}}{2 \sigma_{0}^{2}} \right]} dy,
\end{align}
where $y \equiv \ln{( \bar{\rho} / \bar{\rho} )}$ and $\mu$ is the mean of the distribution \citep[e.g.,][]{1999intu.conf..218N, 2001ApJ...546..980O, 2008ApJ...682L..97L}. 
The constraint that $\langle \rho \rangle = \bar{\rho}$ implies that $\mu= \sigma_{0}^{2} / 2$. 
The volume-weighted PDF, $f_{\rm V}(y)$, is given by the same expression with the simple replacement $\mu \to -\mu$. 

\cite{2008ApJ...688L..79F} showed that the lognormal parameters depend on both the Mach number $\mathcal{M}$ of the turbulence on the driving scale and on the relative importance of solenoidal and compressive driving modes. 
To a good approximation,
\begin{align}
\label{federrath sigma}
\sigma_{0}^{2} \approx \ln{(1 + b^{2} \mathcal{M}^{2})}, 
\end{align}
where $b = 1/3$ for pure solenoidal driving, $b=1$ for pure compressive driving, and intermediate values of $b$ correspond to solenoidal and compressive mixtures. For a random mixture of solenoidal and compressive modes, $b=1/2$. 

\begin{figure}
\includegraphics[width=0.98\columnwidth]{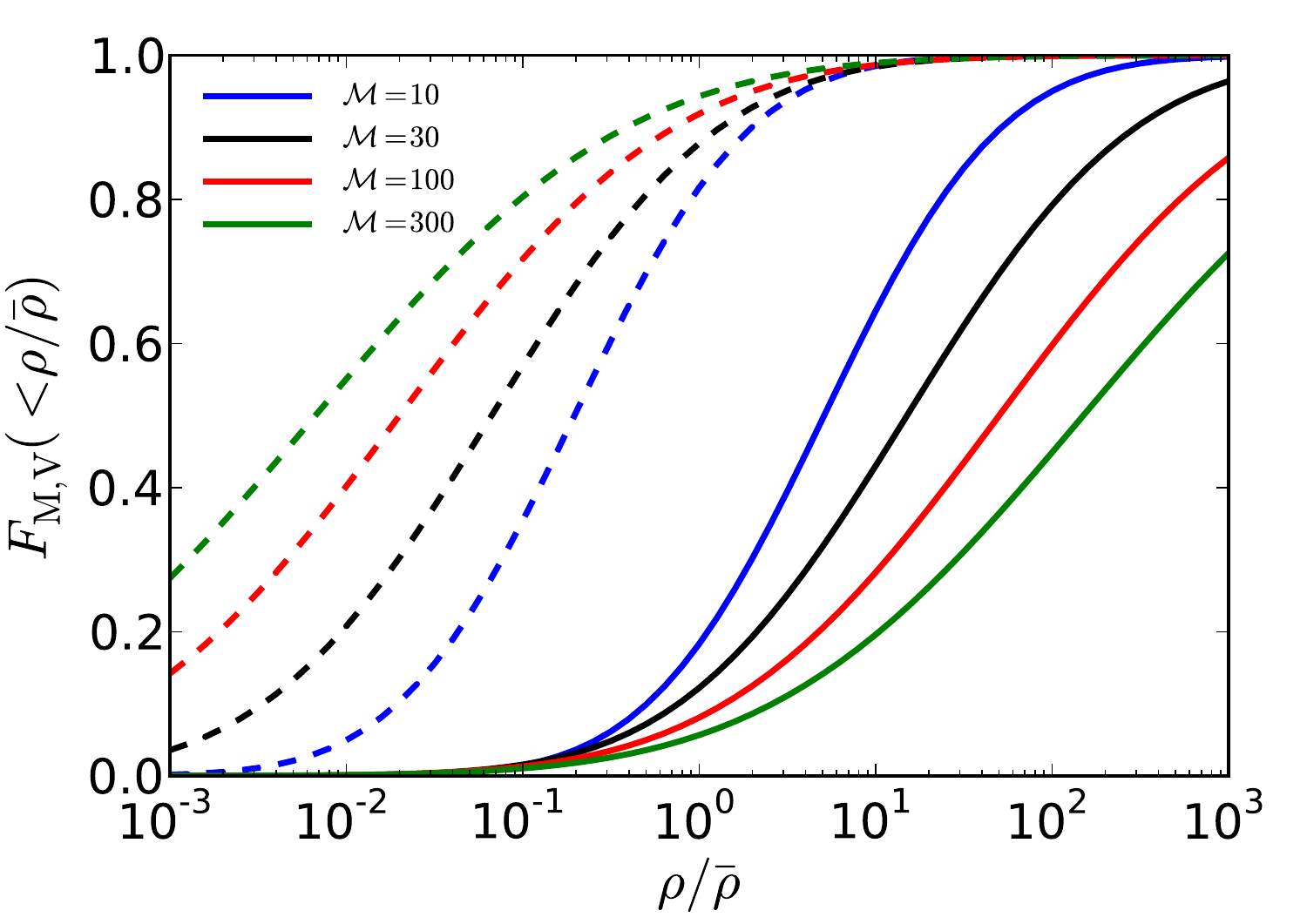}
\caption{
Mass (solid) and volume (dashed) fractions of the gas density distribution for driven isothermal supersonic turbulence for different Mach numbers. 
These values are calculated assuming a lognormal PDF with parameters determined using the prescription of Federrath et al. (2008), for a parameter $b=1/2$ corresponding to a random mixture of solenoidal and compressive driving modes.
}
\label{fig:lognorm_fracs}
\vspace*{0.1in}
\end{figure}

Figure \ref{fig:lognorm_fracs} shows the integrated fractions of the mass and volume contributed by over-densities $< \rho / \bar{\rho}$, as a function of $\rho / \bar{\rho}$.  
We heuristically postulate that SNe acting outside GMCs are embedded in an ISM of effective density $[\rho/\bar{\rho}]_{\rm eff}$ such that $F_{\rm V}(<[\rho/\bar{\rho}]_{\rm eff})=0.5$. 
A simple approximation is that 
\begin{align}
\label{rho eff}
\frac{\rho_{\rm eff}}{\bar{\rho}} \approx  0.06 \left( \frac{\mathcal{M}}{30} \right)^{-1.2}.
\end{align}
For a starburst with Mach number $\mathcal{M}=30$, the effective volume-filling density is therefore about 6\% of the mean ISM density.

\vspace{-0.3cm}
\section{Collapsed fraction in turbulent disc}
\label{collapsed fraction appendix}
We derive here an approximation to the fraction of the gas mass in the turbulent disc that is unstable to gravitational collapse at any time, $f_{\rm coll}$ (see eq. \ref{g def}). 
The approximation is based on the assumption that most of the collapsed mass is contained in the most massive GMCs, corresponding to a spatial scale of order the disc scale height $h$. 
\begin{figure*}
\mbox{
\includegraphics[width=0.999\columnwidth]{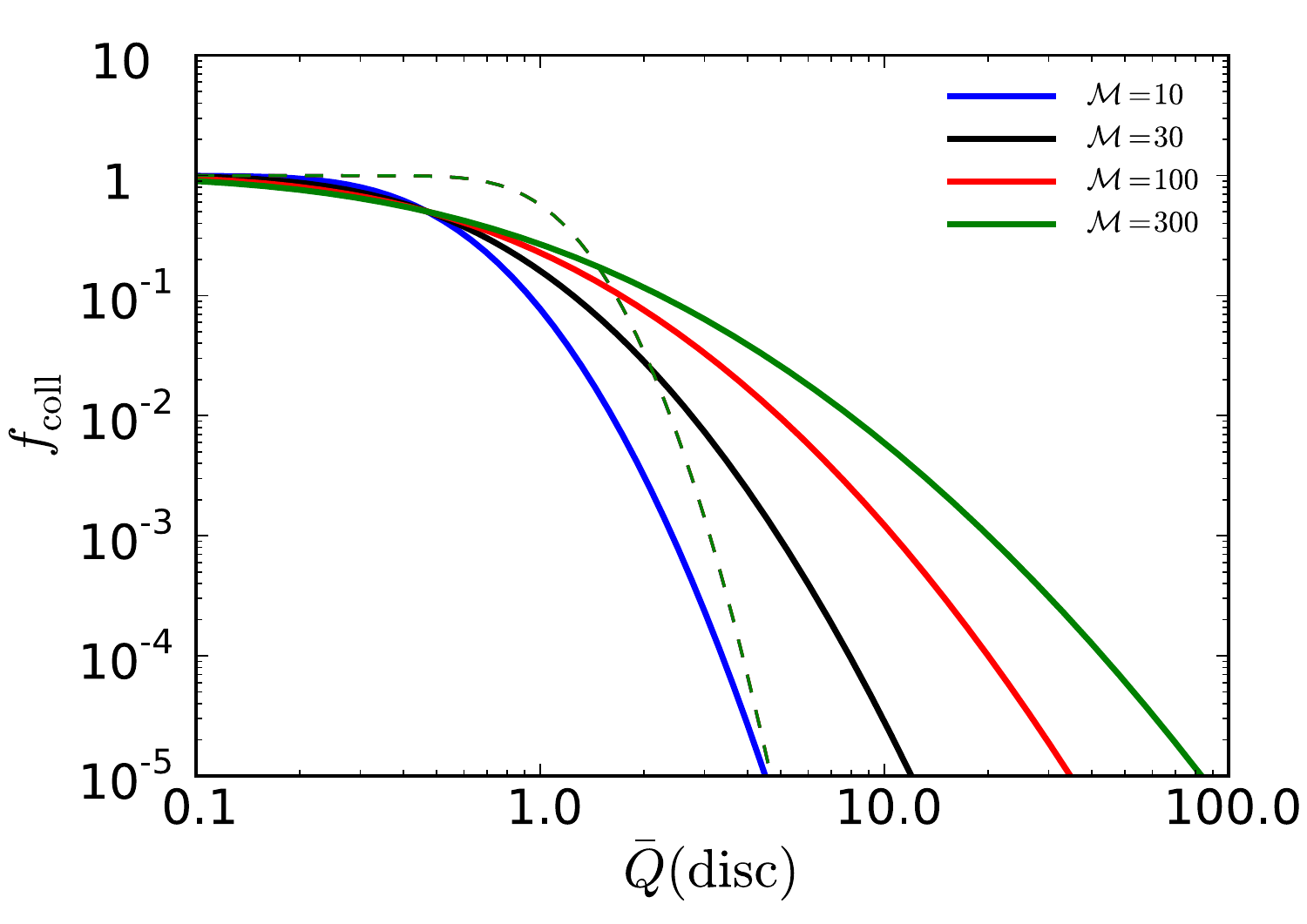}
\includegraphics[width=0.999\columnwidth]{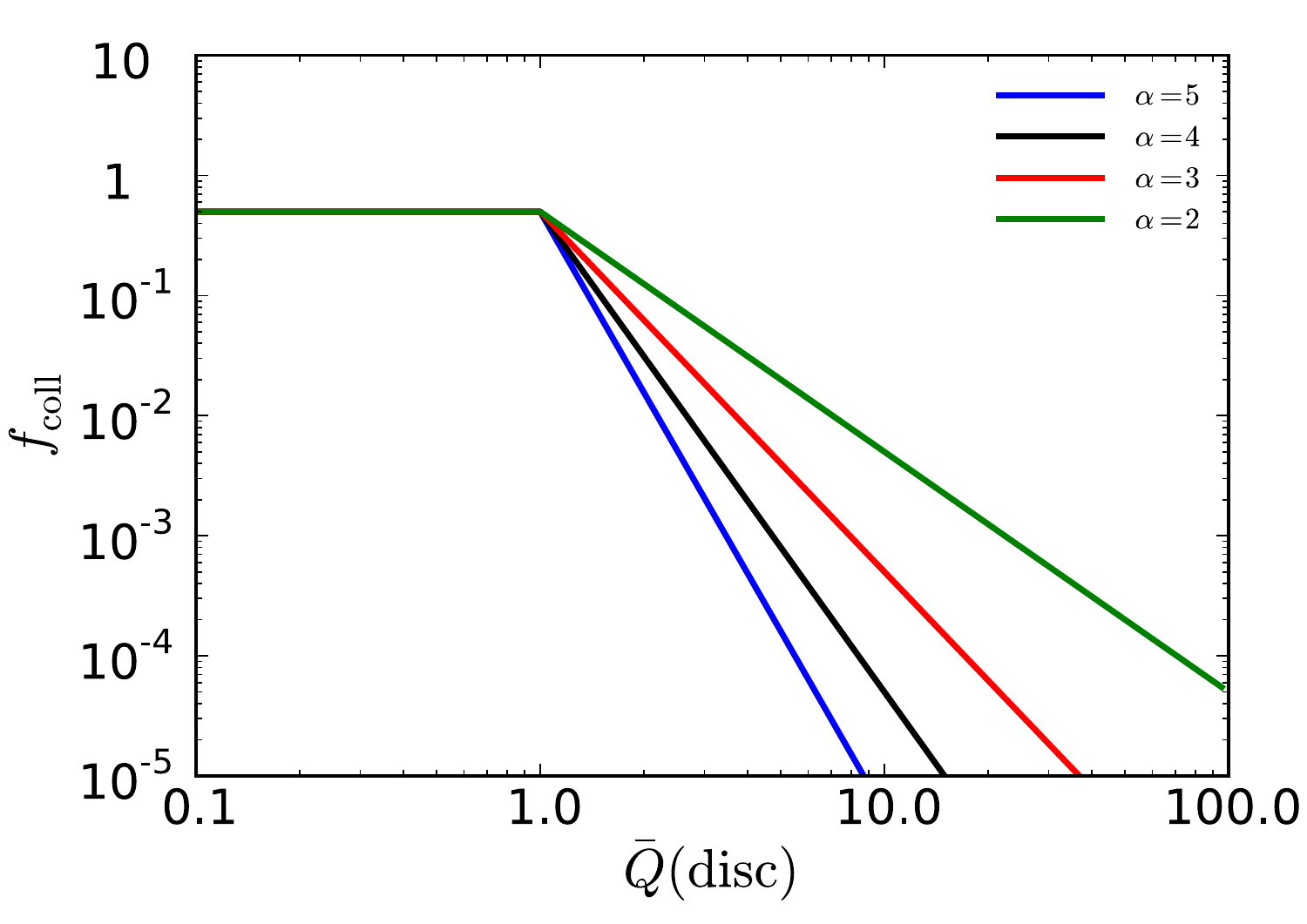}
}
\caption{\emph{Left:} Fraction of the disc gas mass in a turbulent disc that is unstable to gravitational collapse as a function of the disc-averaged Toomre $Q$ parameter. 
The (overlapping) dashed curves show the heuristic approximation  derived in Appendix \ref{collapsed fraction appendix} (eq. \ref{fcoll Toom}) and the solid curves show more accurate results obtained using the excursion set model of Hopkins (2012a) (our eq. \ref{fcoll EPS}).  
These curves assume dimensionless parameters $a=1.5$ and $b=0.5$ in equation \ref{fcoll EPS}. 
\emph{Right:} Power-law approximations used for the analytic estimate in \S \ref{GMC consistency}. These reproduce the more detailed curves in the left panel reasonably well.
}
\label{fig:fcoll}
\vspace*{0.1in}
\end{figure*}

Since $f_{\rm GMC} \sim f_{\rm coll} \tilde{t}_{\rm GMC}$ in a steady-state disc with $Q\sim1$ (eq. \ref{mdot GMC}), $f_{\rm coll} \sim f_{\rm GMC}$ if and only if GMCs survive for a disc free fall time ($\tilde{t}_{\rm GMC} \sim 1$).  
Physically, this arises because if GMCs live longer, then the steady-state mass in GMCs can exceed the instantaneous fraction of the turbulent disc gas mass that is unstable to gravitational collapse.

For a disc with global $\bar{Q}({\rm disc})>1$, we define the collapsed fraction $f_{\rm coll}$ as the fraction of the total gas mass with $\bar{Q}(h)<1$, where $\bar{Q}(h)$ is the $Q$ parameter evaluated after smoothing the gas density distribution on a scale $h$. 
In the main text, $\bar{Q}({\rm disc})$ is abbreviated to simply $Q$. Since there are in general fluctuations on spatial scales $>h$, $\bar{Q}({\rm disc}) \neq \bar{Q}(h)$. 
A more accurate derivation improving the standard Toomre analysis for a turbulent disc is possible using the excursion set formalism \citep[][]{2012arXiv1210.0903H}. 
For our numerical calculations in the main text, we in fact use the analytic approximations derived in \citet[][]{2012arXiv1210.0903H}, but present here a simpler derivation that makes the origin of the qualitative behavior with $\bar{Q}({\rm disc})$ more transparent. 

We must first quantify the gas density fluctuations smoothed over the scale $h$. 
Equation (\ref{federrath sigma}) was derived for simulated turbulence in a static box with periodic boundary conditions and applies to the point PDF. 
In general, it is non-trivial to generalize this expression to the PDF of the density smoothed on a scale $\sim h$ because the latter depends on how the turbulence is driven, and how it is suppressed on large scales by finite mass and rotation effects. 
We adopt a simple model proposed by \cite{2012MNRAS.423.2016H}, in which an analog of equation (\ref{federrath sigma}) is assumed to apply on a $k$-by-$k$ basis:
\begin{align}
\label{sigma k}
\sigma_{k}^{2} = \ln{\left(1 + b^{2} \frac{\mathcal{M}^2(k)}{1 + 2 \mathcal{M}^{2}(k) / |k h|^{2}} \right)},
\end{align}
where $\mathcal{M}(k)$ is the Mach number on scale $k \sim 1 / R$. 
If the turbulence power spectrum $E(k) \propto k^{-p}$, where $c_{\rm T}^{2}(k) \sim k E(k)$, then $\mathcal{M}(k) \propto k^{(1-p)/2}$. 
For a Burgers power spectrum, $p=2$, $\mathcal{M}(k) \propto k^{-0.5}$. 
The real-space variance on scale $R$ is then
\begin{align}
\label{sigma R}
\sigma_{R}^{2} = \int_{0}^{R^{-1}} \frac{dk}{k} \sigma_{k}^{2},
\end{align}
where we assume a top-hat window function in $k-$space. 

The denominator $1 + 2 \mathcal{M}^{2}(k) / |k h|^{2}$ in equation (\ref{sigma k}) was introduced to capture the suppression of turbulent fluctuations on scales $\sim h$ owing to a combination of finite mass and rotation effects. 
It is motivated by the Toomre dispersion relation
\begin{align}
\omega^{2} = \kappa^{2} + c_{\rm T}^{2} k^{2} - 2 \pi G \Sigma_{\rm g} |k|,
\end{align}
which implies that differential rotation in the disc provides an effective pressure $\kappa^2 k^{-2}$ analogous to $c_{\rm T}^{2}$ on scales $\sim h$, so that $\mathcal{M}^{2} = c_{\rm T}^{2} / c_{\rm s}^{2} \to c_{\rm T}^{2} / (c_{\rm s}^{2} + \kappa^2 k^{-2})$. 
The suppression of the density fluctuations on scales $\sim h$ is critical to the strong suppression of $f_{\rm coll}$ for $\bar{Q}({\rm disc})>1$ that we find below. 
However, the exact magnitude and functional form of this suppression is uncertain and an important problem for future work is to quantify this more accurately using numerical simulations. 

The local $Q$ parameter smoothed on a scale $h$ is 
\begin{align}
\bar{Q}(h) = \frac{\sigma c_{\rm T}(h)}{\pi G \bar{\rho}(h) h r}.
\end{align}
Taking $h$ as a disc constant,\footnote{Since $h = [c_{\rm T} / (\sqrt{2 \phi} \sigma)] r$ (eq. \ref{h over r}), this is equivalent to parametrizing the disc by a constant $c_{\rm T}(h)$ and viewing gravitationally collapsed regions as corresponding to upward density fluctuations about the mean $\rho-c_{\rm T}$ relation implied by equation (\ref{sigma R}). In reality, both $c_{\rm T}(h)$ and $\rho(h)$ are likely to fluctuate spatially, possibly with non-trivial correlations. These effects are not captured by our analytic estimate.}
$\bar{Q}(h) < 1$ if and only if
\begin{align}
\frac{\bar{\rho}(h)}{\bar{\rho}({\rm disc})} > \bar{Q}({\rm disc}). 
\end{align}
Thus, the collapsed fraction estimated using this heuristic Toomre-scale argument is
\begin{align}
\label{fcoll Toom}
f_{\rm coll}^{\rm Toom} & \approx f_{\rm M}(y_{h}>\ln{\bar{Q}({\rm disc})}) \\ \notag
& = \frac{1}{2} {\rm erfc} \left[ \frac{\ln{\bar{Q}({\rm disc})} - \mu_{h}}{\sqrt{2} \sigma_{h}} \right],
\end{align}
where the subscript $h$ is used to denote quantities smoothed over a scale $h$.

Using the excursion set (EPS) formalism, \cite{2012arXiv1210.0903H} shows that the ``maximum instability scale'' in a turbulent disc is close to but distinct from $h$. 
\cite{2012arXiv1210.0903H} also obtains a more rigorous estimate of the collapse fraction valid for $\bar{Q}(\rm disc)>1$ and accounting for the full mass spectrum of GMCs: 
\begin{align}
\label{fcoll EPS}
f_{\rm coll}^{\rm EPS} \approx \frac{\mathcal{M}}{2 \sqrt{1 + \mathcal{M}^{2}}} {\rm erfc} \left[ \frac{a \ln{[3\bar{Q}({\rm disc})/\sqrt{2}]}}{\sqrt{2 \ln{[1 + 0.5 b^{2} \mathcal{M}^{2} / \sqrt{2 (1 + \mathcal{M}^{2})}]}}} \right],
\end{align}
where $a \approx 1-2$ is a parameter approximating the complex shape of the collapse barrier. 

The left-hand panel in Figure \ref{fig:fcoll} compares $f_{\rm coll}^{\rm Toom}$ and $f_{\rm coll}^{\rm EPS}$ as a function of $\bar{Q}({\rm disc})$ for several Mach numbers. 
For $\bar{Q}({\rm disc})<1$, the predicted collapsed fraction is $\sim 1$, as expected since the disc is unstable in an average sense. 
For $\bar{Q}({\rm disc})>1$, GMCs only form where turbulent density fluctuations are large enough to locally bring $\bar{Q}(h) < 1$ (in the Toomre heuristic picture). 
Since density fluctuations are suppressed by rotation and finite mass effects on scale $\sim h$, the collapse fraction decreases rapidly as $\bar{Q}({\rm disc})$ is increased above unity. 
The right-hand panel shows power-law approximations used for analytic estimates in \S \ref{GMC consistency}; values of the exponent $\alpha \approx 3-5$ (see eq. \ref{g Q}) provide reasonable approximations for Mach numbers $\mathcal{M}=10-100$. 

\section{Trend of the star formation efficiency with gas fraction}
\label{dependence of eps ff on fg}
 In \S \ref{star formation efficiency observations}, we compiled observations showing that the disc-averaged star formation efficiency increases with increasing gas mass fraction in the disc. 
Since $\epsilon_{\rm ff}^{\rm gal}$ depends on $\Sigma_{\rm g}$, a potential concern is that the observed trend between $\epsilon_{\rm ff}^{\rm gal}$ and $f_{\rm g}$ could be an artifact of scatter in the observational estimates. 
We show here that the observed trend is in fact physical by considering an equivalent relationship between quantities that are measured independently. 

\begin{figure}
\includegraphics[width=0.98\columnwidth]{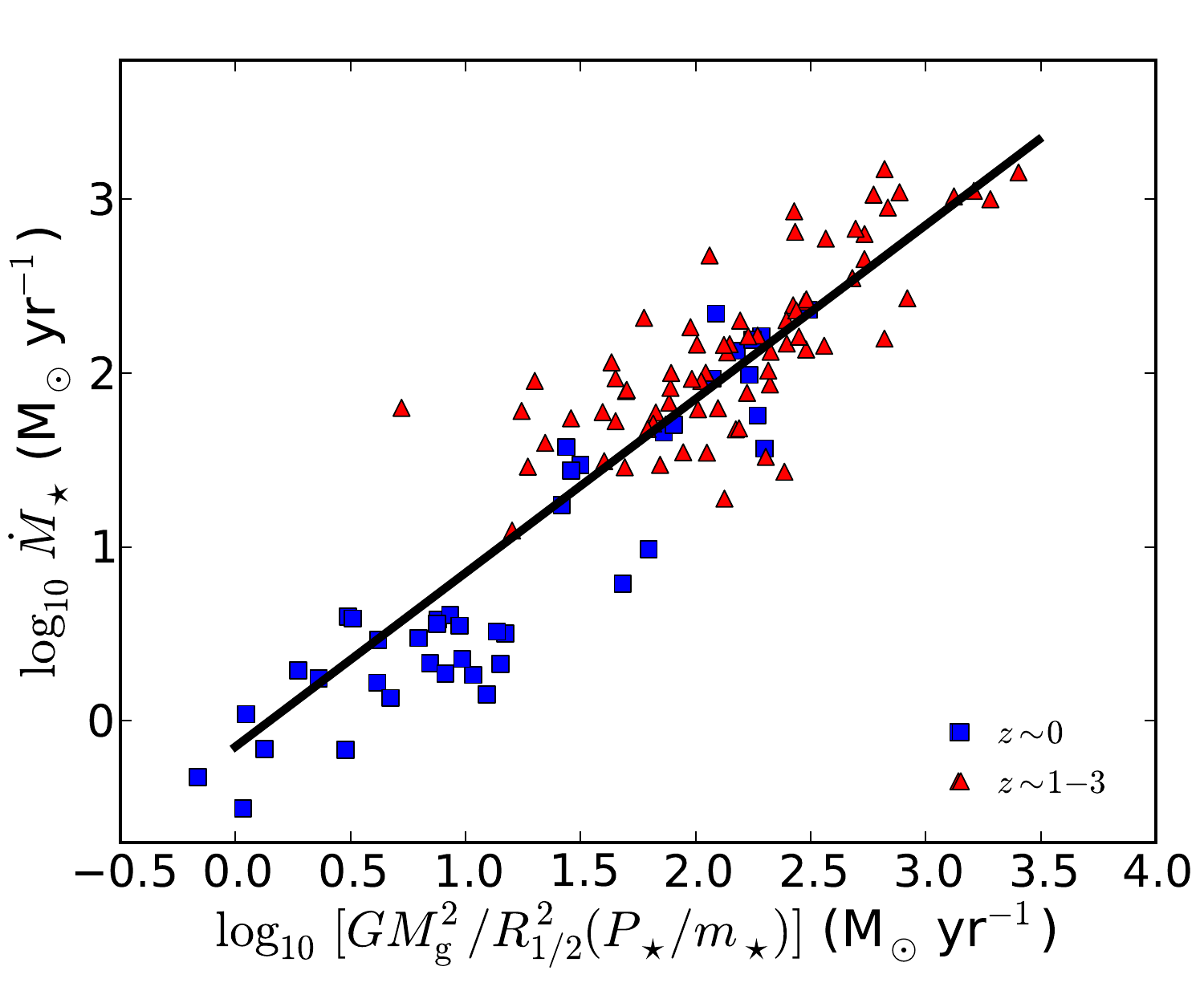}
\caption{Star formation rate  as a function of the independently-measured parameter on the right-hand side of equation (\ref{SFR vs. model eq}), for $P_{\star}/m_{\star}=3,000$ km s$^{-1}$. 
The data are the same as in Figures \ref{fig:eta_comparison} and \ref{fig:eta_comparison2}, and we assume a CO conversion factor varying continuously with $\Sigma_{\rm g}$. 
The solid black curve shows the model prediction for $\mathcal{F}=2$. 
The agreement between the data and this prediction of our feedback-regulated theory confirms that the trend of increasing $\epsilon_{\rm ff}^{\rm gal}$ with increasing $f_{\rm g}$ is not an artifact of common parameter dependencies. 
}
\label{fig:SFR_vs_model}
\vspace*{0.1in}
\end{figure}

We focus on the case $Q=\phi=1$. Then, $t_{\rm ff}^{\rm disc} = 1.14 R_{\rm 1/2} / v_{\rm c}$. Using $M_{\rm tot}  = 2 \sigma^{2} R_{\rm 1/2} / G$, we find that the prediction $\epsilon_{\rm ff}^{\rm gal} \propto f_{\rm g} \sigma$ in equation (\ref{eps ff gal alpha infty}) is equivalent to
\begin{align}
\label{SFR vs. model eq}
\dot{M}_{\star} = \left( \frac{\sqrt{3} \pi}{1.14\times2^{7/4} \mathcal{F}} \right) \frac{G M_{\rm g}^{2}}{R_{\rm 1/2}^{2} (P_{\star} / m_{\star})},
\end{align}
where $\dot{M}_{\star}$, $M_{\rm g}$, and $R_{\rm 1/2}$ are all measured independently. 
Figure \ref{fig:SFR_vs_model} compiles the same observations as in \S \ref{star formation efficiency observations} but for $\dot{M}_{\star}$ as a function of the parameter on the right-hand side of equation (\ref{SFR vs. model eq}). 
The Figure assumes $P_{\star}/m_{\star}=3,000$ km s$^{-1}$, as appropriate for supernova feedback, and a CO conversion factor varying continuously with $\Sigma_{\rm g}$ as before. 
The Figure also shows the model prediction in equation (\ref{SFR vs. model eq}) for $\mathcal{F}=2$, the normalization determined by the observed $\dot{\Sigma}_{\star}-\Sigma_{\rm g}$ relation (Fig. \ref{fig:sf_law}). 
The agreement between the data and this prediction of our feedback-regulated theory confirms that the trend of increasing $\epsilon_{\rm ff}^{\rm gal}$ with increasing $f_{\rm g}$ is not an artifact of common parameter dependencies.

\bibliography{references} 
 
\end{document}